\documentclass[]{emulateapj}

\shorttitle{Evolution of debris disks around F-type stars}

\shortauthors{Mo\'or et al.}

\begin{document}


\title{Structure and evolution of debris disks around F-type stars: \\
       I. Observations, database and basic evolutionary aspects}

\author{A. Mo\'or\altaffilmark{1}}
\email{moor@konkoly.hu}
\author{I. Pascucci\altaffilmark{2}}
\author{\'A. K\'osp\'al\altaffilmark{3}}
\author{P. \'Abrah\'am\altaffilmark{1}}
\author{T. Csengeri\altaffilmark{4}}
\author{L.~L. Kiss\altaffilmark{1,5}}
\author{D. Apai\altaffilmark{2}}
\author{C. Grady\altaffilmark{6,7}} 
\author{Th.~Henning\altaffilmark{8}}
\author{Cs. Kiss\altaffilmark{1}}
\author{D. Bayliss\altaffilmark{9}}
\author{A. Juh\'asz\altaffilmark{8}} 
\author{J. Kov\'acs\altaffilmark{10}}
\author{T. Szalai\altaffilmark{11}}
\altaffiltext{1}{Konkoly Observatory of the Hungarian Academy of Sciences, PO Box 67, H-1525 Budapest, Hungary}
\altaffiltext{2}{Space Telescope Science Institute, 3700 San Martin Drive, Baltimore, MD 21218, USA}
\altaffiltext{3}{Leiden Observatory, Leiden University, Niels Bohrweg 2, NL-2333 CA Leiden, The Netherlands}
\altaffiltext{4}{Laboratoire AIM, CEA/DSM, IRFU/Service d'Astrophysique, 91191 Gif-sur-Yvette Cedex, France}
\altaffiltext{5}{Sydney Institute for Astronomy, School of Physics A28, University of Sydney, NSW 2006, Australia}
\altaffiltext{6}{NASA Goddard Space Flight Center, Code 667, Greenbelt, MD 20771}
\altaffiltext{7}{Eureka Scientific, 2452 Delmer Street, Suite 100, Oakland, CA 94602}
\altaffiltext{8}{Max-Planck-Institut f\"ur Astronomie, K\"onigstuhl 17, 69117 Heidelberg, Germany}
\altaffiltext{9}{Research School of Astronomy and Astrophysics, The Australian National University, Mount Stromlo Observatory, 
Cotter Road, Weston Creek, ACT 2611, Australia}
\altaffiltext{10}{Gothard Astrophysical Observatory, ELTE University, 9707 Szombathely, Hungary}
\altaffiltext{11}{Department of Optics and Quantum Electronics, University of Szeged, 6720 Szeged, D\'om t\'er 9., Hungary}


\begin{abstract}

Although photometric and spectroscopic surveys with the {\sl Spitzer Space Telescope} increased 
remarkably the number of well studied debris disks around A-type and Sun-like stars,  
detailed analyzes of debris disks around F-type stars remained less frequent. 
Using the MIPS camera and the IRS spectrograph we searched for debris dust 
around 82 F-type stars with {\sl Spitzer}. We found 27 stars that harbor debris disks, nine of which are new discoveries. 
 The dust distribution around two of our stars, HD\,50571 and HD\,170773, was found to be marginally extended on the 
70{\micron} MIPS images. Combining the MIPS and IRS measurements with additional infrared and submillimeter data, 
we achieved excellent spectral coverage for most of our debris systems. We have modeled the excess emission of 
22 debris disks using a single temperature dust ring model and of 5 debris systems 
with two-temperature models. The latter systems may contain two dust rings around the star. 
In accordance with the expected trends, the fractional luminosity of the disks declines with time, exhibiting a decay rate consistent with the range of model
predictions. 
We found the distribution of radial dust distances as a function of age to be consistent with the predictions of both the self stirred and the 
planetary stirred disk evolution models. 
A more comprehensive investigation of the evolution of debris disks around F-type stars, partly based on the presented 
data set, will be the subject of an upcoming paper.

\end{abstract}


\keywords{circumstellar matter --- infrared: stars}



\section{Introduction} \label{intro}

Nearly all young stars harbor circumstellar disks, which initially serve as a reservoir for mass accretion, and later can become the 
birthplace of planetary systems. 
During this latter process, the originally submicron-sized dust grains start growing, and their aggregation is
believed to lead to km-size planetesimals \citep[for a review, see][and references therein]{apai2010}. Non-destructive collisions between planetesimals
result in the formation of subsequently larger bodies. These events happen first in the inner disk due to
the shorter collisional timescales, then the process propagates outwards \citep{kenyon2004}.  
The newly formed Pluto-sized protoplanets stir up the motion of leftover smaller bodies in their vicinity,
initializing a collisional cascade. As they become more energetic, collisions result in the erosion of planetesimals and 
the production of small dust grains. An optically thin {\sl debris disk} is formed, in which the second generation, short-lived dust grains
are continuously replenished by collisions and/or evaporation of planetesimals \citep{backman1993,wyatt2008}. 
This {\sl self-stirring} mechanism is not the sole feasible way to incite destructive collisions between minor bodies. 
Giant planets, formed previously in the primordial disk, or stellar companions can also dynamically excite the motion of planetesimals 
via their secular perturbation, even at a significant distance from the planetesimal disk. 
Thus these large bodies can also initiate and govern the formation and evolution of a debris disk \citep{mustill2009}, providing an 
alternative stirring mechanism. 
In a debris disk, mutual collisions grind down planetesimals to small dust grains that are then ejected by radiation pressure, 
or in more tenuous disks removed by the Poynting-Robertson drag \citep{dominik2003,wyatt2005}.
This process is accompanied by the depletion of the reservoir planetesimal belt and eventually leads to the 
decline of the debris production \citep{wyatt2007,loehne2008}.

Due to the strong link between the debris dust and the unseen planetesimals, the investigations of the smallest particles 
of debris systems can lead to a better understanding of the formation and evolution of planetesimal belts and, eventually, 
the formation and evolution of planetary systems. 
The observational verification of the different aspects of planetesimal formation and evolutionary model predictions
requires a detailed study of the incidence of stars with infrared (IR) emission due to debris dust and investigating the
change of debris disk properties (e.g. radius of the dust ring, fractional luminosity) with age. 
The ideal way would be to resolve and observe many debris disks in scattered light or in thermal emission
from optical to millimeter wavelengths with good wavelength coverage. 
In reality, however, the number of resolved disks is very limited 
and the spectral energy distribution (SED) of the dust emission was measured for most debris systems only in a few infrared bands. 
The fundamental parameters of the disks have to be estimated from these sparsely
sampled SEDs. The interpretation of SEDs is ambiguous (e.g. considering the radial location of the dust) but by
handling a debris disk sample as an ensemble, one can obtain a meaningful picture about the basic characteristics of the parent planetesimal belt(s) and 
about the evolutionary trends.

The current theoretical models dealing with the build up of planetesimals \citep{kenyon2004,kenyon2008} and 
with the steady-state collisional evolution 
of the planetesimal belts \citep{dominik2003,wyatt2007,loehne2008} predict how the fundamental properties of 
debris disks evolve with time. 
At a specific radius, the peak of the dust emission is believed to coincide with the formation of 
1000--2000\,km sized planetesimals.
After this stage -- parallel with the depletion of planetesimals -- the dust emission
decreases with time.
The evolution of the disk can be traced both in the variation of incidence of disks with time and 
in the evolution of the brightness of dust emission. The different models predict that the dust fractional luminosity, 
the ratio of the energy radiated by the dust to the stellar luminosity, 
varies with time as $t^{\rm -n}$, where $n=0.3-1$ in disks 
where collisions are the dominant removal process \citep{dominik2003,wyatt2007,loehne2008,kenyon2008}. 
The unique sensitivity of the {\sl Spitzer Space Telescope} in the MIPS 24{\micron} band allowed the detection of stellar photospheres
and a small amount of excess for a large number of field stars \citep[e.g.][]{rieke2005,meyer2008}  and even for 
relatively distant open cluster members \citep[e.g.][]{young2004,gorlova2006,siegler2007,currie2008,balog2009}. 
The latter observations enabled the study of the evolution of warm dust 
around well-dated sample stars.
Investigating early (late B- and A-) type stars \citet{rieke2005} demonstrated a decline of debris disks with age: older stars 
show excess emission less frequently and with lower fractional excess than the younger ones. \citet{siegler2007} found 
similar evolutionary trends for debris disks encircling later type stars (F,G,K). Based on observations of more than 300 Sun-like 
stars with spectral type of F5--K3, \citet{meyer2008} argued that the 24{\micron} excess fraction for this sample is roughly constant 
for ages $\leq$300\,Myr and declines thereafter \citep[see also in][]{carpenter2009}. 
Recently, \citet{gaspar2009} and \citet{carpenter2009b} 
gave a summary of the evolution of 24{\micron} excesses around B7--K5 type stars.
Confirming the previous results they concluded that both the incidence of 24{\micron} excess and the 
excess luminosity monotonically decrease with time at ages $\gtrsim$20\,Myr. 
Utilizing the observations of different infrared space missions ({\sl IRAS, ISO, Spitzer}), the 
predicted evolutionary trend 
in the fractional luminosities was also established \citep[][]{decin2003,su2006,wyatt2007b,rhee2007,carpenter2009}.   
In an extended planetesimal disk both the stirring by Pluto-sized planetesimals that were born in the same belt, 
and the dynamical excitation by secular perturbation of
distinct giant planets is thought to be accompanied by the outward propagation of the dust production site with time \citep{kenyon2008,mustill2009}.
The observational evidence for such a delayed initiation of the collisional cascade as the function of radial 
location is not yet conclusive. Some surveys did not report
any trend in the evolution of the radius with age \citep[e.g. ][]{najita2005}, while studying
debris disks around B- and A-type stars \citet{rhee2007} found some evidence that the radius of dust
belts is increasing with stellar age.

Thanks to the recent photometric and spectroscopic surveys with the {\sl Spitzer Space Telescope}, the  
number of debris disks with detailed spectral energy distribution at mid- and far-IR wavelengths 
has been increased significantly \citep{chen2006,rieke2005,su2006,carpenter2008,rebull2008,trilling2008}. 
This improvement is especially remarkable for disks around 
A-type and Sun-like stars (late F, G, and K-type stars). 
The comparison of these data with the predictions of quasi-steady state evolutionary models showed that most observed trends for 
A-type and Sun-like stars can be reproduced 
adequately \citep[][]{wyatt2008,carpenter2009}.
\citet{kennedy2010} confronted the {\sl Spitzer} observations of A-type stars with an analytic model that also take into account
the effects of the self-stirring on the disk evolution. Utilizing this model they were able to reproduce the observed trends 
and they obtained rough estimates for some initial parameters (e.g. average mass) of disks around A-type stars. 
It was also concluded that debris disks are narrow belts rather than extended disks.      
According to the models, F-type stars are expected to be an intermediate type between the A-type and 
Sun-like stars in terms of debris disk evolution as well: 
1) their disks are predicted to evolve faster than those around main-sequence stars of later types 
(in disks with identical surface density distribution, the timescale of planetesimal formation processes are 
thought to be proportional to $M_{*}^{-1/2}$); 2) F-type stars live much longer than A-type stars (the main-sequence lifetime of an 
1.4\,M$_\sun$ F5-type star 
is 3 times longer than the main-sequence lifetime of a 2.0\,M$_\sun$ A5-type star), making it possible to follow disk
evolution for a significantly longer time. Up to now the number of detailed studies of debris disks around 
F-type stars is modest compared to the A-type and Sun-like samples, preventing us from 
understanding the evolutionary aspects.

In this paper we present the results of a large survey with the {\sl Spitzer Space Telescope} that focuses on debris disks around F-type stars. 
Our main goals are to 1) significantly increase the number of debris disks with detailed SED around F-type stars; 
2) investigate the variations of fundamental properties of the disks, and compare 
the observed trends with the predicted ones; 
3) compare the evolutionary trends obtained for disks around A-, F-, and G/K-type stars.
In the present paper we review the target selection (Sect.~\ref{sample}), observations and data reduction aspects of the 
F-stars program (Sect.~\ref{obsanddatared}). We identify stars with infrared excess, model their SED and
estimate the fundamental properties of the observed debris disks (Sect.~\ref{results}). Using the derived parameters we investigate the diversity of 
the fundamental disk properties, 
and compare the observed trends with the predictions (Sect.~\ref{discussion}).

Four new warm disks -- discovered in the framework of this program -- have already been analyzed and published \citep{moor2009}.
The evolutionary aspects of the current data set -- 
supplemented by the recently discovered four warm debris systems, as well as additional 
debris disks around F-type stars observed by {\sl Spitzer} from the literature
 -- will be further analyzed in 
an upcoming paper (Mo\'or et al. 2010a, in prep.).

\section{Sample selection} \label{sample}

Our primary intention was to study the variations of the disk properties, thus in the sample 
selection we focused on 
stars where previous IR observations hinted on the existence of excess emission.  
Although the formation of planetesimals may last as much as hundred million years in an extended disk, the
most active period of this process is restricted to the first few tens of millions years \citep{kenyon2008}. The possible outward propagation 
of the planetesimal formation can be verified in this period.
Since the age of nearby moving groups overlaps well with this period, they are favorable, 
nearby, and well-dated places for investigations of the debris disk evolution process.
Thus, the above mentioned sample was supplemented by several F-type members of the nearby young kinematic groups.
For these young stars we did not require {\sl a priori} information about the presence of emission in excess to the stellar photosphere.
Because of the selection method the sample is inherently biased with respect to the presence of disks, therefore it can not be used to 
study the incidence of debris disks around F-type stars.

With the aim of constructing a list of F-type main-sequence (and some subgiant) stars, 
where earlier observations indicated the presence of mid- and/or far-infrared excesses, 
we carried out a systematic search using the data of the {\sl IRAS} and {\sl Infrared Space Observatory} satellites.
For the selection of {IRAS}-based candidates, we collected all sources from the {\sl IRAS} Faint Source 
Survey Catalogue \citep[IRAS FSC;][]{moshir1989} and the {\sl IRAS} Serendipitious Survey Catalogue \citep[IRAS SSC;][]{kleinmann1986} 
having at least moderate flux quality at 25 and 60{\micron}. The positions of the IRAS sources were cross correlated 
with the entries of the {\sl Hipparcos} Catalogue (ESA 1997) and the Tycho-2 Spectral Type Catalogue \citep{wright2003} 
and we selected only those objects where the positional match was within 30{\arcsec}. 
Giant stars were  
omitted from the sample on the basis of the spectral information or the absolute magnitude. 
For the identification of stars with excess we used 
the method of \citet{mannings1998}. First, we computed the ratios of measured 25{\micron} and 60{\micron} flux densities to the 
measured flux at 12{\micron} ($R_{\rm 12/25}=\frac{F_{\rm 12}}{F_{\rm 25}}$, $R_{\rm 12/60}=\frac{F_{\rm 12}}{F_{\rm 60}}$) as well as the 
error of the ratios 
($\delta R_{\rm 12/25} = \frac{F_{\rm 12}}{F_{\rm 25}} \sqrt{ (\frac{\delta F_{\rm 12}}{F_{\rm 12}})^2 + (\frac{\delta F_{\rm 25}}{F_{\rm 25}})^2 }$,  
$\delta R_{\rm 12/60} = \frac{F_{\rm 12}}{F_{\rm 60}} \sqrt{ (\frac{\delta F_{\rm 12}}{F_{\rm 12}})^2 + (\frac{\delta F_{\rm 60}}{F_{\rm 60}})^2 }$). 
Then, applying a Kurucz model atmosphere of a typical F-type star 
(a model with effective temperature of 6500\,K, $\log{g}$=4.25, solar metallicity) we derived 
the expected photospheric flux ratios for the specific 
{\sl IRAS} bands ($R_{\rm 12/25}^*, R_{\rm 12/60}^*$). We calculated the significance of the differences between the observed 
and expected ratios as $S_{\rm 12/25} = \frac{R_{\rm 12/25} - R_{\rm 12/25}^*}{\delta R_{\rm 12/25}}$ and 
$S_{\rm 12/60} = \frac{R_{\rm 12/60} - R_{\rm 12/60}^*}{\delta R_{\rm 12/60}}$. Finally, we selected those objects where $S_{\rm 12/25} < -2.5$ or/and 
 $S_{\rm 12/60} < -3.0$.
Due to the low spatial resolution of {\sl IRAS}, many of the positional coincidences 
between a star and an infrared source could be false \citep{moor2006,rhee2007}. In order to reject 
objects that are possibly affected by source confusion we applied the same criteria as those adopted in Sect\,2.3 in \citet{moor2006}. 

Observations with {\sl ISO} confirmed the existence of IR excess and the positional agreement between the optical 
and IR source for several of our IRAS-based candidates.
Moreover, the compiled list was further supplemented by five {\sl ISO}-based discoveries \citep{sprangler2001,decin2003}.

The list of F-type stars belonging to different nearby young stellar kinematic groups (e.g. Tucana-Horologium association, 
AB Dor moving group) was adopted from the catalog  
of \citet{zs04b}. This initial list was amended by adding several new unpublished moving group members (Mo\'or et al., 2010b, in prep.).  

After merging the lists, we excluded all those objects that were reserved by other Spitzer programs. 
Among the finally selected 82 candidates, 27 were based on hints for excess emission at 25{\micron} 
(Warm Disk Candidates, hereafter WDCs), 
46 are suspected to exhibit IR excess at longer wavelengths (Cold Disk Candidates, hereafter CDCs) 
and 9 were selected because of their kinematic group membership (Moving Group Members, MGMs). 
Note that many stars selected on the basis of their suspected excess are also young kinematic group members.
The basic properties of the sample stars as well as the reason for their selection are listed 
in Table~\ref{tab1}.

\subsection{Basic properties of the selected objects} \label{properties}

In order to estimate some basic properties of our stars and to provide photospheric flux predictions 
at relevant mid- and far-IR wavelengths, we modeled the stellar photosphere 
by fitting an ATLAS9 atmosphere model \citep{castelli2003} to the optical and near-IR observations 
(Hipparcos, Tycho2, 2MASS). The surface gravity value was fixed during the fitting 
procedure. If there was no indication that a star has already left the main-sequence 
 -- based on its position in the HR diagram and/or spectroscopic information -- 
we adopted a value of $\log{g}=$4.25 corresponding to the main-sequence stage. 
 For evolved objects the $\log{g}$ values were either taken from literature or computed from available data (see Table~\ref{tab1} for details). 
The metallicity data were also collected from the literature. In the cases where more than one [Fe/H] estimates were available 
for a star, we used the average. 
If no metallicity information was found for a specific star, we adopted solar metallicity.
Most of our stars are located inside the Local Bubble, where the mean extinction is low \citep{lallement2003}. 
Thus, for stars within 80\,pc the visual extinction was neglected ($A_{\rm V}$ set to 0.0).
For more distant objects and for stars without reliable distance information, the $A_{\rm V}$ value was 
a free parameter in fitting the photosphere.
Our sample contains 23 multiple systems (see Table~\ref{tab1}). For close binary systems with separation $<$5{\arcsec}, we 
used the combined photometry of the components, except for HD\,122510 and HD\,199391, where good quality 
photometric data were available separately for each component. In the latter cases the fitting was performed 
for the different components separately and then the obtained photosperic models were co-added.
The fixed ($\log{g}$, [Fe/H]) and fitted ($T_{\rm eff}, A_{\rm V}$) parameters are quoted in Table~\ref{tab1}.

\section{Observations and data reduction} \label{obsanddatared}

We performed observations using the
Multiband Imaging Photometer for Spitzer \citep[MIPS,][]{rieke04} and the Infrared Spectrograph \citep[IRS,][]{houck04}
on the {\it Spitzer Space Telescope} \citep{werner04} in the framework of the programs PID: 3401, 20707, and 40566 (PI: P. \'Abrah\'am).
The MIPS photometry
of \object[HD 35114]{HD\,35114} was taken from the Spitzer Archive (PID: 3600, PI: I. Song).
The MIPS images were obtained in photometric imaging mode (default scale, small-field size). 
All of our objects were measured at 24{\micron} and 70{\micron}, CDCs and most MGMs (in total 52 objects) 
were observed at 160{\micron} as well. In the 24{\micron} band we performed 2 observation cycles with 
integration time of 3\,s providing typically 28 DCE frames 
per source. At 70{\micron} and 160{\micron} 2--10 cycles were made, 
the integration time was set to 3\,s or 10\,s depending on 
the actual source.   
We obtained low-resolution spectra, using standard IRS staring mode observations with
the Short-Low and Long-Low modules, which are sufficient to
outline the continuum and the broad spectral features. Whenever it was possible, peak-up observations in high accuracy IRS or PCRS mode  
were used to place the star at the center of the slit with a radial pointing uncertainty of 0.4{\arcsec} or 0.14{\arcsec}, respectively. 
For most WDCs/MGMs and for some CDCs the full wavelength coverage was requested 
(all four IRS modules were used), while in the remaining cases the shortest wavelength module (SL2, 5.2--7.4\,{\micron}) was dropped.
In the case of HD\,122106 and HD\,125451, where no high accuracy peak-up observations were available, only the longest wavelength 
modules (LL2, LL1) were utilized (at long wavelengths the slit is larger and therefore the probability of loosing flux due to 
inaccurate position is expected to be lower). 
For the two brightest sources (HD\,15745, HD\,170773) low resolution MIPS SED spectra (55--95{\micron}; $\lambda / \Delta \lambda \sim$ 15--25 ) 
were also obtained. 
Six observing cycles were performed in both cases. Each cycle provided six pairs of 10\,s long on- and off-source 
exposures. The on and off positions were separated by 2{\arcmin}.

\subsection{MIPS}

\subsubsection{Image processing} \label{imageprocessing}

The processing of MIPS data started from the Spitzer Science Center (SSC) basic calibrated data (BCD) produced by 
the pipeline version S16.1. 
Additional processing steps, including a flat field correction and a background matching at 24\,{\micron}, were performed 
using MOPEX \citep[MOsaicking and Point source Extraction,][]{makovoz05}. At 70\,{\micron} column mean subtraction and time filtering 
were applied for the images following the steps recommended by \citet{gordon07}. 
Finally, the improved BCD data were coadded and corrected for array distortions. Bad data flagged in the BCD mask files, 
as well as permanently damaged pixels flagged in the 
static pixel mask file, were discarded during the data combination. Outlier pixels were rejected using a 5, 3 and 3\,$\sigma$ 
clipping thresholds at 24{\micron}, 70{\micron} and 160{\micron}, respectively.
Output mosaics had pixels with sizes of 2\farcs5, 4{\arcsec} and 8{\arcsec} at 24{\micron}, 70{\micron}, and 160\,{\micron}, respectively.
The data reduction of the MIPS SED observations was also started with the BCD images (pipeline version S16.1) and the MOPEX was 
utilized to perform the necessary processing steps (combination of data, background removal, application of dispersion solution) and to 
compile the final image with pixel size of 9.8{\arcsec}.

\subsubsection{Photometry} \label{photometry} 

We used a modified version of the IDLPHOT routines to detect sources and to obtain photometry on the mosaic images.

\paragraph{MIPS 24.}

\begin{figure} 
\epsscale{1.}
\plotone{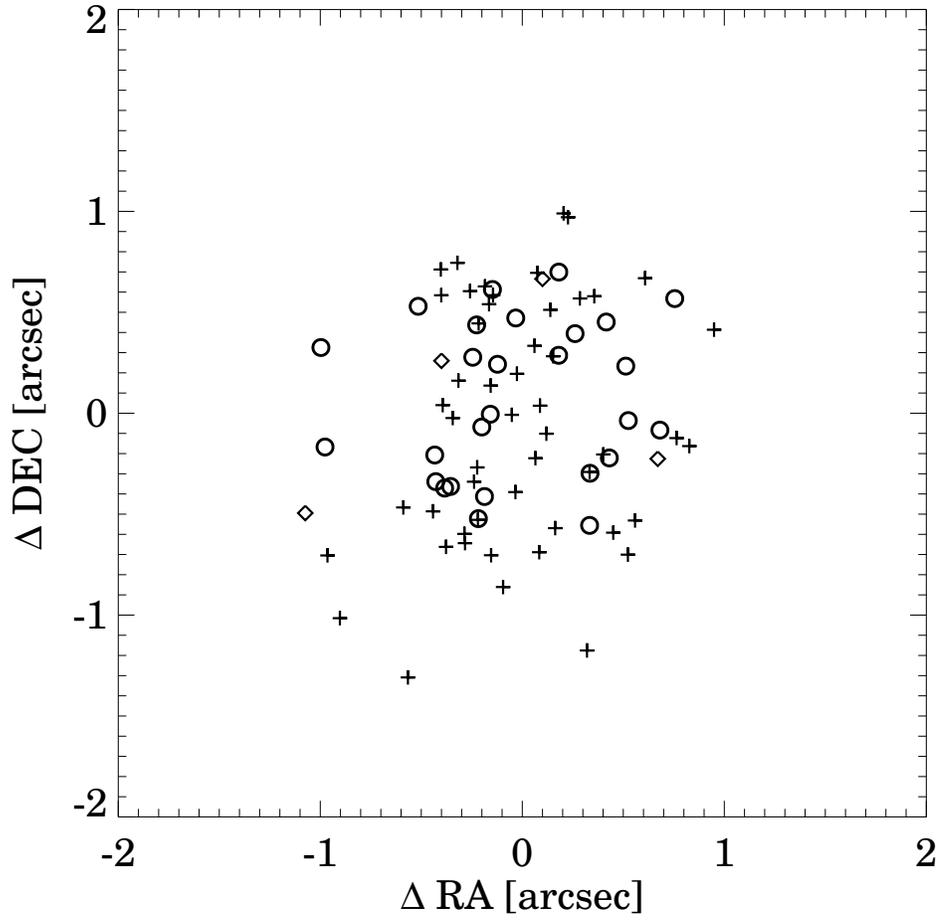}
\caption{Offset between the source positions at 24{\micron} and the 2MASS
position (after correcting for proper motion due to the time difference between the observations). 
Plus signs indicate stars with pure photosperic emission, 
circles show stars exhibiting excess at one or more MIPS wavelengths, while targets surrounded by bright extended nebulosity at 24{\micron} and/or 70{\micron} 
are represented by diamonds. 
The distributions of the offsets for stars with and without excess are in agreement within their formal uncertainties. 
 \label{posMIPS24}}
\end{figure}

All of our sources were detected with good signal-to-noise (S/N$\gtrsim$20) on the final 24{\micron} images. 
Figure~\ref{posMIPS24} shows the positional offsets between the obtained centroids of the identified point sources 
and the 2MASS position (corrected for the proper motion between the epochs of the two observations). 
None of the sources show more than 3$\sigma$ deviations from the averaged position.
Apart from HD\,38905 -- offset of 1.42{\arcsec} from its 2MASS position -- the angular offsets between the different positions 
are within the 1$\sigma$ uncertainty ($\sim$1.4{\arcsec}) of the pointing reconstruction at 24\,{\micron} (see 
MIPS Instrument Handbook). Thus, we verify that the observed 24{\micron} emission is emerging from our targets.

Aperture photometry was performed to estimate the flux densities of the targets. 
The aperture was placed at the measured centroid of the source. 
The aperture radius was set to 13{\arcsec} 
and the background was computed in a sky annulus between 20{\arcsec} and 32{\arcsec}.
In the course of the sky estimates, we used an iterative sigma-clipping method, where the clipping threshold was set 
to 3$\sigma$. 
An aperture correction factor appropriate for a stellar photosphere was taken from \citet{engelbracht2007}.
In some special cases we had to deviate from this standard method. For four sources (HD\,36248, HD\,143840, HD\,185053, HD\,218980) 
that are surrounded by nebulosity at 24{\micron} or/and at 70{\micron}, a smaller aperture with radius of 3\farcs5 was used. 
In cases where a bright nearby source contaminated the aperture photometry (HD\,34739, HD\,38905, HD\,145371, HD\,184169), 
we performed PSF photometry. The model PSF was constructed following \citet{engelbracht2007}.    
The uncertainties of the photometry were computed by adding quadratically the internal error 
and the absolute calibration 
uncertainty of 4\% (MIPS Data Handbook). Due to the fact that our targets were measured  
with high signal-to-noise in this band, the uncertainty of the absolute calibration dominates the error budget.
The resulting 24{\micron} photometry is presented in Table~\ref{mipstable}.

\begin{figure} 
\epsscale{1.}
\plotone{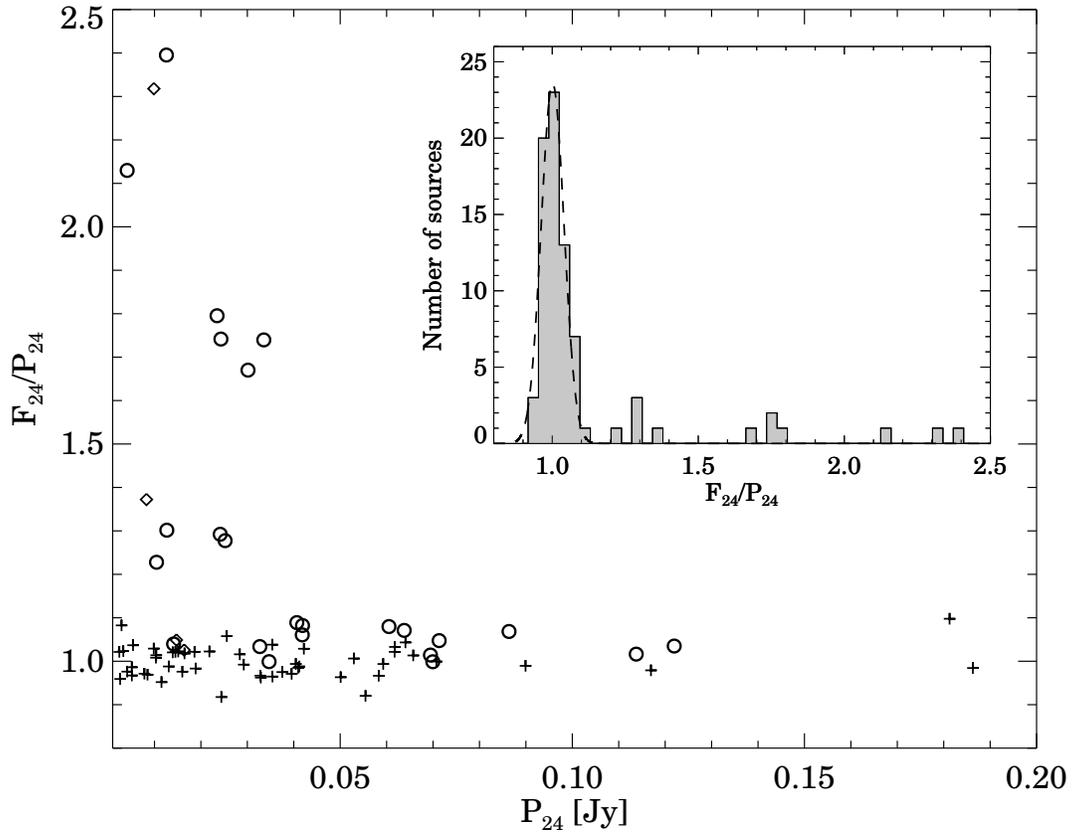}
\caption{Flux ratio of the measured to the predicted flux densitites as a function of the predicted photospheric fluxes 
 for our sample stars measured at
24{\micron} with MIPS. For symbols see the caption of Fig.~\ref{posMIPS24}. 
Small panel: the histogram of the flux ratio. 
The peak at around unity can be fitted by  a Gaussian with a mean of 1.00
and dispersion of 0.038 (dashed line).
 We note that the disks around HD\,15745, HD\,35841 and HD\,221853 with flux ratios of 9.79, 3.39, and 3.87, respectively, are out of the displayed range. 
\label{photMIPS24}}
\end{figure}

Figure~\ref{photMIPS24} shows the distribution of flux ratios -- measured flux densities ($F_{\rm 24}$) relative 
to the predicted fluxes ($P_{\rm 24}$) -- at 24{\micron} as the function of the predicted photospheric fluxes. 
The histogram of the $F_{\rm 24}$/$P_{\rm 24}$ ratios shows a peak at around unity with a mean of 1.00
and dispersion of 0.038. The obtained dispersion is consistent with the quoted absolute calibration uncertainty.
Stars with $F_{\rm 24}$/$P_{\rm 24}>$1.12 have excess emission at this wavelength.

\paragraph{MIPS 70.} 

At 70{\micron} 30 among our 82 targets were detected at  $\geq 4\sigma$ level.   

Aperture photometry with an aperture of 16{\arcsec} and sky annulus between 39{\arcsec} and 65{\arcsec} in radius was performed 
for all 82 targets.
The sky level was estimated using the 3$\sigma$ clipped mean. 
For the detected sources the aperture was placed around the derived centroid of the object, 
while in the remaining cases the 24{\micron} positions were used as the target coordinates.
An aperture correction factor appropriate for a stellar photosphere was taken from \citet{gordon07}. 
After identifying stars with 70{\micron} excess, we recalculated their fluxes with an aperture
correction appropriate for a 60\,K blackbody (characteristic temperature of our disks).
There are eleven images where bright nearby sources contaminated 
the photometry of our targets (see Notes in Table~\ref{mipstable}). 
In order to remove the contribution of these background objects, 
we fitted a PSF to them and then subtracted their emission 
\citep[the PSF was constructed based on the method described by][]{gordon07}. 
For stars surrounded by bright extended nebulosity 
in the 70{\micron} images (HD\,36248, HD\,143840, HD\,185053, HD\,218980), no photometry was derived. 
Due to the resampling of 70{\micron} mosaics, the noise between the adjacent pixels became correlated.
In order to take into account this effect, in the course of internal photometric error estimation, we 
followed the method described by \citet{carpenter2008}.
The resulting 70{\micron} photometry is presented in Table~\ref{mipstable}. 

\begin{figure} 
\epsscale{1.}
\plotone{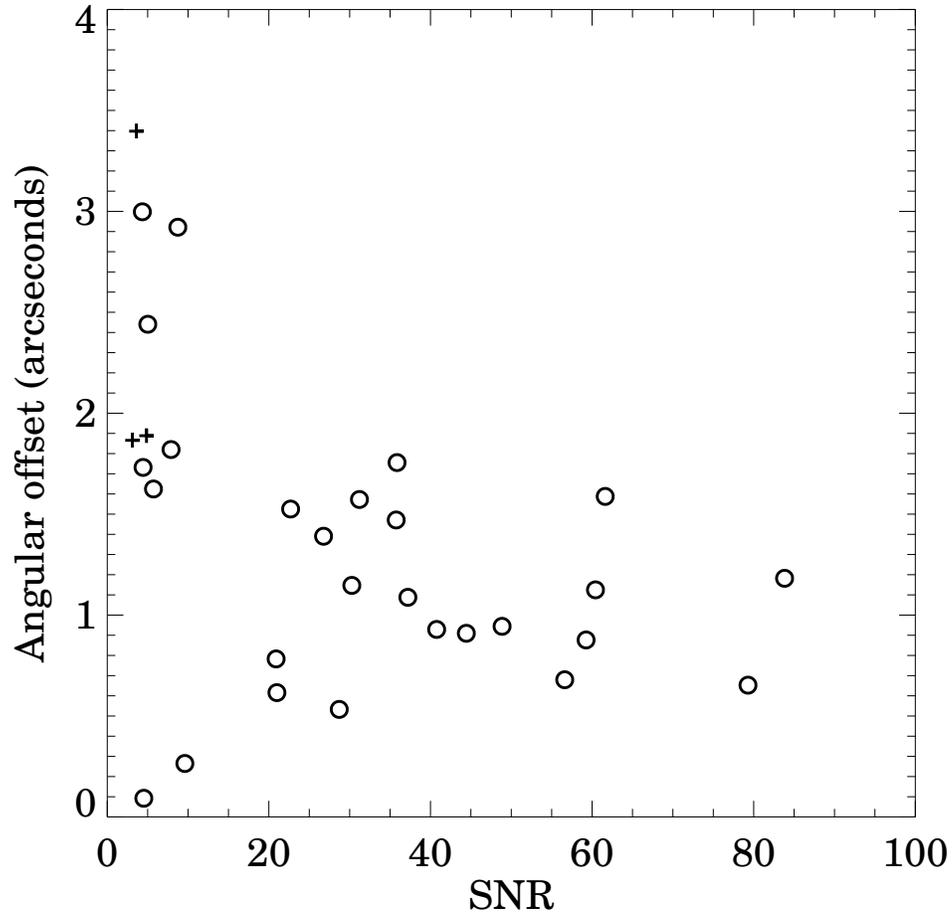}
\caption{Positional offsets between the centroids of point sources detected at 70{\micron}
and the 2MASS position (after correcting for proper motion due to the time difference between the observations) 
as a function of the signal-to-noise ratio measured at 70{\micron}. For symbols see the caption of Fig.~\ref{posMIPS24}.
Apart from HD\,14691, HD\,48391 and HD\,86146 all of the detected sources exhibit excess emission at 70{\micron}. 
\label{posMIPS70}}
\end{figure}

Positional offsets between the obtained centroids of the detected sources  
and their 2MASS positions as a function of signal-to-noise ratio ($F_{\rm 70}$/$\sigma_{\rm 70}^{\rm int}$) measured at 70{\micron}
are displayed in Figure~\ref{posMIPS70}. Most of our detected 70{\micron} sources are within 1.7{\arcsec} 
($\sim$1$\sigma$ pointing reconstruction accuracy at 70{\micron}, MIPS Instrument Handbook)
of the corresponding stellar 
positions. It is worth noting that even in the case of HD\,14691 -- which shows the largest angular offset -- 
the measured flux density is in good agreement with the predicted photospheric flux density at 70{\micron}, confirming the association 
between the 70{\micron} source and the star. It is thus probable that all of the detected sources are likely to be 
associated with our target stars.

\begin{figure} 
\epsscale{1.}
\plotone{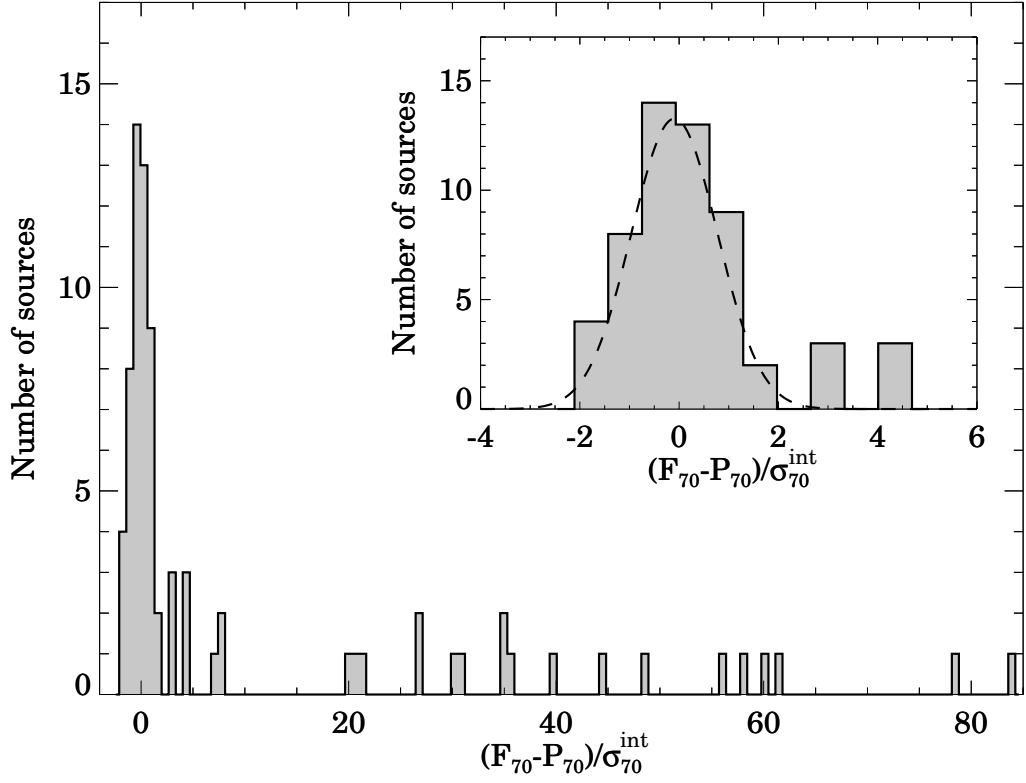}
\caption{Histogram of the significances of the differences between the measured and predicted photospheric 
flux densitites, defined as $(F_{\rm 70}-P_{\rm 70})/\sigma_{\rm 70}^{\rm int}$, for targets measured at 70{\micron}. 
Small panel: a zoom for the peak at zero. A Gaussian fit to the peak provides a mean of $-$0.09 and 
a dispersion of 0.87, in good agreement with the expectations. \label{photMIPS70}}
\end{figure}

Figure~\ref{photMIPS70} shows the histogram of the significances of the differences between the measured and predicted photospheric 
flux densities, defined as $(F_{\rm 70}-P_{\rm 70})/\sigma_{\rm 70}^{\rm int}$, for all of our 78 targets (nebulous objects were excluded). 
The peak around zero significance level corresponds to 
either stars that have not been 
detected at this wavelength or objects that show pure photospheric emission.  
A Gaussian fit to this peak yields a mean of -0.09 and 
a dispersion of 0.87, and these parameters are in good agreement with the expectations 
(mean of 0.0 with dispersion of 1.0).
 
The total noise in the photometry was derived as the quadratic sum of the internal and the absolute calibration uncertainties 
(7\%, see the MIPS Data Handbook).  

\paragraph{MIPS 160.}

MIPS 160{\micron} observations suffer from a spectral leak resulting 
in a ghost image at a certain offset from the nominal position of the original target \citep[e.g.][]{stansberry2007}.
According to the MIPS Data Handbook (ver.~3.2) the ghost is bright enough to appear above the 
confusion noise only for sources brighter than 5.5 mag in $J$ band.
For those of our targets (10 objects, see Table~\ref{mipstable} for details) that are brighter than this limit, 
we applied the procedure proposed by \citet{tanner2009} to minimize the contamination of the leak image.
We downloaded from the Spitzer Archive 24{\micron} and 160{\micron} BCD level data for 
13 bright stars listed in Table\,2 of \citet{tanner2009} that have no excess 
in MIPS bands. 
These data were processed identically to our F-stars observations (see Sect.~\ref{imageprocessing}).
On the final images we determined the average offset of the ghost images from the position of the stars (obtained 
at 24{\micron} images), yielding 5 and 0.2 \,pixels 
in the X and Y directions, respectively, in good agreement with \citet{tanner2009}. Then we compiled the typical PSF of the leak image using 
the ghost images of the brightest four stars. 
As a final step, utilizing the leak PSF and the estimated offset position of the ghost image, we 
applied an iterative cleaning method to remove the leak from the 160{\micron} images of our brightest F stars.  
We performed aperture photometry to obtain flux measurements on MIPS 160{\micron} images.
The aperture was centered on the source positions obtained on the 24{\micron} images. 
The aperture radius was set to 32{\arcsec}, the background level was estimated using an iterative clipping procedure 
in an annulus extending from 64{\arcsec} to 128{\arcsec}. We used an aperture correction factor 
corresponding to a 50\,K temperature blackbody \citep{stansberry2007}. 
For those stars where extended nebulosity or a nearby bright source contaminated the area of the aperture we give
 no photometry at 160{\micron} in Table~\ref{mipstable}. 
In some cases, nearby bright sources appear in the background annulus. 
To minimize their influence, we placed a 4\,pixel radius mask on these sources before we estimated 
the sky level (see Table~\ref{mipstable} for the affected sources).

\paragraph{MIPS SED}

Two of our targets, HD\,50571 and HD\,170773 were observed in MIPS SED mode as well.
The spectra were extracted from the sky-subtracted mosaic images using a five-pixel-wide aperture.
Aperture correction was needed to correct for flux losses. 
Although HD\,170773 is slightly extended at 70{\micron} (see Sect.~\ref{spatext}), 
our studies based on smoothed {\it Spitzer} TinyTim models of the MIPS SED PSF \citep{krist2002,lu2008} showed that 
the difference in aperture corrections between a point source and a point source convolved by a Gaussian with FWHM of 10{\arcsec} 
is insignificant. Thus we applied aperture correction factors valid for point sources \citep{lu2008} for both stars.    
We discarded the longest wavelength part of both spectra ($\lambda>90\mu$m) because of their very low signal to noise. 
As a final step of the data processing, we scaled our spectra to the photometry obtained in the 70{\micron} band as follows 
(the MIPS SED and 
MIPS photometric observations were performed on the same day).
We extrapolated the spectra towards both shorter and longer wavelengths based on IRS and 160{\micron} photometry, in order to completely cover the transmission 
curve of the MIPS 70{\micron} filter. 
We computed synthetic 70{\micron} photometry and derived the ratio of the synthetic to the measured flux density. 
According to the resulting ratios, 
the MIPS SED of HD\,15745 and HD\,170773 were scaled down by 1.11 and 1.16, respectively. 
  
\subsection{IRS}
 
The reduction of the IRS observations started with intermediate {\sl droopres} products of 
the SSC pipeline (version S15.3). 
We used the SMART reduction package \citep{higdon04}, in combination with IDL routines
developed for the FEPS Spitzer Science Legacy program \citep{meyer06}.  
We first subtracted the pairs of imaged spectra acquired along the
spatial direction of the slit to correct for background emission, stray light, and pixels with
anomalous dark current. We replaced bad pixels by interpolating over neighboring good pixels. 
Spectra were extracted from the background-subtracted pixel-corrected images using a 6-pixel
fixed-width aperture in the spatial direction centered at the position derived by the FEPS
Legacy program.  After extracting the spectra for each order, nod, and cycle, we computed a
mean spectrum for each order and as uncertainty we quoted the 1$\sigma$ standard deviation
of the distribution of the flux densities measured at the given wavelength. We converted our
output spectra to flux units by applying the spectral response function derived by the FEPS
Legacy team \citep{bouwman08}, and propagate the calibration error into the quoted uncertainties.

\begin{figure} 
\epsscale{1.}
\plotone{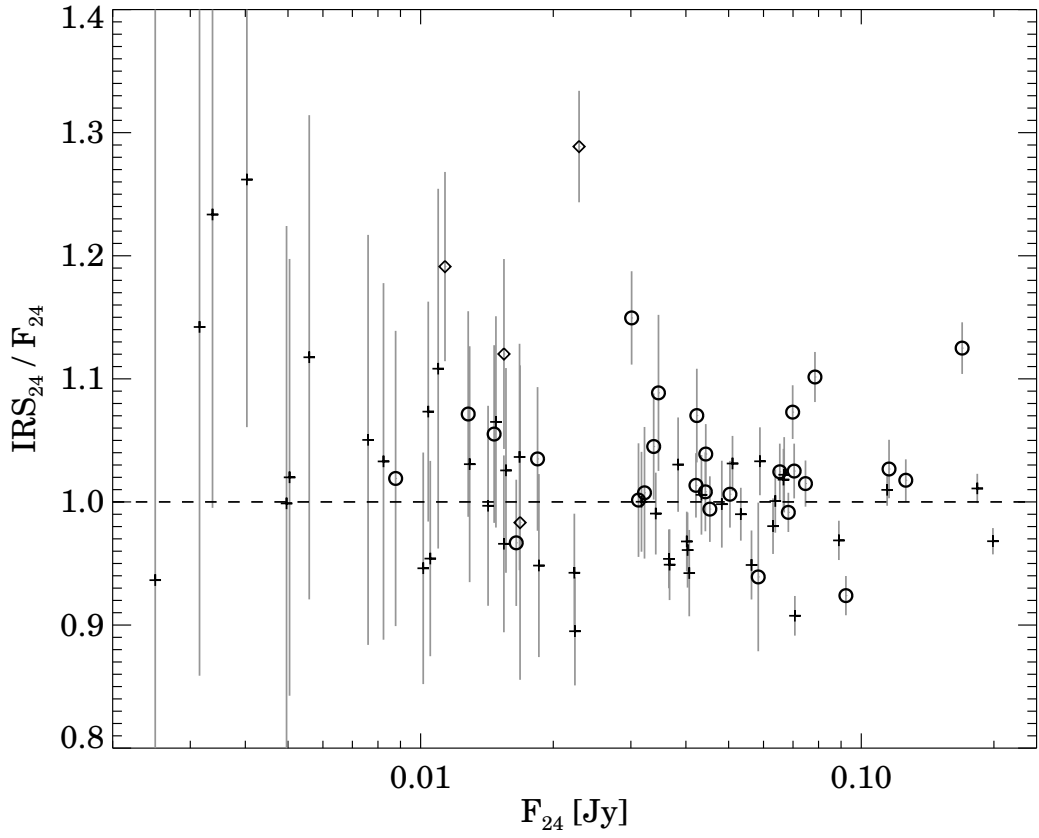}
\caption{The ratio of the synthetic IRS 24{\micron} photometry to the MIPS photometry at 24{\micron} as a function of the MIPS flux densities. 
For symbols see the caption of Fig.~\ref{posMIPS24}. The mean of the $IRS_{\rm 24}/F_{\rm 24}$ ratios is 1.02 with a dispersion of 0.06. 
The dashed line corresponds to the $IRS_{\rm 24}/F_{\rm 24}$ ratio of 1. 
 \label{MIPS24IRS24}}
\end{figure}

Using the filter transmission curve of the MIPS 24{\micron} band we computed synthetic photometry 
from the IRS spectra ($IRS_{\rm 24}$) for all of the targets. Figure~\ref{MIPS24IRS24} shows a comparison between the synthetic 
IRS 24{\micron} and 
the measured MIPS 24{\micron} flux densities ($F_{\rm 24}$). The calibration of the two instruments shows good agreement in general,  
the mean of the $IRS_{\rm 24}/F_{\rm 24}$ ratios is 1.02 with a dispersion of 0.06. The largest discrepancy between 
the two instruments ($\sim$30\%) was found at HD\,185053, a source surrounded with extended nebulosity. 
 
\subsection{Additional data}

For the targets where the {\sl Spitzer} observations pointed to the existence 
of excess emission (Sect.~\ref{excess}) we collected additional infrared and submillimeter data from the literature.
IRAS 60{\micron} and 100{\micron} flux densities and their uncertainties were taken from the IRAS FSC \citep{moshir1989}.
\citet{williams2006} detected three of our targets (HD\,15115, HD\,127821, HD\,206893) at 850{\micron}
using the JCMT/SCUBA instrument, while \citet{nilsson2010} observed three sample stars (HD\,17390, HD\,30447, HD\,170773) 
at 870{\micron} with the LABOCA/APEX instrument.

Observations obtained with {\sl ISOPHOT}, the photometer on-board 
the {\sl Infrared Space Observatory} were available for ten sources. We observed three northern stars with the IRAM 30-m 
telescope at millimeter wavelengths.  
Moreover, for some targets we performed optical spectroscopy. 
In the following we describe the details of the data processing.

\subsubsection{ISO/ISOPHOT}

Ten out of the 27 objects with confirmed infrared excess
were observed with {\sl ISOPHOT} as well. For eight of these ten stars, there are published ISOPHOT fluxes in
\citet{moor2006}.
For the remaining two objects, HD\,151044, HD\,213617, we processed the ISOPHOT data in the same way as described in that paper and the results are 
given in Table~\ref{addphottable}.
HD\,170773 was included in \citet{moor2006}, however, the photometric results were extracted
assuming a point-like source. The analysis of our MIPS data showed that the disk around HD\,170773 is spatially extended 
with an extent of $\sim$10{\arcsec} (Sect.~\ref{spatext}). Assuming that the spatial extent of the disk is similar at 60{\micron} and 90{\micron} 
we re-analyzed the {\sl ISOPHOT} data by convolving the appropriate {\sl ISOPHOT} 
beam profiles by a Gaussian with FWHM of $\sim$10{\arcsec} and then using this new profile in the course of the flux extraction.
The new photometry is given in Table~\ref{addphottable}.

\subsubsection{IRAM/MAMBO2}

We observed three stars (HD\,15745, HD \,25570, HD\,113337) at 1.2\,mm using the IRAM 30-m telescope at Pico
Veleta with the 117-element MAMBO2 bolometer (proposal ID: 195/07, PI: A. Mo\'or). None of them have been observed at millimeter wavelengths before.
The observations were carried out between December 2007 and April 2008, using the standard on-off observing mode.
The targets were always positioned on pixel\,20 (the most sensitive bolometer pixel).  
Observations of Mars were used to establish the absolute flux calibration.
The value of the zenith opacity at 1.2\,mm ranged between 0.17 and 0.48 during our observations.
We performed 3, 3 and 11 of 20\,minutes long ON-OFF scan blocks for 
HD\,15745, HD \,25570, HD\,113337, respectively.  
We utilized the MOPSIC software package (R. Zylka) to perform the data processing using the standard scripts developed for
the reduction of on-off observation data with sky noise subtraction.
None of our targets were detected at 3\,$\sigma$ level. The obtained flux densities and their uncertainties are quoted in 
Table~\ref{addphottable}. 
  
\subsubsection{Ground based spectroscopy}

In July and August 2009 we obtained new high-resolution optical spectroscopy for stars in Table~\ref{spectable}, using the
2.3-m telescope and the Echelle spectrograph of the Australian National University.
The total integration time per object ranged from 30\,s to 1800 s,
depending on the target brightness. The spectra covered the whole visual range in 27 echelle orders between 3900 \AA\ and 6720 \AA\,, with only small
gaps between the three reddest orders. The nominal spectral resolution is $\lambda/\Delta \lambda\approx$ 23~000 at  the H$\alpha$ line, with typical
signal-to-noise ratios of about 100.

All data were reduced with standard IRAF\footnote{IRAF is distributed by the National Optical Astronomy  Observatories, which are operated by the
Association of Universities for Research in Astronomy, Inc., under cooperative  agreement with the National Science Foundation.} tasks, including
bias and flat-field corrections, cosmic ray removal,  extraction of the 27 individual orders of the echelle spectra, wavelength calibration, and
continuum normalization. ThAr  spectral lamp exposures were regularly taken before and after every object spectrum to monitor the wavelength shifts
of  the spectra on the CCD. We also obtained spectra for the telluric standard HD~177724 and IAU radial velocity (RV) standards $\beta$~Vir (sp. type F9V)
and HD~223311 (K4III).

The spectroscopic data analysis consisted of two main steps. First, we measured radial velocities by cross-correlating the target spectra (using the
IRAF task {\it fxcor}) with that of the RV standard that matched the spectral type of the target -- $\beta$~Vir was used for the F-type targets, 
HD~223311 for the lone late G-type target (SAO\,232842). Each spectral order was treated separately and the resulting velocities and the
estimated uncertainties were calculated as the means and the standard deviations of the velocities from the individual orders. For most of the
targets, the two IAU standards gave velocities within 0.1--0.5 km/s, which is an independent measure of the absolute uncertainties. 
The equivalent width of the 6708\AA\, Li were measured with the IRAF task {\sl splot}.
The projected rotational velocity of the targets was determined via fitting theoretical models
\citep{munari2005} to the observed spectra with the $\chi^{2}$ method.
Table~\ref{spectable} summarizes the derived properties of the observed stars. 

\section{Results} \label{results}

\subsection{Identification of stars with infrared excess} \label{excess}

We used our {\sl Spitzer} data to identify stars exhibiting excess at infrared wavelengths.
First, the predicted photospheric flux densities 
were determined at the relevant wavelengths using the best-fit Kurucz models of the stars (see Sect.~\ref{properties}).
The average accuracy of the predicted far-infrared fluxes is estimated to be
around 3\%.
The predicted flux densities of the stars for the MIPS bands ($P_{\rm 24}$, $P_{\rm 70}$, $P_{\rm 160}$) are listed in Table~\ref{mipstable}.
The significance level of the infrared excess was calculated in each 
photometric band using the following formula:
\begin{equation}
\chi_{\rm \nu} = \frac{F_{\rm \nu} - P_{\rm \nu}}{\sigma_{\rm \nu}^{\rm tot}}, 
\end{equation}
where $F_{\rm \nu}$ is the measured flux density, $P_{\rm \nu}$ is the predicted stellar flux, while $\sigma_{\rm \nu}^{\rm tot}$ is the 
quadratic sum of the uncertainty of the measured flux density and the uncertainty of the predicted flux density 
in the specific band.
When $\chi_{\rm \nu}$ was greater than 3 in any of the MIPS bands, the object was selected as a star with excess emission.
Applying this criterion, we identified 28 stars that exhibit IR excess.
One of our targets, HD\,199391, shows excess only at 160{\micron}. Since at this wavelength the position measured on the
{24\micron} image was adopted in the course of photometry, 
we cannot exclude the possibility that the excess emission is related to a nearby background source.
Thus, HD\,199391 was excluded from the list of stars with excess emission. 
Future observations with higher spatial resolution in far-infrared bands should reveal the true nature 
of the excess emission observed 
towards this object.
Among the 27 remaining systems 15 show excess at 24{\micron}, all 27 exhibit excess at 70{\micron} and we found 
excess at 160{\micron} in 17 cases.
All of our targets that exhibit excess at 24{\micron} also show excess at 
 70{\micron} as well.

\begin{figure*} 
\includegraphics[angle=90,scale=0.7]{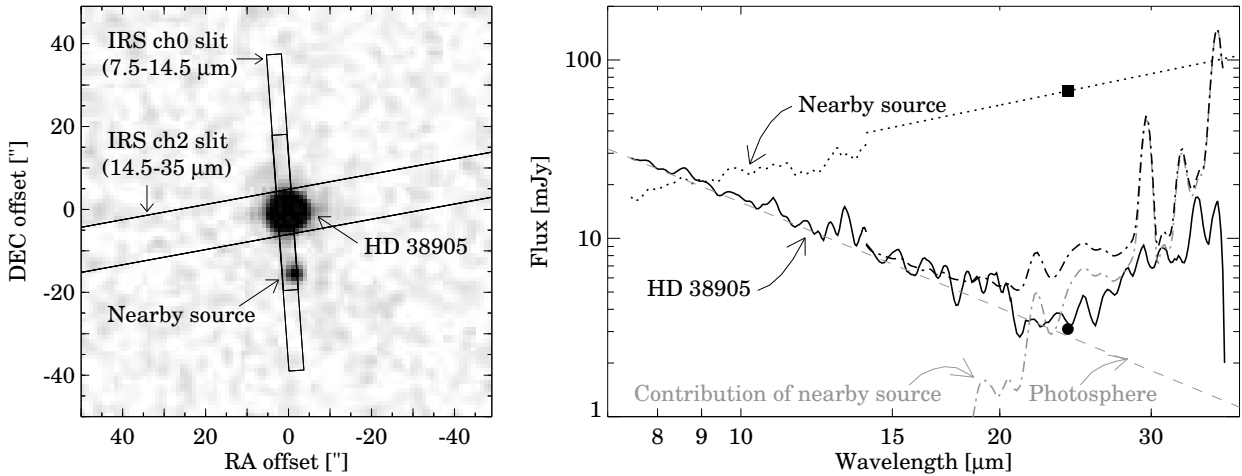}
\caption{{\it Left:} \small{2MASS K$_{\rm s}$ image of HD\,38905 and its
surroundings. The position of the two Spitzer/IRS slits are
overplotted. A nearby source at a distance of 13.5{\arcsec} is included in
the ch0 slit. At these wavelengths, the two sources are well resolved,
and a separate spectrum for each source can be extracted. The ch2 slit
is nearly perpendicular to the ch0 slit, but due to the longer
wavelength, the PSF is wider, thus the nearby source might have a
contribution to the ch2 spectrum extracted for HD\,38905. {\it Right:}
the solid black line indicates the spectrum extracted at the position
of HD\,38905, and the black dot is a MIPS 24$\,\mu$m photometric point
for HD\,38905. The black dotted line indicates the spectrum of the
nearby source: for $\lambda<14.5\,\mu$m, it is observed and resolved
by IRS, and the black square is also a resolved MIPS 24$\,\mu$m
photometric point. The straight line above 14.5$\,\mu$m is a linear
extrapolation. Using the IRS beam profiles and the spectrum of the
nearby source, we estimated the contribution this source has in the
ch2 slit (gray dash-dotted line). The gray dashed line represents the
stellar photosphere of HD\,38905. The black dash dotted line is the
sum of the photosphere and the contribution of the nearby source. Our
conclusion is that the excess emission with respect to the stellar
photosphere observed at the position of HD\,38905 can be well
explained by the contamination from the nearby source. The shape of
the nearby source's spectrum ($F_{\rm \nu}\sim\lambda$) indicates that it
is probably a background galaxy (see e.g.~Wu et al.~2009 or Buchanan
et al.~2006). A similar analysis was done for HD\,34739, HD\,145371 and
HD\,184169, although in those cases, no resolved spectroscopy is
available for the nearby sources. Supposing that these sources are
also background galaxies, a spectral shape of $F_{\rm \nu}\sim\lambda$ is
assumed and absolute brightness level was scaled to resolved
24$\,\mu$m MIPS photometry. Our analysis indicates that apart from
these four stars, no other targets have suffered contamination
by nearby sources.}
\label{irsdemonstrate} }
\end{figure*}

Although in most cases the shape of the obtained IRS spectra could be well fitted by the photospheric model of 
the specific target, 32 among the 82 spectra showed significant deviations. 
Out of the 27 objects where excess was indicated by the MIPS photometry, 25 also showed excess in the IRS 
spectra. In the case of HD\,15060 and HD\,213429 the IRS spectra were consistent with the predicted photospheric emission 
and these two objects showed excess only at 70{\micron}. 
Seven additional excess sources were revealed by the IRS data.
In three of the seven cases (HD\,143840, HD\,185053 and HD\,218980) the MIPS images show bright nebulosity around 
the stars with a spatial extent of 50-60{\arcsec} (see also Sect.~\ref{photometry}). 
At the distance of these stars the estimated angular extents correspond to a size of 4000--7000\,AU.
Thus in these cases the observed emission is likely to be of interstellar, 
rather than circumstellar origin. 
The remaining four stars (HD\,34739, HD\,38905, HD\,145371, HD\,184169) 
are located close to background sources that are brighter at mid- and far-IR wavelengths than the original target. 
In these cases we investigated 
-- using the beam profiles of the different IRS modules and the known position of the background sources (derived on the MIPS images) -- 
 the possibility that the observed excess emission was associated with these bright nearby objects. We found that in all four cases the 
 apparent excess is likely to be related to the nearby sources (Fig.~\ref{irsdemonstrate} demonstrates the main steps of our analysis for HD\,38905).    
As a consequence, these additional seven candidates were discarded from the further analysis.   

A significant fraction of our candidates that were selected based on previous IR observations turned out to be misidentifications 
in the light of the new {\sl Spitzer} data. Several earlier works showed that IRAS-based debris candidate lists are strongly contaminated 
by false identifications (bogus debris disks) especially due to the low spatial resolution of IRAS observations at far-IR
wavelengths \citep{kalas2002,moor2006,rhee2007}. The most common reasons of the misidentification are: 
1) confusion with background sources, where the IR emission is associated with a nearby object; 2) the
presence of extended nebulosity where the IR emission is of interstellar rather than of circumstellar origin; 3) 
erroneous infrared photometric measurements.
The last reason played an especially important role in the case of warm disk candidates.

In total 27 stars have been identified where the observed excess emission may originate from circumstellar dust grains (see Table~\ref{diskprop}). 
Nine out of the 27 stars exhibiting IR excess are new discoveries (see in Table~\ref{diskprop}).

\subsection{Spectral features in the IRS spectra}

No prominent features have been identified in the IRS spectra of the 27 stars with excess emission. 
This finding is consistent with the results of previous IRS observations 
related to debris disks around solar-like and more massive 
A-type stars: although many debris systems show significant excess in the wavelength range between 5 and {35\micron} the majority of these systems
do not possess spectral features \citep{chen2006,carpenter2009,lawler2009,morales2009}. 

\subsection{Spatially extended sources at 70{\micron}} \label{spatext}

\begin{figure} 
\epsscale{1.}
\plotone{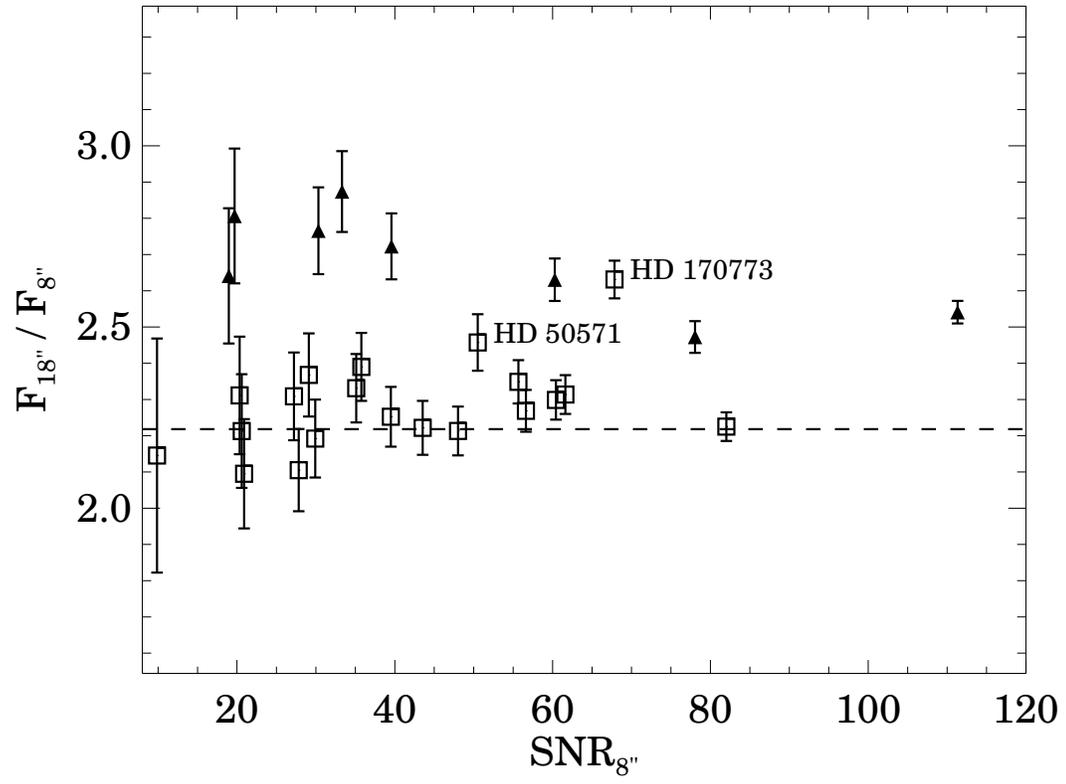}
\caption{ The ratio of the flux density measured in apertures with radii of 18{\arcsec} and 8{\arcsec} as a function of the SNR 
obtained in the smaller aperture. Squares indicate our stars that exhibit excess at 70{\micron}, triangles correspond to 
debris disks (HD\,10647, HD\,38858, HD\,48682, HD\,105211, HD\,115617, HD\,109085, HD\,139664, HD\,207129) that found 
to be marginally resolved at this wavelength by \citet{bryden2006}. \label{MIPS70spat}}
\end{figure}

Observations with the MIPS detector at 70{\micron} revealed several (marginally) resolved debris disks even at relatively large 
distances from the Sun \citep[e.g.][]{bryden2006,su2009}. In order to identify sources with extended emission, in Figure~\ref{MIPS70spat} we plotted 
the ratio of the flux density measured in apertures with radius of 8{\arcsec} and 18{\arcsec} as the function of the SNR 
obtained in the smaller aperture. 
Similar ratios derived for some known resolved debris disks \citep[based on the list of][]{bryden2006} were also plotted. 
MIPS 70{\micron} data for these objects were downloaded from the 
{\sl Spitzer} Archive and processed using the same method applied to our targets. 
As expected, objects with known extended disks show significantly larger flux ratios. 
Utilizing the {\it Spitzer} TinyTim software \citep{krist2002} and applying the smoothing procedure 
proposed by \citet{gordon07} we calculated the PSF of a 60\,K blackbody source that was used to derive 
the expected flux ratio plotted in Fig.~\ref{MIPS70spat}. 
By comparing the obtained flux ratios to the expected one, we found that two of our 
sources -- HD 50571, HD 170773 -- show significant ($>$3$\sigma$ level) deviation and may be spatially extended at 
70{\micron}. In order to model the observed profiles we convolved the PSF 
with different Gaussian profiles. The minor- and major-axis FWHMs as well as the position angle of the major axis of 
the Gaussians were varied to find the best-fitting model profile. 
We found that the measured profile for HD 50571 can be well fitted using a Gaussian broadened by 9.5{\arcsec} only in one direction along a 
position angle of 91{\degr} (measured from north to east).
In the case of HD\,170773, a convolution by a Gaussian with minor- and major-axis FWHMs of 9{\arcsec} and 10{\arcsec} with position angle of 
110{\degr} provided the best fit.
MIPS {70\micron} photometry for these sources, which takes into account the spatial extent, is given 
in Table~\ref{mipstable}. 

It is a common expectation that the rotational spin axis of the star is aligned with the orbital spin axis of the planetesimals. 
HD\,50571 has an unusually high projected rotational velocity of 60\,kms$^{-1}$ 
\citep{holmberg07}. When comparing this value to the $v\sin{i}$ of 1566 stars with similar effective temperatures ($T_{\rm eff}=6480\pm100$\,K) 
included in the same catalog \citep{holmberg07} we found that the projected velocity of HD\,50571 is higher than 99.4\% of 
that of the catalog sample. This implies a very high inclination of the spin axis and  
consistent with our finding that the disk around this star may be seen nearly edge-on (it is resolved only in one direction).
The derived size of the debris disk around HD\,170773 indicates that it might be seen close to pole-on. 
It is interesting that the projected rotational velocity is remarkably high (see Table~\ref{spectable}) for this orientation.

\subsection{Modeling the observed infrared excess} \label{modeling}

\begin{figure} 
\epsscale{0.86}
\plotone{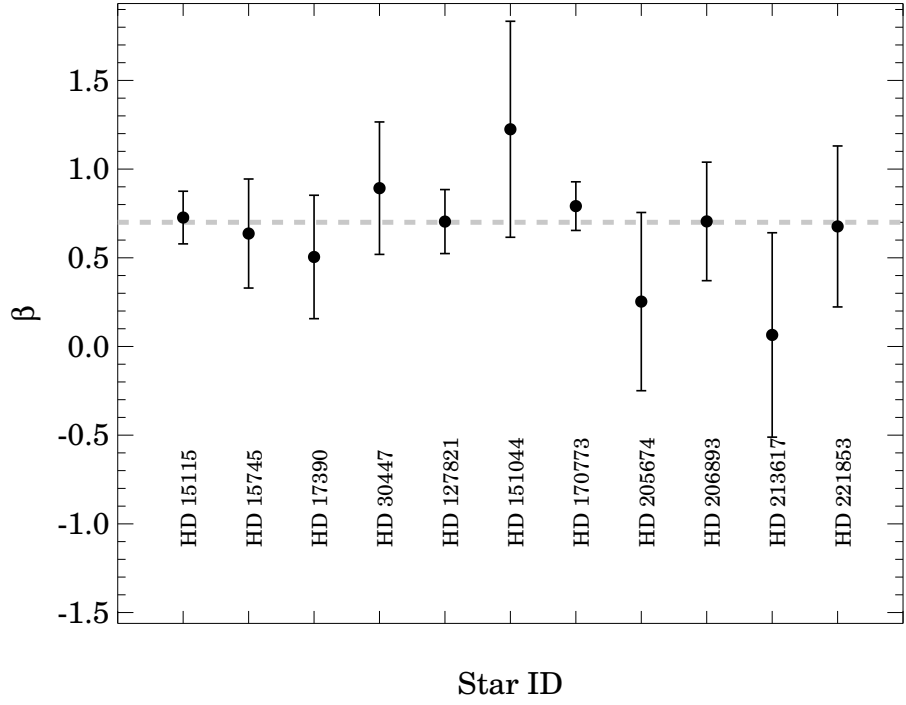}
\caption{Derived $\beta$ values for some selected disks (see Sect.~\ref{modeling}).
 \label{beta}}
\end{figure}

The infrared excess emission of F-type main-sequence stars is generally believed to be attributed to the 
optically thin thermal emission of second generation circumstellar 
dust grains heated by the central star. 
Since no resolved images are available for our targets, the fundamental parameters 
of these dust disks have to be derived by modeling the SEDs of the systems. 
Four newly discovered disks have already been modeled in a previous paper \citep{moor2009}.
For the remaining 27 stars exhibiting excess emission we compiled the SED 
from the data listed in Sect.~\ref{obsanddatared}. 
For the fitting process the IRS spectra were sampled in 11 adjacent bins 
(for the center and width of the bins, as well as for the obtained flux densities see Table~\ref{irstable}).
We used a simple model to characterize the disk
properties based on the excess emission. 
We assume that the dust grains are distributed around a single radius and we adopt the same temperature for all particles
 within this ring. Then the excesses  
are fitted by a single temperature modified blackbody, where 
the emissivity is equal to 1 at $\lambda \leq \lambda_0$ and vary as $(\lambda/\lambda_0)^{\rm -\beta}$ at 
$\lambda > \lambda_0$ wavelengths. We used this modified blackbody model in order to account for the falloff
in the emission spectrum at longer wavelengths, that is faster than in the case of
a blackbody. 
Following \citet{williams2006} we fixed ${\lambda_0}$ to 100{\micron} consistent with the lack of spectral features in the IRS spectra 
which indicates a relatively large average grain size. 
 Since $\beta$ cannot be reliably determined in all systems due to the lack 
of long wavelength data, we decided to estimate a characteristic $\beta$ value based on those disks
where the excess emission at $\lambda > 50${\micron} was measured with $\chi_{\rm \nu}>3$ in at least 4 different bands
and were successfully detected at $\lambda>100${\micron} as well.
A Levenberg-Marquardt algorithm was used to fit the model to the measured data
and an iterative method was used to compute and apply color corrections for the photometric data 
during the fitting process \citep[see e.g.][]{moor2006}.
Figure~\ref{beta} shows the obtained $\beta$ parameters for the selected 11 disks. 
A characteristic $\beta$ value of 0.7, derived by computing the weighted average of these values, seems to represent 
very well the whole sample. 
 Thus, in the following we fixed ${\beta}$ equal to 0.7 and ${\lambda_0}$ to 100{\micron} and repeated the fitting process for all debris 
systems. In two cases when the excess was detected only in the MIPS 70$\mu$m band,
we determined the highest possible dust temperature which was still consistent with the IRS data.
The derived dust temperatures ($T_{\rm dust}$) and the reduced chi-square values of the best fits are quoted in Table~\ref{diskprop}.
We note that using a simple blackbody model instead of the modified blackbody would change 
the dust temperatures less than their formal uncertainties, but would yield worse fitting for the long wavelength data
in most cases when we have $\lambda >$100{\micron} measurements.

\begin{figure*} 
\epsscale{0.76}
\plotone{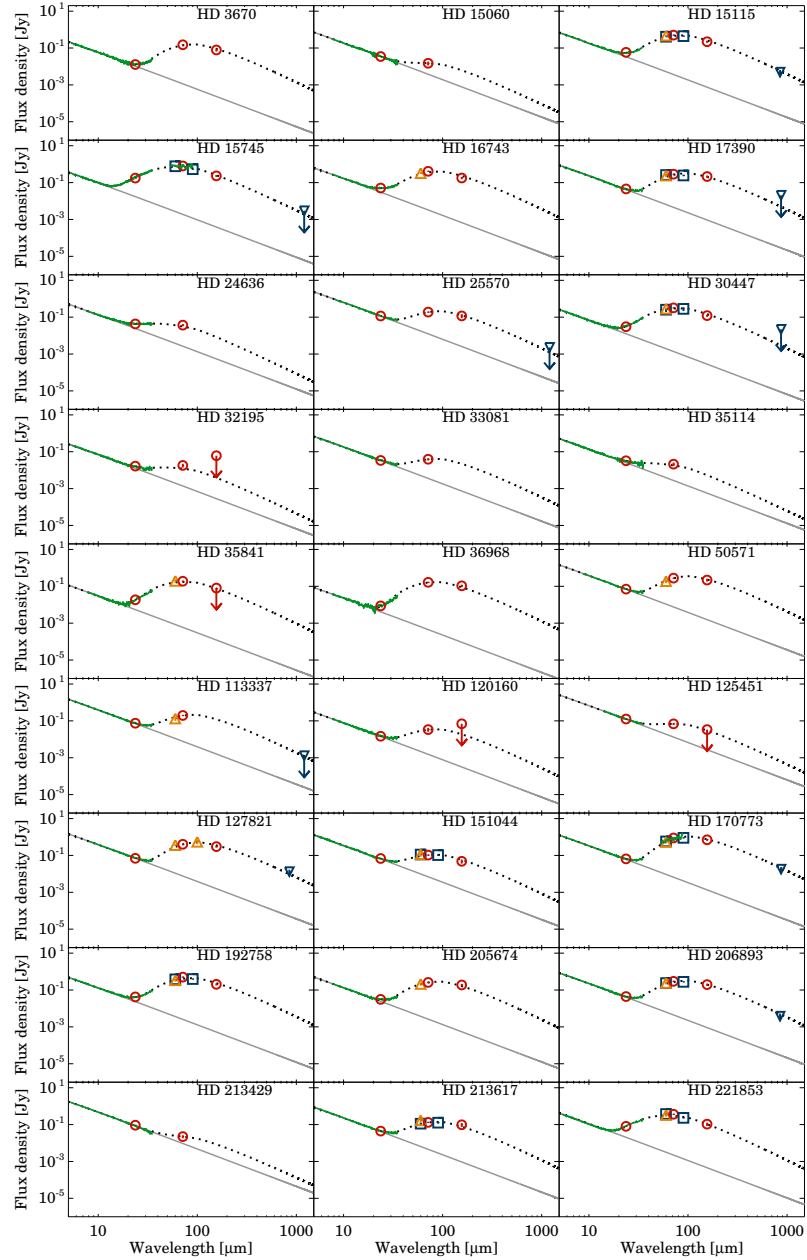}
\caption{\small{Spectral energy distributions for stars exhibiting IR excess in our sample. 
The different symbols represent the following photometric data - 
red circles: MIPS, blue squares: ISOPHOT, yellow triangles: IRAS, blue upside down triangles: submillimeter/millimeter observations.
The IRS and MIPS SED spectra are displayed with green lines. The photospheric models and the disk models are shown 
by solid grey lines and dotted black lines, respectively. For HD\,15115, HD\,15745, HD\,16743, HD\,30447 and HD\,192758 
the two-component disk models are displayed.}  
\label{sedplot}}
\end{figure*}

Most of our SEDs can be fitted well using the simple model described above. 
However, in several cases systematic deviations between the model and measured fluxes can be seen, especially at 
shorter wavelengths, where the model typically underestimates the observed excess. 
For example, for HD\,16743 and HD\,192758 the high reduced chi-square values indicate poor model fit.
It is a general trend among the deviating cases that fitting only the IRS spectra yields higher dust temperature estimates than 
fits based on the long wavelength photometric points. We also found that the extrapolation of IRS-based fits to 
longer wavelengths underpredicts the excess measured in the IRAS, ISOPHOT and MIPS bands.
Several authors reported similar findings in different debris disk samples 
\citep{hillenbrand2008,carpenter2009,morales2009}. One possible explanation for the observed discrepancy
is that dust grains in these systems are distributed in two spatially separated rings similarly to 
our Solar System, where the majority of dust grains are thought to be co-located with the main asteroid 
and with the Kuiper belt. Assuming that the dust is concentrated in two distinct narrow rings, 
we used a two-component model, where grains in the warmer component act like blackbodies, while 
the emission of the outer ring can be described by the modified blackbody as defined above.  
To decide whether a single or a two-component model should be used for a certain target, 
we used a variant of the Akaike Information Criterion, the so-called AICu, proposed 
by \citet{mcquarrie}. The value of AICu can be calculated as:
\begin{equation}
AICu = \ln{\frac{SSE_k}{n-k}} + \frac{2(k+1)}{n-k-2},
\end{equation} where $n$ is the number of observations, $k$ is the number of parameters in the model, 
while $SSE_k$ is the usual sum of squared errors. 
Besides the fact that the Akaike Information Criterion take into account the goodness of fit, it penalizes the usage of 
unnecessary additional model parameters. This test can be used to rank the competing models, from which 
the best one gives the lowest AICu value.
We found the two- component model to be better in five cases 
(HD\,15115, HD\,15745, HD\,16743, HD\,30447, HD\,192758).
The derived parameters of these five disks are presented 
in Table~\ref{diskprop2}. 
The SED and the best-fit models for each of our targets are plotted in Figure~\ref{sedplot}. 

The fractional luminosity of the disks was computed as 
$f_{\rm dust} = {L_{\rm dust}}/{L_{\rm bol}}$. 
The integrated dust emission was derived based on the fitted model while the star's luminosity 
was calculated from the best-fit Kurucz model.
We estimated the radius of the dust ring (or rings) using the following formula \citep{backman1993}:
\begin{equation}
\frac{R_{\rm dust}}{AU} = \left(\frac{L_{\rm star}}{L_{\sun}}\right)^{0.5} \left(\frac{278\,K}{T_{\rm dust}}\right)^2
\end{equation}
Because this formula assumes blackbody-like grains,  
the resulting $R_{\rm dust}$ corresponds to a minimum possible radius. 
The obtained fundamental disk properties are listed in Tables~\ref{diskprop} and \ref{diskprop2}.

We made several simplifying assumptions in the applied model. Collisions among planetesimals in debris disks produce a
collisional cascade, in which collisions gradually grind large bodies into smaller ones that are removed by radiation forces.
This process is thought to result in a characteristic dust grain size distribution and copious amounts of small 
dust \citep[][and references therein]{wyatt2008}.
Even if planetesimals are distributed in a narrow ring, the radiation pressure pushes the smallest grains into more eccentric orbits 
extending the dust disk outward. 
Moreover, the planetesimal ring(s) can be extended. 
Our model assumes relatively large grains that act like a blackbody 
at $\lambda \leq 100${\micron} 
located in one (or two, see above) ring(s) of dust.
Both the existence of smaller dust grains -- that are ineffective emitters and therefore have higher temperature 
than large grains 
at the same radial distant from the star -- and the finite radial extension lead to multi-temperature distribution.

Two of the multiple temperature disks from Table~\ref{diskprop2} were resolved or marginally resolved in scattered light.
Note, however, that different observation techniques are sensitive to different populations of dust grains.
The scattered light images are expected to trace very small grains which may show different spatial distribution than 
those grains which dominate the mid- and far-IR emission of the disk \citep{wyatt2006}, because a significant fraction of 
small grains could be blown out from the system outside the planetesimal ring.  
Using coronographic images, \citet{kalas2007a} and \citet{debes2008} successfully resolved a very extended circumstellar disk around HD\,15115 at optical and 
near-IR wavelengths, revealing a strongly asymmetric disk structure. 
The lobes of the disk can be traced inward to $\sim$31\,AU. 
It could be consistent both with our single/two-ring models where the outer ring is located at $\sim$40\,AU.
HD\,15745 was also resolved in scattered light using the Advanced Camera for Surveys aboard the {\sl Hubble Space Telescope} \citep{kalas2007b}.
The circumstellar disk is detected between $\sim$128-480\,AU radius. The detection at the inner part of the disk is limited by PSF subtraction artifacts,  
thus this image does not provide further information about the disk morphology in the inner regions. 
Our model with an outer ring at 21\,AU disk is not inconsistent with this result.
Since the existing data do not allow to localize the warm dust unambiguously, in the following 
analysis of the five disks in Table~\ref{diskprop2} we assume that the warm emission originates from an inner ring.

Two additional disks in our sample were marginally resolved in the 70{\micron} images (Sect.~\ref{spatext}).
The measured broadening suggests a dust structure size of 320\,AU and 370\,AU for HD\,50571 and HD\,170773, respectively.
Although both stars harbor relatively cold and extended dust rings,  
the minimum diameter of 132\,AU and 146\,AU, derived from our simple model, are significantly lower than the 
measured extensions. One possible explanation would be a dust structure similar to that of the F5/F6V star HD\,181327 
(not included in our sample), where \citet{schneider2006} discovered a dust ring
using NICMOS coronagraphic observations. The extension of this ring ($\sim$86\,AU) 
significantly exceeds the modeled dust radius of 22\,AU
derived from the SED of the object assuming blackbody grains \citep{schneider2006}. 
They propose that a large amount of small dust particles in the ring 
can explain the observed discrepancy since small grains are hotter than large grains at the 
same location.
An analogy with Vega can provide another explanation. Using MIPS observations that resolved the source, 
\citet{su2005} found that the radius of the disk at 70{\micron} exceeds significantly the size of the disk seen at 
submillimeter wavelengths. The discrepancy probably comes from an extended cloud of small particles blown 
away from a planetesimal ring by the radiation pressure of the star. 
Due to the marginal resolution of the Spitzer 70{\micron} observations of HD\,50571 and HD\,170773, we cannot 
decide between the two scenarios. 
In the further analysis of these two objects we adopt the parameters listed in Table~\ref{diskprop}.  

 Using the derived disk parameters in Table~\ref{diskprop} and \ref{diskprop2} we estimated both the Poynting-Robertson ($\tau_{\rm PR}$) 
and the collision timescales ($\tau_{\rm coll}$) for grains with radii ranging between the grain size corresponding to the blowout limit 
and 1000{\micron}. The blowout limit was computed using the equation presented by \citet{hillenbrand2008}: 
\begin{equation}
a_{\rm blow} = 0.52 \frac{2.5\,g\,cm^{-3}}{\rho} \frac{L_*/L_\sun}{(T_*/5780)}, 
\end{equation} 
where $\rho$ is the density of the grain (assumed to be 2.7\,gcm$^{-3}$).
The Poynting-Robertson timescales were estimated based on equation~14 in \citet{backman1993}, 
while the collisional timescales were computed using the semi-empirical formulae (equations~7-8) derived by \citet{thebault2007}. 
We found that for our disks $\tau_{\rm coll}\ll\tau_{\rm PR}$ for all grain sizes and the obtained 
timescales are short with respect to the ages of the stars, implying that the grains have
second generation ('debris') nature and their evolution is mainly governed by collisions.


In a debris disk the mutual collisions continually grind down the larger planetesimals into smaller fragments 
that can be removed by the Poynting-Robertson drag and by the radiation pressure. 
Since the estimated collisional timescale in our disks is significantly shorter than the PR-drag lifetime, the dust removal processes 
in these systems may be collisionally dominated. In such a disk, the frequent collisions shatter dust grains more rapidly  
to sizes below the blowout limit before the effect of the PR drag can be manifested. 
Due to the stellar radiation pressure, grains smaller than the blowout limit are ejected on a short timescale, 
while somewhat larger grains are pushed into a more eccentric orbit. Thus, in a collisionally dominated disk, 
the dust grains extend outward from their birth ring, where the planetesimals are located. However, since    
the lifetime of blowout grains is significantly shorter than the normal grains' lifetime, it is a 
plausible assumption that the dominant part of the dust mass is co-located with the parent planetesimals.
Thus in the further analysis we assume that the derived radii of the dust rings can be considered as the 
size of the underlying planetesimal belts as well. 

\subsection{Age determination} \label{agedet}

In order to estimate the age of the 27 debris systems, we use the the following general strategy.
The most accurate and reliable dating can be derived via cluster membership. Thus, if 
a specific target can be assigned to a stellar kinematic group then we adopt the age of the 
group for the star. Beside the nine previously known cluster members, 
we classify five new young moving group members among our targets (for more details related to the new 
assignments see below).
For field stars, isochrone fitting combined with diagnostics of rotation-driven activity indicators and  
lithium content (especially that of late-type companions) are considered in the age estimates.
In the analysis of chromospheric and  coronal activity indicators, the calibration derived by \citet{mamajek2008} is applied 
whenever it was applicable (i.e. the $B-V$ color indices fall in the appropriate range). 
In the case of lithium content, our age estimates are based on  
comparisons with the distribution of similar properties in well-dated open clusters \citep{sestito2005} and young moving groups \citep{mentuch2008,dasilva2009}.
Pre-main sequence evolutionary models are used if the lithium content or the activity indicators measured in the target or in 
its late-type companion indicated that the specific system may be in a pre-main sequence evolutionary stage.    
We use literature data for several stars. 
Among the 27 debris systems, 13 are younger than 100\,Myr.
Table~\ref{agetable} summarizes the age-related data for stars with debris disks.

\paragraph{HD\,3670.}
The derived galactic velocity components of HD\,3670, U=$-$12.6, V=$-$22.6, W=$-$4.6\,kms$^{-1}$, 
are consistent with the characteristic motion of the 30\,Myr old Columba association \citep[see][]{torres2008}.
HD\,3670 has a ROSAT counterpart
with fractional X-ray luminosity of $\rm \log({L_{\rm x}}/L_{\rm bol}) = -4.39,$ 
which is comparable with the fractional X-ray luminosity of those stars with similar spectral type in the Columba association and supports 
the youthfulness of the star.

\paragraph{HD\,15745.} Based on our new radial velocity measurements (Table~\ref{spectable}), we recomputed the 
galactic space velocities of this star, obtaining $-$10.4, $-$15.3, $-$7.9\,kms$^{-1}$ for the U, V, W components, respectively. 
This space velocity corresponds well to the characteristic space motion of the $\beta$ Pic moving 
group \citep[see][]{torres2008}. 

\begin{figure*} 
\epsscale{1.1}
\plotone{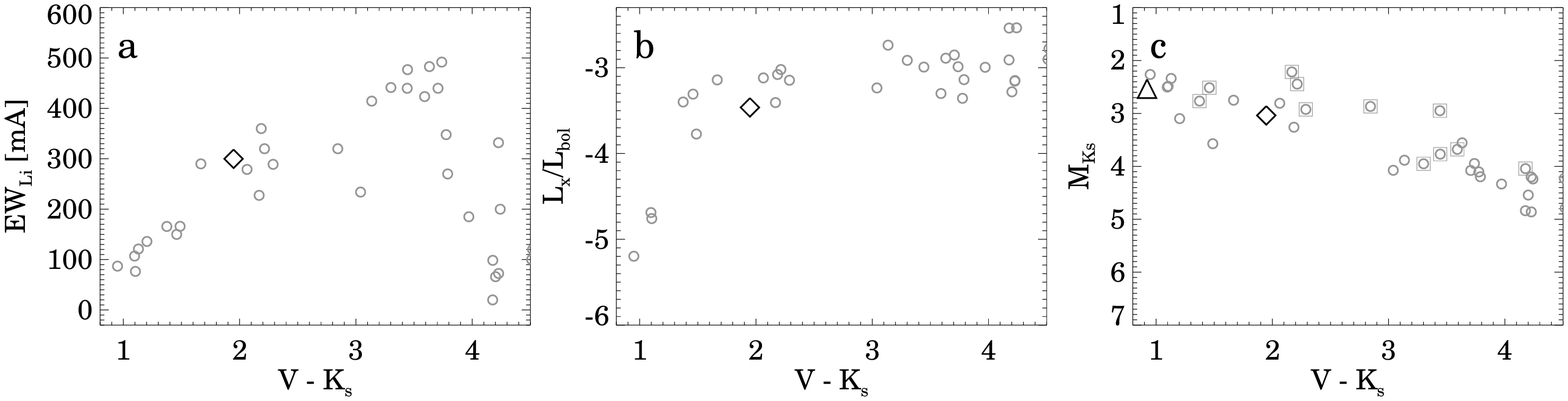}
\caption{a) The distribution of Li equivalent widths as a function of $V-K_s$ color indices. 
b) The distribution of fractional x-ray luminosities as a function of $V-K_s$ color indices. 
c) Color-magnitude diagram. Grey circles represent the known members of the $\beta$\,Pic moving group. 
Grey squares represent binary stars in the CMD.
The black diamond represents BD+45{\degr} 598, the black triangle shows HD\,15745.   
\label{HD15745}}
\end{figure*}

In the framework of a recent high-resolution spectroscopic survey focusing on optical counterparts of X-ray 
sources, \citet{guillout2009} discovered several new post-T Tauri stars located on the northern hemisphere.
One of them, BD+45{\degr} 598 -- which is offset by $\sim$9{\degr} from HD\,15745  -- shows very similar 
proper motion and radial velocity ($\mu_{\rm {\alpha}\cos{\delta}}=44.7\pm1.1$\,mas, $\mu_{\rm \delta}=-44.3\pm1.0$\,mas, 
$v_{\rm r}=-0.77\pm0.97$\,kms$^{-1}$) to that of our target ($\mu_{\rm {\alpha}\cos{\delta}}=45.8\pm0.6$\,mas, $\mu_{\rm \delta}=-47.9\pm0.5$\,mas, 
$v_{\rm r}=+2.5\pm3.3$\,kms$^{-1}$). 
BD+45{\degr} 598 was classified as a K1 type star by \citet{guillout2009}.
The measured lithium equivalent width and the fractional X-ray luminosity of this star are consistent 
with the similar properties of the known $\beta$\,Pic members (see Fig.~\ref{HD15745}~a,~b). 
Assuming that BD+45{\degr} 598 also belongs to the $\beta$\,Pic moving group, a kinematic distance of 
70\,pc can be estimated for it, obtaining $-$10.4, $-$16.2, $-$8.1\,kms$^{-1}$ as its galactic space motion. 
The position of the two stars on the color-magnitude diagram of the $\beta$ Pic moving group also confirms 
their membership status (see Fig.~\ref{HD15745}~c). 
The fact that HD\,15745 harbors a debris disk with very large fractional dust luminosity also supports this assignment.

Therefore, we propose that both HD\,15745 and BD+45{\degr} 598 are new members of the $\beta$\,Pic moving 
group and we adopt an age of 12\,Myr for HD\,15745.

\paragraph{HD\,16743}
Both the proper motion ($\mu_{\rm {\alpha}\cos{\delta}}=73.12\pm0.27$\,mas, $\mu_{\rm \delta}=49.65\pm0.3$\,mas) 
and the trigonometric parallax ($\pi=16.99\pm0.31$\,mas) of HD\,16743 are in good agreement 
with the corresponding astrometric properties of HD\,16699AB 
($\mu_{\rm {\alpha}\cos{\delta}}=72.33\pm0.86$\,mas, $\mu_{\rm \delta}=48.56\pm0.87$\,mas, $\pi=16.54\pm0.99$\,mas), 
that is also a multiple system itself with a separation of 8\farcs7 (HD\,16699+SAO\,232842). 
The similarities of the measured radial velocities of the three stars (see Table~\ref{spectable}) also confirm  
that they may form a wide multiple system.
This offers a good opportunity to improve the age determination of HD\,16743 by combining the result of different
age diagnostic methods for the three members of the system.

The ROSAT source J023845.4-525710 is located close to both HD\,16699 (with separation of 12\farcs6) and  
SAO\,232842 (3\farcs9). This X-ray source is also present in the XMM-Newton slew survey Source Catalogue
 \citep[XMMSL1 J023845.1-525708, ][]{saxton2008}, located 1{\arcsec} away from the position of 
 SAO\,232842. Due to the better positional accuracy of the 
XMM catalog, the latter data make it clear that the X-ray source corresponds to SAO\,232842. 
The high fractional X-ray luminosity of the source, $L_{\rm x}/L_{\rm bol}=-3.32$, is comparable with that of 
stars with similar spectral type in the Pleiades and in young nearby moving groups confirming 
a young age for this object.  

The age estimates based on the lithium abundance measured in HD\,16699 and SAO\,232842 are somewhat controversial.
The high lithium abundance in SAO\,232842 suggests a very young age, it exceeds the upper envelope of the 
distribution of lithium equivalent width measured in Pleiades stars and consistent with 
the lithium content of late-type G stars belonging to very young ($<$40\,Myr old) stellar kinematic groups and 
open clusters. On the other hand, the lithium equivalent width measured in HD\,16699 suggests an older age,   
since it is somewhat lower than for stars with comparable effective temperatures in the Hyades and Coma Berenice open clusters (age $\sim 500-600$\,Myr). 

\begin{figure*} 
\epsscale{1.0}
\plotone{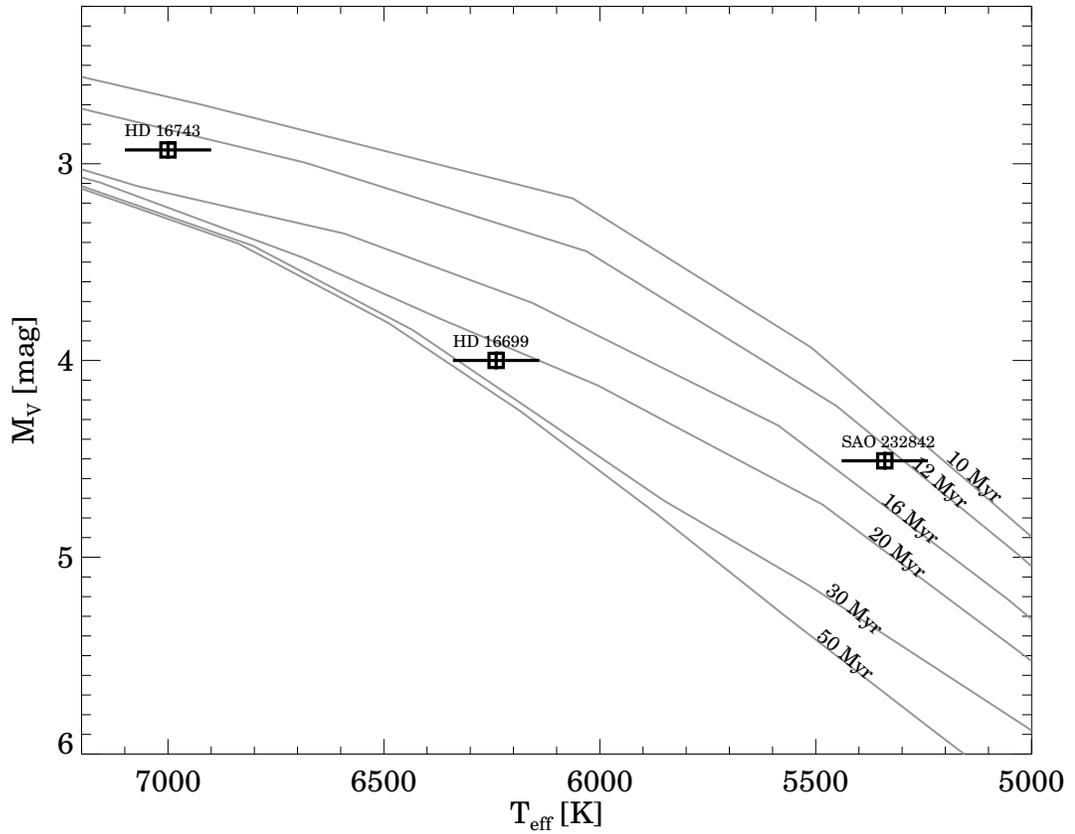}
\caption{H-R diagram for the HD\,16743 system overplotted by isochrones for different ages between 10\,Myr and 100\,Myr. 
The isochrones are taken from \citet{siess2000}. The effective temperature of HD\,16699 and SAO\,232842 was estimated  
using the same method as described in Sect.~\ref{properties}.
 \label{HD16743}}
\end{figure*}

Figure~\ref{HD16743} shows the positions of the three components in the H-R diagram overplotted
by isochrone models from \citet{siess2000} for ages between 10\,Myr and 50\,Myr for a metallicity of Z=0.014.  
Assuming, primarily based on the characteristics of SAO\,232842, that the system is in a pre-main sequence evolutionary stage, 
the positions of the three components suggest an age between 10\,Myr and 50\,Myr.
 
HD\,16743 and HD\,16699AB belong to a very wide multiple system with a minimum separation of $\sim$12700\,AU.
Gravitational perturbations due to the subsequent encounters with other stars and giant molecular clouds 
during the galactic orbit would detach such wide binaries. 
Thus the existence of this system is also an additional argument concerning the young age.

Taking into account age estimates obtained from the different dating methods we propose an age of 
10--50\,Myr for the HD\,16743 system.

Both HD\,16699 and SAO\,232842 are detected and their fluxes are derived via PSF fitting on 
the MIPS 24{\micron} image of HD\,16743.  
The obtained flux densities of 16.4$\pm$0.7\,mJy and 18.8$\pm$0.8\,mJy correspond well to the predicted 
photosperic flux densities in this band for both targets.

\paragraph{HD\,24636.}
The derived galactic space motion of HD\,24636 (see Table~\ref{spectable}) 
is consistent with the characteristic space motion of the 30\,Myr old Tucana-Horologium association \citep[see][]{torres2008}.
HD\,24636 is located quite close to two other Tucana-Horologium stars from our sample (HD\,32195, HD\,53842), 
and all of these objects are within a sphere 14\,pc across. 
Interestingly, all three stars exhibit relatively warm 
infrared excess \citep[$\rm T_{\rm dust} \gtrsim$ 90\,K, see Table~\ref{diskprop} and ][]{moor2009}.

\paragraph{HD\,25570.}
\citet{eggen1982} classified HD\,25570 as a star that belongs to the Hyades group. 
\citet{perryman1998} found that HD\,25570 
 is only just outside their 3$\sigma$ membership criteria and speculated that may be an example 
 of an object just moving away from the Hyades but still close to the tidal radius.
For the further analysis we consider HD\,25570 as a Hyades member and adopt an age of 625\,Myr 
for the star.

\paragraph{HD\,36968.}
Both the galactic space motion ([U,V,W]=[$-$14.8,$-$6.6,$-$8.5]\,km$^{-1}$) and the galactic position of HD\,36968 
([X,Y,Z]=[$-$51,$-$108,$-$73]\,pc) are consistent with those of 
the recently discovered $\sim20$\,Myr old Octans Association \citep[see Fig.~17 in ][]{torres2008}. 
HD\,36968 is the first member of this association where the existence of a debris disk was 
revealed by our results. 
The very high fractional luminosity of this debris disk is also consistent with this young age. 

\paragraph{HD\,50571} is an F7-type star that shows both moderate chromospheric \citep[$R'_{\rm HK}$=$-$4.55, ][]{2006AJ....132..161G} and coronal activity 
($\log{{\frac{L_{\rm x}}{L_{\rm bol}}}}$=$-$5.27). Its 
galactic motion (U\,=\,$-$16.6$\pm$0.3, V\,=\,$-$22.3$\pm$0.7, W\,=\,$-$4.4$\pm$0.3\,kms$^{-1}$) 
as well as its galactic position are consistent with those of the "B3" subgroup of the 
Local Association, implying an age of 300$\pm$120\,Myr \citep{asiain1999}. 
Using an independent age estimating method, \citet{rhee2007} also derived 300\,Myr as the age of the object.

\begin{figure} 
\epsscale{1.}
\plotone{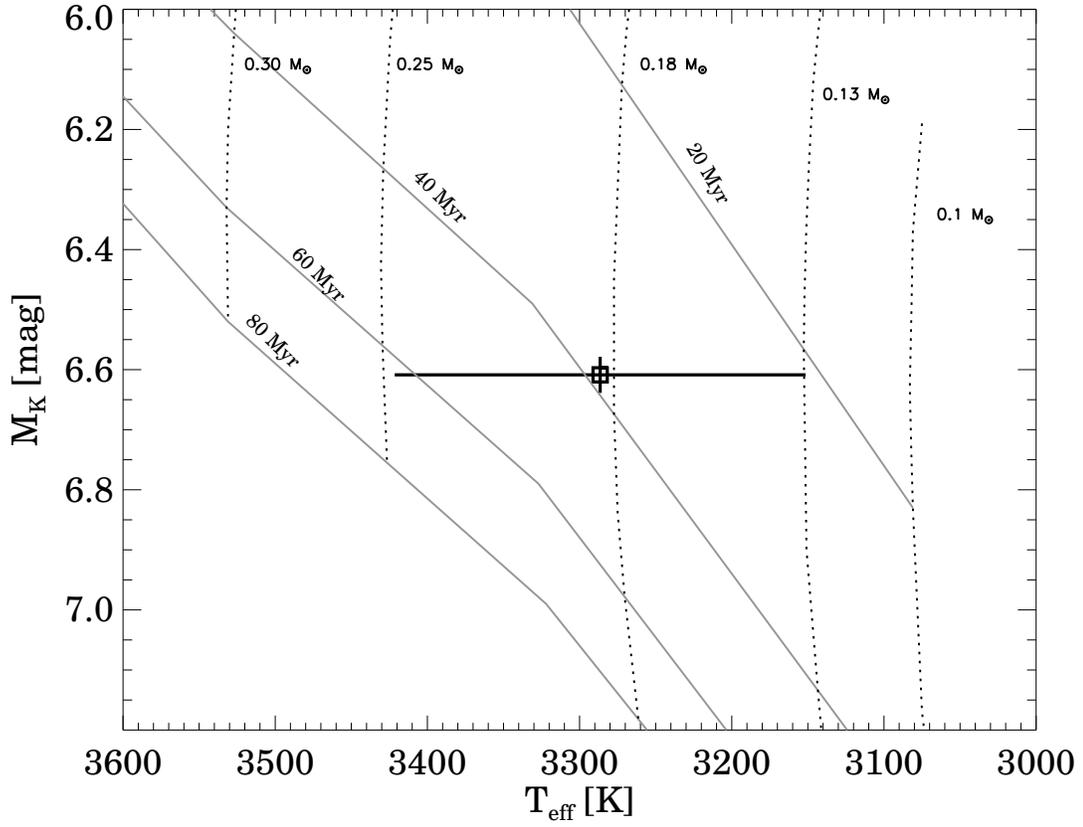}
\caption{H-R diagram for HD 113337B. The effective temperature of the star and its uncertainty are estimated 
based on spectral class information (1 subclass uncertainty is assumed),
the absolute magnitude in K-band is computed by assuming that it is located at the same distance as the primary 
component. Mass tracks and isochrones computed by \citet{siess2000} for the evolution of stars with solar metallicity 
are overplotted. Based on these evolutionary tracks we find
the age of the HD 113337B to be 40$\pm$20\,Myr and the mass is $\sim$0.13-0.25\,M$_{\sun}$.
 \label{HD113337}}
\end{figure}

\paragraph{HD\,113337} has a late-type companion with a separation of $\sim${120{\arcsec}}.
The companion's proper motion ($\mu_{\rm {\alpha}\cos{\delta}}=-174\pm4$\,mas, $\mu_{\rm \delta}=26\pm1$\,mas)
is consistent with that of HD\,113337 ($\mu_{\rm {\alpha}\cos{\delta}}=-171.7\pm0.3$\,mas, $\mu_{\rm \delta}=25.4\pm0.2$\,mas).
Based on medium resolution spectroscopic observations,
\citet{reid2007} classified the companion as an M3.5 dwarf with strong $H_{\alpha}$ emission   
($\log{{\frac{f_{\rm H_{\rm \alpha}}}{f_{\rm bol}}}}$=$-$3.31). 
The companion is also included in the 2MASS and SDSS (Sloan Digital Sky Survey) catalogs. 
Utilizing the spectroscopic analysis of a large number of M-type stars 
from the SDSS, \citet{west2008} provided 
characteristic median SDSS/2MASS colors ($r-i, i-z, z-J, J-H, H-K$) for 
different spectral subclasses (see their Table\,1). Comparing the 
measured colors of HD\,113337B with the quoted median values, we classify it 
as an M4-type star, confirming the result of \citet{reid2007}.  
Using the spectral class information we estimated the effective temperature of 
the star utilizing the formula of \citet{gray2009}, $T_{\rm eff} [K] = 3759 - 135x$, 
where $x$ is the optically defined spectral type valid for optical types M0 through L8 with $x=0$ 
for an M0 type, $x=4$ for an M4 type star, etc. Adopting 3.5 for the value of $x$ this equation yields
$\sim$3290\,K as the effective temperature of HD\,113337B.
The age of the companion is estimated by comparing its position in the 
H-R diagram with the evolutionary tracks compiled by \citet{siess2000}, yielding $\sim$40$\pm$20\,Myr (see Fig.~\ref{HD113337}). 
This age estimate is adopted for the whole system 
\citep[a similar age estimate was proposed by ][]{rhee2007}. 

\begin{figure} 
\epsscale{1.}
\plotone{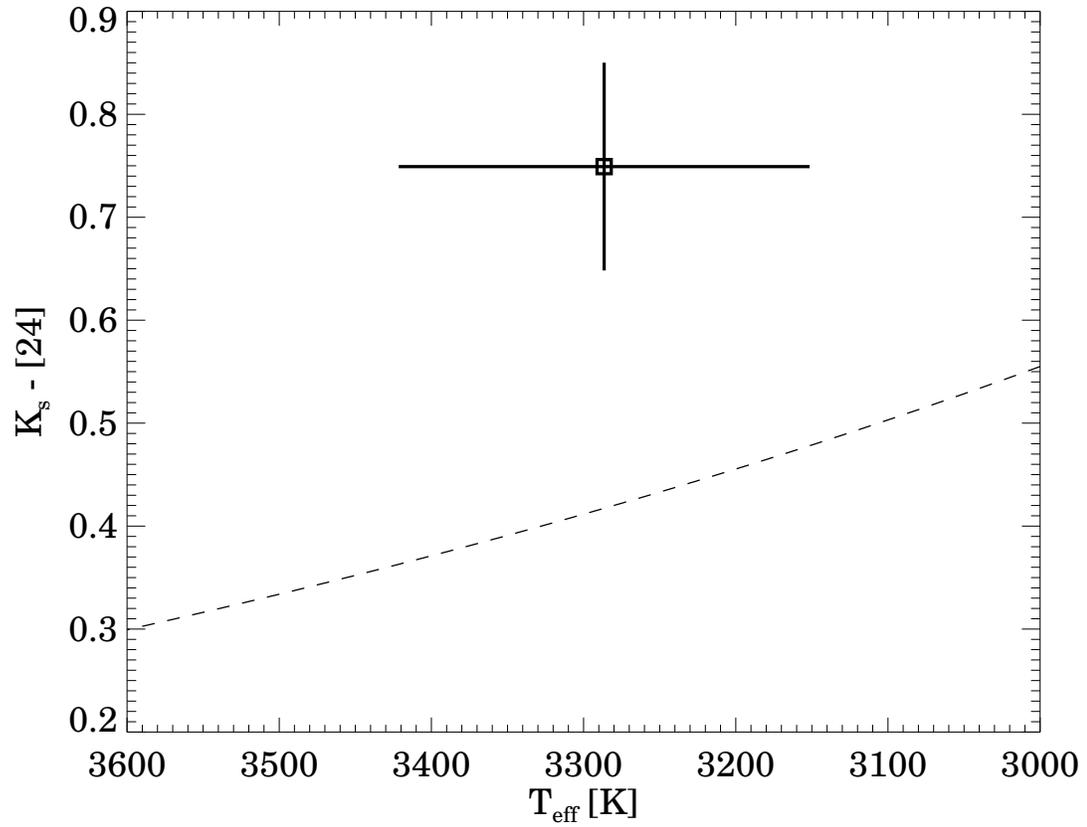}
\caption{$K_{\rm s}$-[24] color versus the effective temperature. The square shows the position of HD\,113337B, the dashed line
represents the locus of stellar photospheric colors determined
by \citet{gautier2007}. 
 \label{gautier}}
\end{figure}

Aperture photometry for HD\,113337B on the {24\micron} image yields a flux density of 
2.38$\pm$0.22\,mJy. After a conversion of the {24\micron} MIPS flux to magnitude we can compare 
the measured $K_s-[24]$ color of the star to the stellar photospheric colors determined
from a study of nearby M-type stars by \citet{gautier2007}. 
On the basis of this  comparison (see Fig.~\ref{gautier}) 
HD\,113337B shows $>$3$\sigma$ excess at {24\micron}.
The presence of circumstellar dust may explain the observed excess. 
However, since the significance of the excess detection 
is just above the 3$\sigma$ level, confirmation at additional wavelengths with higher sensitivity and spatial resolution 
is desirable. These observations can also help to exclude alternative explanations of the 
apparent excess, like the possibility of contamination by a background galaxy or the presence 
of an unresolved low mass companion.
Assuming that HD\,113337B hosts a debris disk, we note that the number of known debris disks around 
M-type stars is very limited \citep{forbrich2008} and all sources in this small sample showing excess 
at {24\micron} are younger than 50\,Myr.

\paragraph{HD\,205674.}

Apart from the star's galactic space velocity component toward the Galactic center, that differs from the average U velocity component 
of the AB\,Dor moving group more than 2$\sigma$, the other velocity and space components are in good agreement with the similar properties 
of this kinematic assemblage \citep[see Table~1 and Fig.~21 in ][]{torres2008}.  HD\,205674 fits well to the locus of AB\,Dor stars in the
color-magnitude diagram. 
It has an X-ray counterpart as well (see Table~\ref{agetable}). 
\citet{rhee2007} proposed an age of 300\,Myr. Since currently no reliable age-dating criteria 
for the youth of HD\,205674 are available and the membership status is questionable because of the deviation of 
U space velocity component from the cluster center, we adopt an age range which 
covers both the age of the moving group \citep[70--150\,Myr, ][]{luhman2005,torres2008}
and the 300\,Myr.

\section{Discussion} \label{discussion}

Our investigation of 82 F-type stars with the {\sl Spitzer Space Telescope} resulted in the detection of 
27 debris disks, out of which 9 are new discoveries. 
In the following we analyze the parameters of these disks with special attention to disk evolution and  
host star properties.

\subsection{Metallicity}

Table~\ref{tab1} lists metallicity estimates for 24 out of the 27 disk bearing stars. 
The average metallicity value in this sample is -0.09$\pm$0.09. For comparison we selected 9138 stars from the Geneva-Copenhagen Survey of Solar neighbourhood 
\citep{holmberg07} with effective temperature falling 
in the range spanned by our sample. Their average metallicity, -0.11$\pm$0.22, is in good agreement with the result for our 
stars. Narrowing the comparison sample to stars with similar age range would not change the conclusion.
Thus, our sample is similar to stars located in the Solar vicinity in terms of metallicity. 
This is in accordance with findings that the incidence of debris disks does not correlate with stellar metallicity \citep{beichman2006,greaves2006}.       
The lack of any such correlation may suggest that the formation of planetesimals is not sensitive to the metallicity in the protoplanetary disks.

\subsection{Multiplicity}

The effect of binarity on the presence of debris disks was studied by \citet{trilling2007}.  
They found that the incidence of debris disks is $\sim$50\% in systems with small ($<$3\,AU) or wide ($>$50\,AU) separations, 
even higher than the corresponding value among single systems.
In our 82-star sample there are 23 known multiple systems (13 have known separation). 
Three of the multiple systems harbor debris disks. Two disks are associated with the widest binaries with separation $>$4400\,AU, 
where the components practically can be regarded as isolated stars. It is interesting to note that in the case of HD\,113337, the 
secondary component might also harbor a debris disk based on the 24{\micron} image (Sect.~\ref{agedet}). 
The third disk encircles HD\,213429, which is a spectroscopic binary whose orbital solution \citep{2000A&AS..145..215P} indicates the smallest known separation 
in our sample (1.8\,AU), probably forming a circumbinary structure.   
Thus, all three binaries fall in groups of multiple systems where the incidence of debris disks 
is high according to \citet{trilling2007}. 

\subsection{Disk temperature}

\begin{figure} 
\epsscale{1.}
\plotone{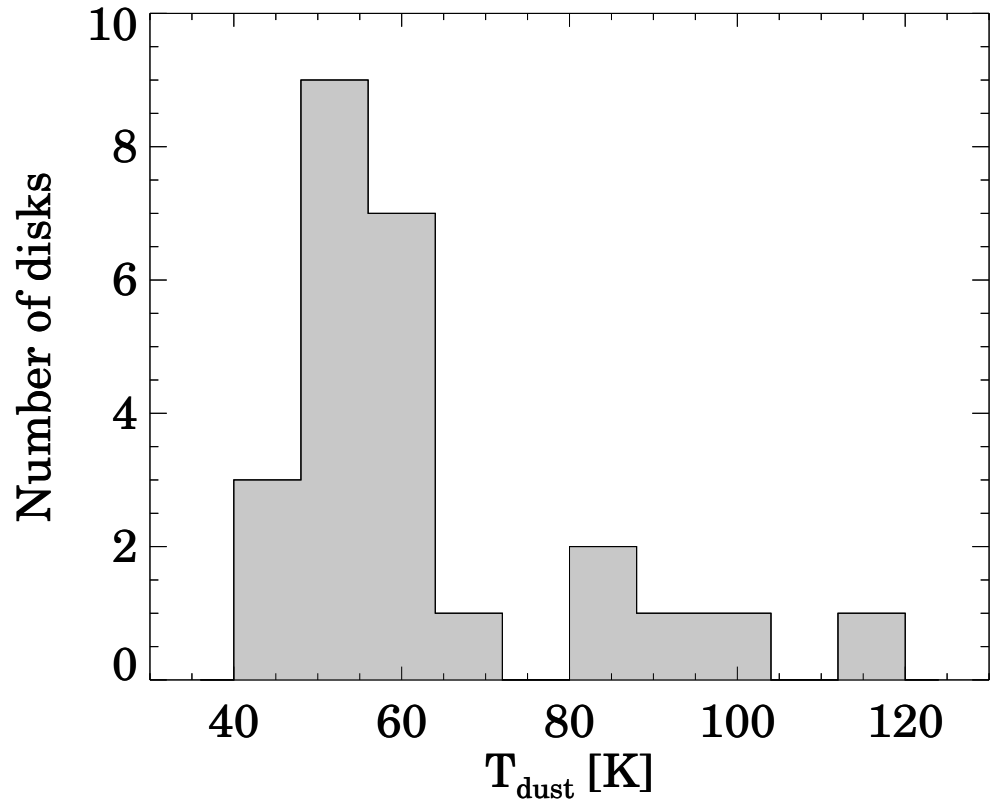}
\caption{Histogram of the derived dust temperatures. In those cases where the presence of two separated dust rings are assumed  
only the temperature of the colder ring has been taken into account. 
 \label{tdust}}
\end{figure}

The dust temperature provided by our modeling (Sect.~\ref{modeling}) is a fundamental disk parameter, whose computation 
includes only a few assumptions and can be determined with confidence. 
In Fig.~\ref{tdust} we plot a histogram of the derived dust temperature values. 
In those five cases when the SED was modeled by multiple dust rings (Table~\ref{diskprop2}) only the colder component 
was taken into account. Disks with an upper limit for the temperature were also omitted.
Most of the disks in Fig.~\ref{tdust} have temperatures falling in the range of 40--70\,K 
(in the Solar System the Kuiper-belt exhibits similar temperature). 
A smaller sample shows temperatures 70$<T_{\rm dust}<$120\,K. Note that our complete sample of F-type stars 
contains also four disks of even higher temperature (135--200\,K) published in our previous paper \citep{moor2009}.  
While the age of the disk host stars belonging to the colder group ranges between 12\,Myr and 3,100\,Myr, 
all warmer disks encircle stars with age $<$150\,Myr.   
Using the formulae~1-5 in \citet{grigorieva2007}, we computed the sublimation time for possible ice grains with radii range between 
the blowout limit and  10{\micron} using the derived 
grain temperatures.
Apart from the case of HD\,24636 the timescale of sublimation is significantly longer than the age of 
the system.
Thus the presence of icy grains cannot be excluded in most of our disks. Note, however, that in such disks the photo-desorption can 
remove icy mantles of grains more effectively than the sublimation \citep[][]{grigorieva2007}. Moreover, grains are thought to 
be produced in collisions that can provide enough heat to sublimate the icy mantles of the particles 
\citep{czechowski2007}.     
In HD\,24636, the sublimation timescale is lower than the age of the system even for icy grains with a radius of a few centimeters. 
Taking into account the uncertainties of the timescale estimates and the derived parameters in this disk, the sublimation and the collisional 
timescales for the smallest grains are in the same order. If the grains are icy in this system, this finding 
suggests that sublimation may have a role in the removal of small grains.

\subsection{Fractional luminosity} \label{fdustsect}

\begin{figure} 
\epsscale{1.}
\plotone{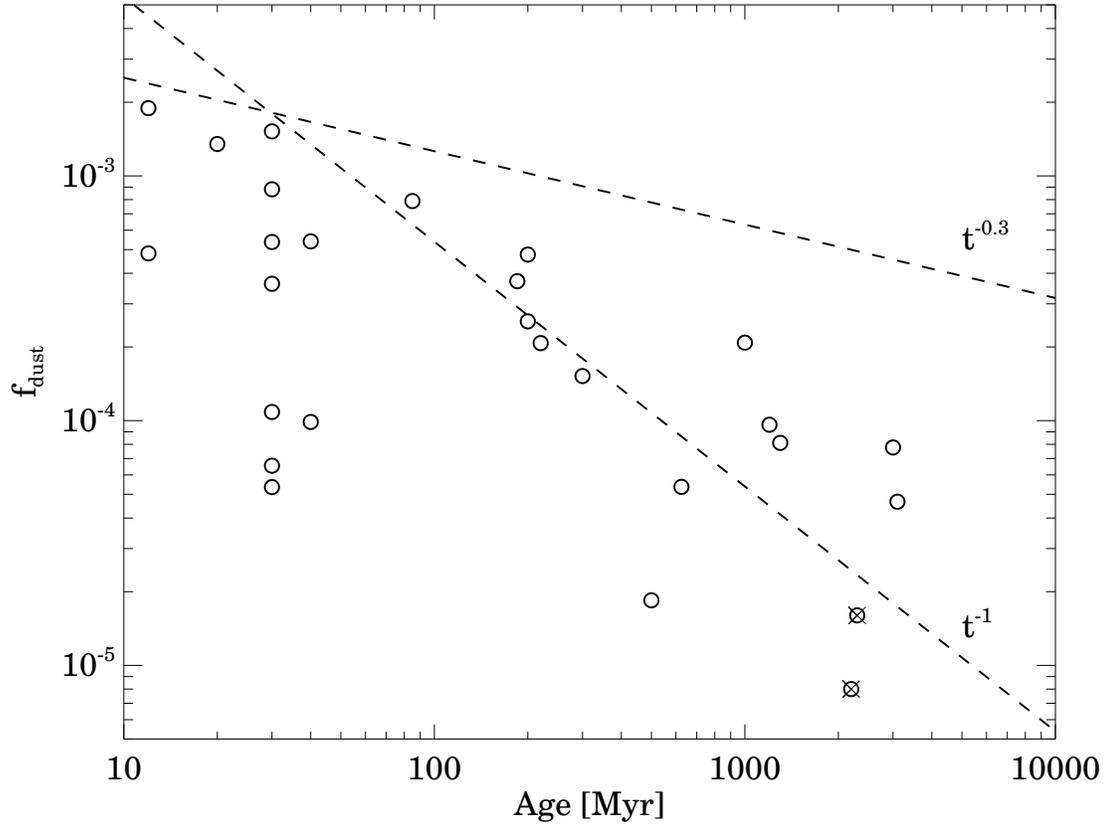}
\caption{Fractional luminosity of the infrared excess as a function of age. Typical uncertainties 
in fractional luminosity range from 0.02\,dex to 0.12\,dex. 
Crosses mark disks where the excess emission was measured only in one IR band, thus they have less 
reliable fractional luminosities. 
Models of debris disk evolution predict that the decay of fractional luminosity is proportional 
to t$^{-\alpha}$ where $\alpha$ ranges between 0.3 and 1.0 (see Sect.\,\ref{fdustsect}).
For comparison with our data, we plotted the two extremes of the evolutionary models with 
arbitrary normalization (dashed lines). 
The distribution of the data points, in particular their upper envelope, seems to suggest a decay rate
halfway between the two extremes. 
\label{fdust}}
\end{figure}

Fractional luminosity is a fundamental observable parameter of debris disks that is frequently used to characterize 
the amount of circumstellar matter. Fig.~\ref{fdust} shows the derived fractional luminosities for our disks as a function of 
the system's age. Although the dust fractional luminosity shows a large dispersion at any given age, a clear decline with time 
can be recognized. 
Following the evolution of a narrow debris ring co-located with a planetesimal belt 
in an analytical steady state collisional evolution model, \citet{wyatt2007} found that the fractional luminosity of the disk varies with 
time as  $\sim t^{-1}$. In the framework of a more realistic numerical evolution model, \citet{loehne2008} lifted some 
simplifications of the above mentioned analytical model by taking into account that planetesimal strength depends on their size and the 
complex grain size distribution close to the blowout limit. 
They predicted that the decay of dust luminosity is proportional to 
$t^{\rm -\alpha}$, with $\alpha = 0.3$ to $0.4$. \citet{kenyon2008} dealt with the formation and evolution of an extended 
planetesimal disk using a numerical model. 
They found that the maximum in the dust emission coincides with the formation of 1000-2000\,km 
size planetesimals at the inner edge of the disk. 
Therefore, the fractional luminosity of a disk starts to rise when 
large planetesimals appear at the inner edge of the disk
and stir the motion of smaller bodies, initiating their destructive collisions. 
After the rise to a peak brightness, a slow, roughly power-law decline is predicted, 
$f_{\rm dust}\propto t^{\rm -\alpha}$ with $\alpha\sim 0.6-1.0$. 
In Fig.~\ref{fdust} we plotted the two extremes of these model predictions ($\alpha$=0.3 and $\alpha$=1.0). 
The distribution of the data points, in particular their upper envelope, seems to suggest a decay rate halfway between the two extremes.

The steady state evolution model
of \citet{wyatt2007} predicted that in the course of the evolution of a narrow debris ring, at any given age there is a maximum fractional dust luminosity ($f_{\rm max}$), 
since originally more massive disks eat up their mass faster. 
Disks with fractional luminosity of $\gg$\,$f_{\rm max}$ could be the result of a transient event that increases the dust production 
for a short period \citep[for possible transient effects, see][]{wyatt2008}. 
Using their formula \citep[eq~20. in][]{wyatt2007} and adopting their fixed model parameters 
(belt width: $dr/r=0.5$, planetesimal strength: 
$Q^*_D$\,=\,200\,J\,kg$^{-1}$, 
planetesimal eccentricity: $e=0.05$, diameter of largest planetesimal in cascade: $D_c=2000$\,km) we computed the $f_{\rm max}$ values for our debris 
systems. By comparing the measured fractional luminosities (see Tables~\ref{diskprop} and \ref{diskprop2}) with the calculated maxima we conclude
that all our cold disks can be consistent with a steady-state evolutionary scenario within the uncertainties of the model. 
In reality, transient processes might be present but their influence is undetectable with this simple comparison.
Note that this model uses a pre-stirred planetesimal belt (i.e. the stirring is initiated at $t=0$) 
and that in the numerical model of \citet{loehne2008} 
the dust luminosity depends on the initial mass even at late evolutionary stages. Nevertheless, taking into account the delayed stirring or the 
dependence on initial disk mass do not affect our conclusions since both alterations would increase the $f_{\rm max}$ at a specific age.

Due to the decline in dust mass with time, high fractional luminosity debris disks are associated mainly with young stars. 
The rare exceptions tend to harbor very hot debris dust, in which  
the ejection of a small amount of transient dust can cause a significant rise in the fractional luminosities (because of the disk's proximity to the star).   
\citet{moor2006} hypothesized that all stars with $f_{\rm dust} > 5\cdot10^{-4}$ are younger than 100\,Myr, therefore a high 
$f_{\rm dust}$ values can be used as an age indicator \citep[see also in][but with a limit of $f_{\rm dust} > 10^{-3}$]{zs04}. 
Among our new discoveries there are two debris disks with fractional luminosity exceeding the limit of $f_{\rm dust} > 5\cdot10^{-4}$, 
HD\,3670 and HD\,36968. 
Both objects are proposed to belong to young moving groups (age $\leq$ 30\,Myr) in agreement with the hypothesis.

\subsection{Disk radii} \label{drad}

In the course of disk radius estimates we assumed the dust grains to be confined to a narrow ring and that they
interact with the stellar radiation as a blackbody. The resulting $R_{\rm dust}$ values 
correspond to minimum possible radii and possibly underestimate the ``true'' radii, 
meaning that the real size of a specific debris disk could depart significantly from the derived $R_{\rm dust}$.
However, if the disks are composed of similar dust grain populations then
the differences between the real dust distribution and the assumed one would 
shift the computed values in a similar way \citep{wyatt2008}, i.e. the relative radii of the disks 
are better constrained than the absolute values.
Thus, in the following analysis we assume that the derived values can be used to study general trends in the disk radii distribution.

Destructive collisions between planetesimals can occur when the collision velocity exceeds a critical value 
that requires a dynamically excited (stirred) disk. In {\sl self-stirring} models the formation of large planetesimals 
in collisional coagulation among smaller planetesimals naturally leads to the formation of a debris ring 
as well. These oligarchs can stir up the motion of the leftover smaller bodies initializing a collisional cascade. 
According to the models of \citet{kenyon2008}, the maximum of the dust production via these collisions 
coincides roughly with the formation of $\sim$1000\,km planetesimals in the same region. 
Since the formation of such large bodies requires longer time at larger radial locations, the site of the dust production in an 
extended planetesimal disk is thought to propagate outward. Secular perturbations by giant planets -- formed previously 
in the inner regions of the protoplanetary disk -- can also initialize a collisional cascade in a planetesimal disk. 
\citet{mustill2009} concluded that planetary stirring can also eventuate in an outwardly propagating dust ring.  
In some regions the time-scale of this process can be even shorter than the growth time of $\sim$1000\,km planetesimals \citep{mustill2009}.
Stellar flybys can also initiate more energetic collisions in a planetesimal disk. However, such rare events are not likely 
to be responsible for large numbers of debris systems. 

\begin{figure} 
\epsscale{1.}
\plotone{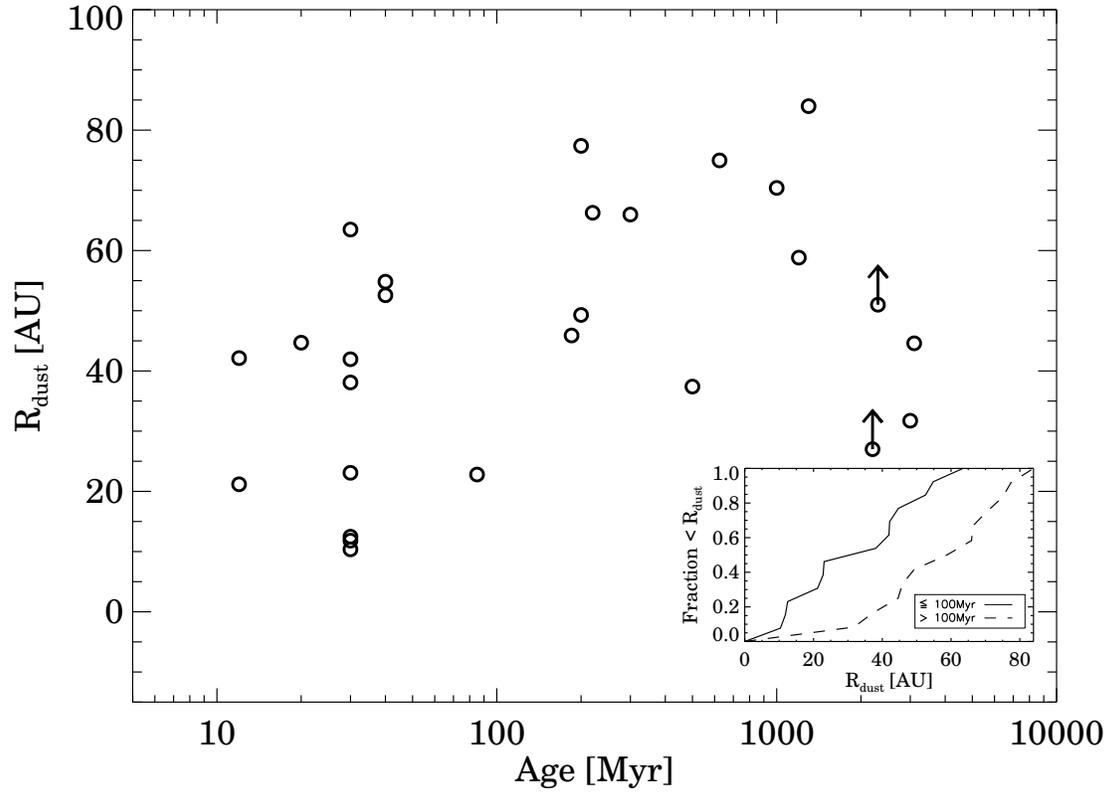}
\caption{Derived dust radii as a function of age. Typical formal uncertainties in the derived disk radii range from 5\% to 20\%. 
The inset shows a comparison between the cumulative distribution of disks' radii around stars with age $<$100\,Myr and 
stars with age $>$100\,Myr (disks with lower limits for radius are not included). 
The inset does not cover any symbols.   
 \label{rdust}}
\end{figure}

Figure~\ref{rdust} shows the derived radii of the dust rings as a function of age.
The radii of the rings show large dispersion at any given age. 
The data points seem to suggest an increase of the upper envelope of the distribution with increasing age. 
It is even more salient that while older systems (age $>$ 100\,Myr) harbor dust rings located at radii of $>$30\,AU, 
around younger systems there are several disks at radial location between 10\,AU and 30\,AU in a region
where Saturn, Uranus, and Neptune orbit in the current configuration of our Solar System. 
The inset in Figure~\ref{rdust} shows a comparison between the cumulative distribution of disk radii around stars with age $<$100\,Myr and 
stars with age $>$100\,Myr (disks with lower limit for radius are not included). 
This comparison suggests a significant difference between the two distributions. 
Based on a Wilcoxon test, the null hypothesis that the two samples come from identical populations can be rejected on a 99.9\% confidence level.

The lack of dust rings with small radii at larger ages as well as the hint for an increase of the upper envelope of the distribution 
are in good accordance with the predicted outward propagation of the dust production site as the result of self- or planetary stirring. 
\citet{rhee2007} also reported increasing radii at larger ages for a sample of late B- and A-type stars. 
The latter authors estimated disk radii identically to our approach (Sect.~\ref{modeling}), thus direct comparison 
with our results is meaningful. 
Apart from seven very extended disks (where the radius estimate was based on IRAS data only) the general distribution of 
points in Fig.~7 of \citet{rhee2007} is very similar to our results shown in Fig.~\ref{rdust}. 
 
\begin{figure} 
\epsscale{1.}
\plotone{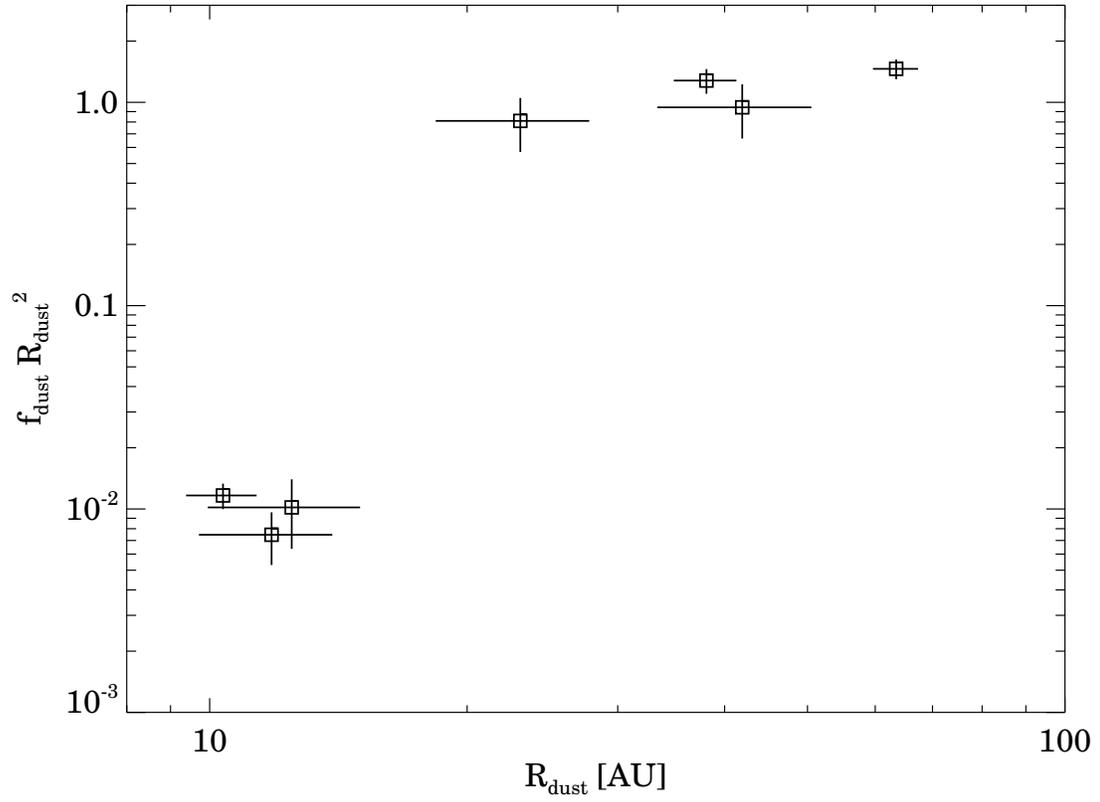}
\caption{The relative dust masses ($f_{\rm dust}\cdot R_{\rm dust}^2$ see Sect~\ref{drad}) for the disks with ages of 30\,Myr as a function of 
the derived radii. 
 \label{rdust2}}
\end{figure}

\citet{kenyon2008} predicted that the pace of the outward propagation in a disk depends on the disk mass: the more massive the disk the faster the 
spread outwards. Thus, during the active period of self-stirring evolution (when the expanding ring reaches the outer boundary of the disk), 
in an initally more massive disk, the bright ring associated with the formation of Pluto-sized planetesimals is located at
larger radius at any given age. 
This effect offers a good explanation for the large scatter in dust ring radii we observe for younger stars ($<$50\,Myr). 
In order to test this hypothesis,
we estimated the relative dust masses of the disks as $M_{\rm dust}\propto f_{\rm dust} R_{\rm dust}^2$ \citep[see eq.~7 in][]{wyatt2008} for 
the seven 30\,Myr-old disks.
We assume that the relative disk mass distribution 
does not change during disk evolution, thus the current relative masses reflect the initial mass distribution.  
For these disks we displayed the relative dust masses as the function of the estimated radii of the dust ring in Fig.~\ref{rdust2}.
The obvious trend is fully consistent with predictions of \citet{kenyon2008}: more massive disks are located farther from the star.

\begin{figure} 
\epsscale{1.}
\plotone{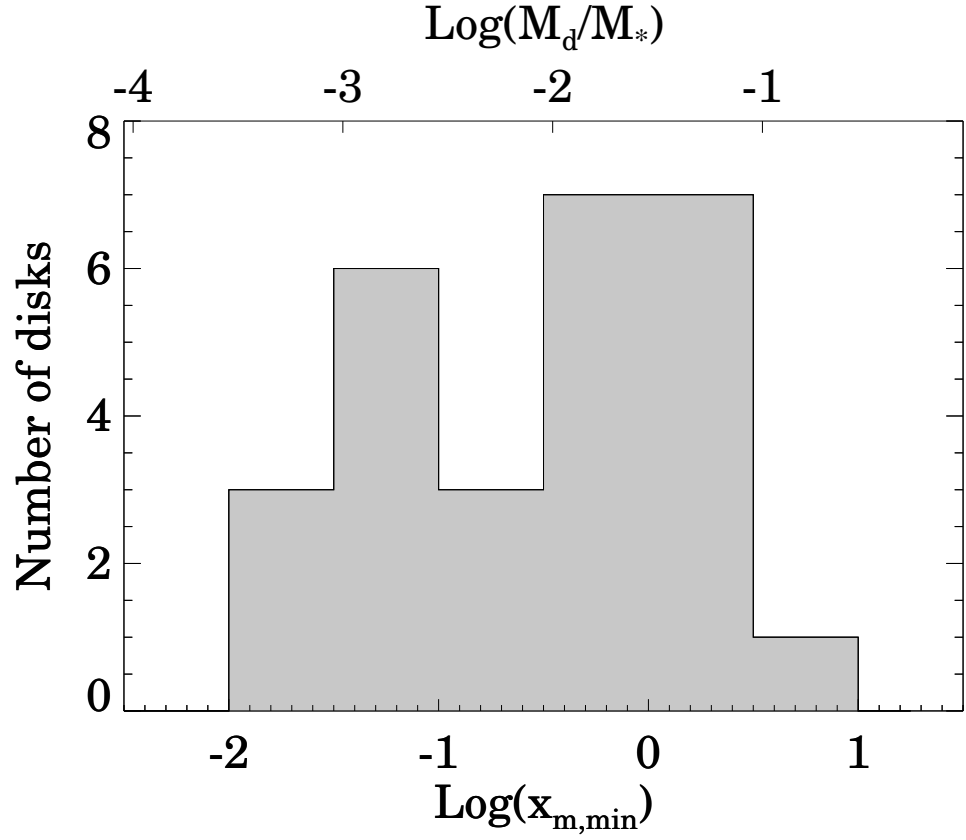}
\caption{Histogram of the minimum $x_{\rm m,min}$ values that are required for the observed disk to become self stirred in $t<t_{\rm system}$. 
Initial disk-to-star mass ratios corresponding to the computed $x_{\rm m,min}$ are indicated on the top of the graph (Sect.~\ref{drad}). 
  \label{diskmass}}
\end{figure}

\citet{kenyon2008} derived the characteristic timescales of planetesimal disk formation and evolution processes in 
a disk with an initial surface density distribution of  
\begin{equation}
\Sigma = \Sigma_0(M_*) x_m (a/a_0)^{-3/2},
\end{equation} where $\Sigma_0$ is the reference surface density at a radius of $a_0$ = 30\,AU, while $x_m$ is a scaling factor. 
The reference surface density was scaled with the stellar mass as $\Sigma_0(M_*) = 0.18(M_*/M_{\sun})$\,g\,cm$^{-2}$ 
($\Sigma_0=0.18$\,g\,cm$^{-2}$ corresponds to the minimum mass solar nebula density at the radius of 30\,AU).
According to their results, the timescale for the formation of the first 1000\,km icy planetesimals at a radius $a$
can be computed as: 
\begin{equation} 
t_{\rm 1000} = 145 x_{\rm m}^{-1.15} (a/80\,{\rm AU})^3 (2M_{\sun}/M_*)^{3/2} [\rm Myr].
\end{equation}
Adopting the radii of dust rings as the radii of the planetesimal belts and taking into account the derived ages of 
the systems, we can estimate a minimum $x_{\rm m}$ value (practically a minimum initial surface density) that is necessary for 
the formation of Pluto-sized planetesimals 
occurring at the given radial location. Assuming a fix disk geometry ($R_{\rm in}=10$\,AU, $R_{\rm out}=100$\,AU), an estimated 
initial disk mass ($M_{\rm d}$) 
and an initial disk-to-star mass ratio ($M_{\rm d}/M_*$) can also be computed (adopting the canonical gas-to-dust ratio of 100). 
Fig.~\ref{diskmass} shows the derived minimum $x_{\rm m,min}$ values and the minimum initial disk-to-star mass ratios for our sample. 
Eight disks (HD\,3670, HD\,15115, HD\,16743, HD\,30447, HD\,36968, HD\,113337, HD\,170773, HD\,192758) require an initial surface 
density higher than that of the minimum solar mass nebula. The largest $x_{\rm m,min}$ value (3.3) was obtained
for HD\,16743. Assuming that self-stirring occurred in these systems and comparing the 
derived $M_{\rm d}/M_*$ ratios with the cumulative distribution 
of the corresponding ratios for protoplanetary disks in the Taurus and Ophiucus star forming regions 
\citep[Fig. 7,][]{andrews2007}, we concluded  
that these eight disks might represent the high end of the disk mass distribution. 
\citet{mustill2009} argued that a disk with $x_{\rm m,min}\geq 10$ is unlikely to be the site of a collisonal cascade ignited by 
self-stirring. 
Since the highest $x_{\rm m,min}$ in our sample is 3.3, all our disks could be the result of self-stirring. 
Note, however, that in the calculation of dust ring radii 
we assumed the presence of large blackbody grains. The presence of smaller grains that are ineffective emitters would have resulted in larger radial locations. 
Due to the strong dependence of the minimum initial surface density on the radial location, the uncertainty of $R_{\rm dust}$ 
can affect (usually increase) the value of $x_{\rm m,min}$. 

Six out of the eight disks where our calculations implied high initial surface density are younger than 60\,Myr (see Table~\ref{agetable}) 
and are located within 80\,pc (Table~\ref{tab1}). 
These systems might be good targets for future planet searching surveys via direct imaging, because 1) according to the 
models, disks with high initial surface density are favorable for planet formation; 2) young ages make giant planets more easily detectable
(since they are still bright).  
We note that HR\,8799, Fomalhaut, and $\beta$\,Pic, systems hosting massive outer planets, harbor massive bright debris disks as well.

     In the present Solar System, planetesimals reside mainly in two spatially separated belts, the Kuiper belt and the main asteroid belt.  
Infrared and submillimeter observations of debris disks implied that stars harbouring multiple component disks might be 
common \citep[][and references therein]{chen2009,smith2010}.
The dust replenishment processes may also be similar to those in our Solar System: collisions between planetesimals are thought to lead to the production of 
fresh dust in the outer cold rings, while besides the collisions, sublimation of icy planetesimals (scattered from the outer reservoir) 
can also contribute to the maintenance of the inner warm dust belt.  
Several of these systems are younger than 100\,Myr and their age coincides well 
with the era of terrestrial planet formation in our Solar System \citep[see][for a review]{apailauretta2010}, a process believed to be accompanied with the 
release of a huge amount of 
dust grains due to collisions between large protoplanets \citep{kenyon2004b}. 
The warm dust around these young stars might either originate from terrestrial planet formation  
\citep[e.g.][]{lisse2008} or from the collisional grinding of a rocky asteroidal belt. 
In Sect.~\ref{modeling} we found five debris systems where the SED can be fitted better with a two-component temperature model than 
with a single temperature one.
Assuming that the warm component is associated with inner dust rings 
we propose that these five stars harbor two spatially separated dust rings co-located with two distinct planetesimal belts.
All of these systems are younger than 40\,Myr. 
Taking into account the above mentioned arguments, these five warm dust rings might also be linked to the formation of terrestrial planets 
or collisional evolution of an asteroid belt. 
Since the characteristic temperature of the dust grains (150--180\,K) significantly exceeds the sublimation 
temperature of comets \citep[$\sim$110\,K,][]{wyatt2008}, 
cometary activity can also play a role in the replenishment of warm dust particles.
\citet{moor2009} found two young F-type members of the Tucana-Horologium association (HD\,13246 and HD\,53842) that harbor warm debris rings with  
properties ($T_{\rm dust}\sim$150--170\,K, $f_{\rm dust}\sim$0.5--1.7$\times10^{-4}$) very similar to those of the currently analyzed five disks. 
The inner zone of the two Tucana-Horologium stars 
might also be sites of intense planet formation, although, no cold excess emission has been seen toward them. 
The lack of observable cold excess may suggest a relatively dust-free outer disk perhaps caused by disk truncation.

\section{Summary}

With the aim of investigating the properties and evolutionary trends of debris disks around F-type stars, 
we observed 82 targets with the {\sl Spitzer Space Telescope} using the {\sl IRS} and {\sl MIPS} instruments 
to obtain infrared spectroscopy and photometry.  The core sample was selected based on hints for excess emission in earlier infrared observations, 
and this sample was supplemented by F-type members of some young kinematic groups. 
We found 27 stars that harbor debris disks, nine of which are new discoveries.  
Apart from two stars, all of the observed disks exhibit excess in the IRS spectra, and 17 disks were detected even at 160{\micron}. 
In two cases, the emission was found to be marginally extended on the 70{\micron} MIPS images. 
Combining the MIPS and IRS measurements with additional data taken from former infrared space missions (IRAS, ISO) and ground-based submillimeter/millimeter (SCUBA, IRAM) 
observations, we compiled the spectral energy distribution of the targets. 
Thanks to the large number of measurements at different wavelengths we achieved excellent spectral coverage for most of 
our debris systems. 
We have modeled the excess emission of 22 debris disks using a single temperature dust ring model and 5 debris systems 
with a two-temperature model. The latter systems 
may contain two dust rings around the star. 
All of our disks were found to be collision dominated.

Among the 27 disks, 15 encircle stars that likely belong to young moving groups, offering 
accurate age estimates for these systems. We identified five new moving group members. HD\,15745, that exhibit one of the largest fractional luminosities 
among northern debris disks, is a likely member of the $\beta$\,Pic moving group. HD\,36968, a system with high fractional luminosity, can be 
assigned to the recently discovered Octans kinematic group \citep{torres2008}, being the first member where the existence of a 
debris disk was discovered.        

In accordance with the expected trends, the fractional luminosity of the disks declines with time. 
The distribution of data points, in particular their upper envelope, seems to suggest a decay rate falling within the 
range of the model predictions \citep[$f_{\rm dust}\propto t^{-0.3 \ldots -1.0}$,][]{dominik2003,wyatt2007,loehne2008,kenyon2008}.  
All disks in the sample seem to be consistent with 
quasi steady-state evolutionary scenarios of \citet{wyatt2007}.  

Assuming blackbody grains, we computed the radial location of the cold dust rings in all debris systems.   
We found a hint for an increase of the upper envelope of the radius distribution with increasing age.
While older systems (age $>$ 100\,Myr) harbor dust rings located at radii of $>$30\,AU, 
around younger systems there are several disks at radial location between 10 and 30\,AU. 
Both findings are in accordance with the predictions of self- or planetary stirring theories of 
\citet{kenyon2008} and \citet{mustill2009}.

\section{Acknowledgment}

We thank an anonymous referee for his/her careful comments which improved the manuscript.
Support for this work was provided
by NASA through contract 1311495 to Eureka Scientific.
This work was partly supported by the Hungarian Research Fund OTKA K81966.
I.P. and D.A. acknowledge support through the Spitzer NASA/RSA contract number 1351891.
The research of \'A. K. is supported by the Nederlands Organization for Scientific Research.
L.L.K. has been supported by the Australian Research Council, the University of Sydney, the 'Lend\"ulet' 
Young Researchers Program of the Hungarian Academy of Sciences, and the Hungarian OTKA grants
K76816 and MB0C 81013. T. Cs. acknowledges support from the FP6 Marie-Curie Research Training Network Constellation: The origin of
stellar masses (MRTN-CT-2006-035890).
This work is based in part on observations made with the Spitzer Space Telescope, which is operated by the 
Jet Propulsion Laboratory, California Institute of Technology under a contract with NASA. 
Partly based on observations carried out with the IRAM 30m Telescope. IRAM is supported by INSU/CNRS (France), MPG (Germany) and IGN (Spain).
We are grateful to the IRAM staff for help provided during the observations.
This research has made use of the VizieR catalogue access tool, CDS, Strasbourg, France.
The publication makes use of data products from the Two Micron All Sky Survey, which is a joint project of the University of Massachusetts and the Infrared Processing and Analysis
Center/California Institute of Technology, funded by the National Aeronautics and Space Administration and the National Science Foundation.

{\it Facilities:} \facility{Spitzer ()}, \facility{ISO ()}, \facility{IRAS ()}, \facility{IRAM:30m ()}.

\appendix

\clearpage



\LongTables
\begin{deluxetable}{lcccccccccccc}                                                                                                                    
\tablecaption{Basic properties of our target list \label{tab1}}                                                                                  
\tablewidth{0pt}                                                                                                                                      
\tablehead{ 
\colhead{(1)} & \colhead{(2)} & \colhead{(3)} &                                                                                                          
 \colhead{(4)} & \colhead{(5)} & \colhead{(6)} &  \colhead{(7)}                                                                              
& \colhead{(8)} & \colhead{(9)} & \colhead{(10)} & \colhead{(11)} &                                                                  
\colhead{(12)} &\colhead{(13)}  \\                                                                                                                                                             
\colhead{ID} & \colhead{SpT} & \colhead{V} &                                                                                                          
 \colhead{D} & \colhead{[Fe/H]} & \colhead{Ref.} &  \colhead{$\rm \log{g}$}                                                                              
& \colhead{$\rm T_{eff}$} & \colhead{$\rm A_V$} & \colhead{Mult.} & \colhead{Sep.} &                                                                  
\colhead{Ref.} &\colhead{Sel. cr.}  \\                                                                                                                
\colhead{} &  \colhead{} & \colhead{[mag]} &                                                                                                          
 \colhead{[pc]} & \colhead{} & \colhead{} & \colhead{}                                                                                                
& \colhead{[K]} & \colhead{[mag]} & \colhead{} & \colhead{[\arcsec]} &                                                                                
\colhead{}                                                                                                                                            
}                                                                                                                                                     
\startdata                                                                                                                                            
   HD 3670 &          F5V &    8.23 &  (76.0) &   -0.13 &		  1 &	 4.25 &    6480 &    0.00 &   N &   \ldots &   \ldots &        2 \\
   HD 14691 &          F0V &    5.43 &   29.7  &   -0.12 &		1,2 &	 4.25 &    6800 &    0.00 &   N &   \ldots &   \ldots &        1 \\
   HD 15060 &          F5V &    7.02 &   76.0  &   -0.14 &          1 &	 4.00 &    6260 &    0.00 &   N &   \ldots &   \ldots &        1 \\
   HD 15115 &           F2 &    6.79 &   45.2  &   -0.06 &		  1 &	 4.25 &    6780 &    0.00 &   N &   \ldots &   \ldots &        1 \\
   HD 15745 &           F0 &    7.47 &   63.5  &   -0.10 &		  1 &	 4.25 &    6860 &    0.00 &   N &   \ldots &   \ldots &        1 \\
   HD 16743 &  F0/F2III/IV &    6.78 &   58.9  &   -0.13 &		  1 &	 4.25 &    7000 &    0.00 &   Y &    216.2 &	    1 &        1 \\
   HD 17390 &       F3IV/V &    6.48 &   48.0  &    0.03 &		  1 &	 4.25 &    6840 &    0.00 &   N &   \ldots &   \ldots &        1 \\
  BD+49{\degr} 896 &          F4V &    9.68 & (175.0) &   -0.15 &		  3 &	 4.25 &    6740 &    0.06 &   N &   \ldots &   \ldots &        1 \\
   HD 20759 &          F5V &    7.70 &   76.8  &   -0.48 &		  1 &	 4.00 &    6280 &    0.00 &   N &   \ldots &   \ldots &        1 \\
   HD 24636 &       F3IV/V &    7.13 &   54.1  &   -0.11 &		  1 &	 4.25 &    6820 &    0.00 &   N &   \ldots &   \ldots &        2 \\
   HD 25570 &          F2V &    5.45 &   34.9  &   -0.29 &		  1 &	 4.00 &    6760 &    0.00 &   N &   \ldots &   \ldots &        1 \\
   HD 25953 &           F5 &    7.83 &   55.2  &   -0.22 &		  1 &	 4.25 &    6240 &    0.00 &   N &   \ldots &   \ldots &        2 \\
   HD 27429 &      F3:V... &    6.11 &   48.3  &    0.01 &		  1 &	 4.00 &    6720 &    0.00 &   Y &   \ldots &	    2 &        1 \\
   HD 30447 &          F3V &    7.85 &   80.3  &   -0.21 &		  1 &	 4.25 &    6800 &    0.00 &   N &   \ldots &   \ldots &      1,2 \\
   HD 32195 &          F7V &    8.14 &   61.0  &   -0.13 &		  1 &	 4.25 &    6180 &    0.00 &   N &   \ldots &   \ldots &        2 \\
   HD 30743 &       F3/F5V &    6.27 &   33.8  &   -0.38 &	    1,2,4,5 &	 4.25 &    6440 &    0.00 &   N &   \ldots &   \ldots &        1 \\
   HD 33081 &          F7V &    7.04 &   50.6  &   -0.19 &		  1 &	 4.25 &    6360 &    0.00 &   N &   \ldots &   \ldots &        1 \\
   HD 33276 &         F2IV &    4.81 &  165.0  &    0.27 &		6,7 &	 3.25 &    6920 &    0.00 &   Y &    0.300 &	    3 &        1 \\
   HD 34739 &       F7IV/V &    9.33 & (121.0) &  \ldots &	     \ldots &	 4.25 &    6280 &    0.00 &   N &   \ldots &   \ldots &        1 \\
   HD 35114 &          F6V &    7.39 &   48.3  &  -0.18 &		  1 &	 4.25 &    6200 &    0.00 &   N &   \ldots &   \ldots &        2 \\
   HD 35841 &          F5V &    8.91 & ( 96.0) &  \ldots &	     \ldots &	 4.25 &    6460 &    0.00 &   N &   \ldots &   \ldots &      1,2 \\
   HD 36248 &           F8 &    8.05 &   73.6  &    0.11 &		  1 &	 4.25 &    5960 &    0.06 &   Y &    3.800 &	    3 &        1 \\
   HD 36968 &          F2V &    9.02 & (140.0) &  \ldots &	     \ldots &	 4.25 &    6880 &    0.00 &   N &   \ldots &   \ldots &      1,2 \\
   HD 37402 &          F6V &    8.38 &   78.2  &   -0.15 &		  1 &	 4.25 &    6160 &    0.00 &   N &   \ldots &   \ldots &        2 \\
   HD 38905 &       F6/F7V &    9.73 & (140.0) &  \ldots &	     \ldots &	 4.25 &    6240 &    0.04 &   N &   \ldots &   \ldots &        1 \\
   HD 47412 &           F2 &    6.82 &  111.0  &   -0.03 &		  1 &	 4.00 &    6300 &    0.00 &   Y &    0.500 &	    3 &        1 \\
   HD 48391 &           F5 &    7.89 &   58.3  &   -0.20 &		  1 &	 4.25 &    6160 &    0.00 &   N &   \ldots &   \ldots &        2 \\
   HD 50571 &     F7III-IV &    6.11 &   33.6  &   -0.02 &		1,2 &	 4.25 &    6480 &    0.00 &   N &   \ldots &   \ldots &        1 \\
   HD 55003 &           F2 &    7.04 &   60.1  &  -0.02 &		  1 &	 4.00 &    6340 &    0.00 &   Y &   \ldots &	    2 &        1 \\
   HD 56099 &           F8 &    7.62 &   86.7  &   -0.03 &		  1 &	 4.00 &    6060 &    0.00 &   Y &    0.130 &	    4 &        1 \\
   HD 58853 &          F5V &    9.07 &  122.4  &  \ldots &	     \ldots &	 4.25 &    6280 &    0.00 &   N &   \ldots &   \ldots &        1 \\
   HD 61518 &          F5V &    7.88 &   62.0  &   -0.19 &		  1 &	 4.25 &    6340 &    0.00 &   N &   \ldots &   \ldots &        2 \\
   HD 67587 &           F8 &    6.65 &   46.8  &   -0.19 &		  1 &	 4.00 &    5980 &    0.00 &   Y &   \ldots &	    2 &        1 \\
   HD 69351 &           F8 &    7.16 &   80.0  &    0.08 &		  1 &	 4.00 &    5980 &    0.00 &   Y &    1.400 &	    3 &        1 \\
   HD 79873 &           F5 &    6.73 &   68.3  &    0.13 &		  1 &	 4.00 &    6400 &    0.00 &   Y &    2.100 &	    3 &        1 \\
   HD 82821 &           F8 &    8.69 &   74.7  &  \ldots &	     \ldots &	 4.25 &    6140 &    0.00 &   Y &   \ldots &	    2 &        1 \\
   HD 86146 &         F6Vs &    5.11 &   28.1  &    0.02 &		  1 &	 4.00 &    6360 &    0.00 &   Y &   \ldots &	    5 &        1 \\
   PPM 7774 &           F5 &    8.96 & (116.0) &  \ldots &	     \ldots &	 4.25 &    6520 &    0.00 &   N &   \ldots &   \ldots &        1 \\
  HD 103257 &          F2V &    6.62 &   63.4  &   -0.39 &		  1 &	 4.00 &    6980 &    0.00 &   N &   \ldots &   \ldots &        1 \\
  HD 107067 &        F8... &    8.69 &   66.0  &   -0.09 &	      1,8,9 &	 4.25 &    6060 &    0.00 &   N &   \ldots &   \ldots &        1 \\
  HD 108102 &        F8... &    8.12 &   95.1  &   -0.13 &		  1 &	 4.00 &    6060 &    0.00 &   N &   \ldots &   \ldots &        1 \\
  HD 113337 &          F6V &    6.01 &   36.9  &    0.06 &	       1,10 &	 4.25 &    6600 &    0.00 &   Y &    119.7 &	    3 &        1 \\
  HD 114905 &          F7V &    6.83 &   61.4  &   -0.21 &		  1 &	 4.00 &    6280 &    0.00 &   N &   \ldots &   \ldots &        1 \\
  HD 117360 &          F6V &    6.52 &   35.2  &   -0.34 &	        1,2 &	 4.25 &    6400 &    0.00 &   Y &    22.40 &	    3 &        1 \\
  HD 120160 &       F0IV/V &    7.67 &  138.3  &    0.02 &		  1 &	 3.75 &    6760 &    0.00 &   N &   \ldots &   \ldots &        1 \\
  HD 122106 &          F8V &    6.36 &   77.5  &    0.08 &	       1,11 &	 3.75 &    6280 &    0.00 &   N &   \ldots &   \ldots &        1 \\
  HD 122510 &          F6V &    6.18 &   38.2  &   -0.11 &		1,2 &	 4.25 &    6600 &    0.00 &   Y &    1.900 &	    3 &        1 \\
  HD 124988 &           F0 &    6.88 &   96.0  &   -0.09 &		  1 &	 4.00 &    6880 &    0.00 &   N &   \ldots &   \ldots &        1 \\
  HD 125451 &         F5IV &    5.41 &   26.1  &   -0.00 &	       1,12 &	 4.25 &    6660 &    0.00 &   N &   \ldots &   \ldots &        1 \\
  HD 127821 &         F4IV &    6.10 &   31.8  &   -0.18 &		  1 &	 4.25 &    6660 &    0.00 &   N &   \ldots &   \ldots &        1 \\
  HD 131495 &           F2 &    6.87 &   72.3  &   -0.13 &		  1 &	 4.00 &    6460 &    0.00 &   N &   \ldots &   \ldots &        1 \\
  HD 134150 &           F8 &    9.84 & (166.0) &  \ldots &	     \ldots &	 4.25 &    6460 &    0.03 &   N &   \ldots &   \ldots &        1 \\
  HD 136580 &           F5 &    6.90 &   41.0  &   -0.15 &	       1,13 &	 4.25 &    6180 &    0.00 &   Y &   \ldots &	    2 &        1 \\
  HD 136407 &          F2V &    6.14 &   56.7  &   -0.08 &		  1 &	 4.00 &    6640 &    0.00 &   Y &    44.40 &	    3 &        1 \\
  HD 138100 &           F0 &    6.69 &   57.8  &   -0.07 &		  1 &	 4.00 &    6720 &    0.00 &   Y &   \ldots &	    2 &        1 \\
  HD 139798 &          F2V &    5.76 &   35.7  &   -0.22 &	       1,12 &	 4.00 &    6700 &    0.00 &   Y &   \ldots &	    2 &        1 \\
  HD 143840 &          F1V &    8.11 &  132.3  &   -0.03 &		  1 &	 4.00 &    6680 &    0.47 &   N &   \ldots &   \ldots &        1 \\
  HD 145371 &           G0 &    9.46 & (144.0) &  \ldots &	     \ldots &	 4.25 &    6480 &    0.00 &   N &   \ldots &   \ldots &        1 \\
  HD 151044 &          F8V &    6.48 &   29.3  &   -0.02 &  1,9,12,13,14,15 &	 4.25 &    6060 &    0.00 &   N &   \ldots &   \ldots &        1 \\
  HD 153377 &           F2 &    7.55 &   64.0  &   -0.16 &		  1 &	 4.25 &    6620 &    0.00 &   N &   \ldots &   \ldots &        1 \\
  HD 155990 &           F8 &    8.07 &   57.7  &   -0.18 &		  1 &	 4.25 &    6100 &    0.00 &   N &   \ldots &   \ldots &        2 \\
  HD 156635 &           F8 &    6.66 &   40.3  &   -0.09 &	       1,16 &	 4.25 &    6160 &    0.00 &   Y &   \ldots &	    2 &        1 \\
  HD 170773 &          F5V &    6.22 &   37.0  &   -0.05 &		1,2 &	 4.25 &    6640 &    0.00 &   N &   \ldots &   \ldots &        1 \\
  HD 184169 &           F2 &    8.20 &  (83.0) &  \ldots &	     \ldots &	 4.25 &    6540 &    0.00 &   N &   \ldots &   \ldots &        1 \\
  HD 183577 &          F6V &    6.48 &   41.6  &   -0.24 &		  1 &	 4.00 &    6100 &    0.00 &   Y &   \ldots &	    2 &        1 \\
  HD 185053 &       F5/F6V &    8.83 &  (67.0) &  \ldots &	     \ldots &	 4.25 &    5960 &    0.24 &   N &   \ldots &   \ldots &        1 \\
  HD 189207 &           F2 &    8.08 &  116.3  &   -0.06 &		  1 &	 4.00 &    6780 &    0.00 &   N &   \ldots &   \ldots &        1 \\
  HD 192486 &          F2V &    6.55 &   44.7  &   -0.18 &		  1 &	 4.25 &    6820 &    0.00 &   N &   \ldots &   \ldots &        1 \\
  HD 192758 &          F0V &    7.03 &  (62.0) &   -0.06 &		 17 &	 4.25 &    7080 &    0.00 &   N &   \ldots &   \ldots &        1 \\
  HD 195952 &          F3V &    8.12 &  153.4  &   -0.06 &		  1 &	 3.75 &    6240 &    0.06 &   N &   \ldots &   \ldots &        1 \\
  HD 199391 &      F0/F2IV &    7.12 &   78.7  &  \ldots &	     \ldots &	 4.00 &    7160 &    0.00 &   Y &    4.800 &	    3 &        1 \\
 PPM 171537 &           F8 &    9.23 &  (90.0) &  \ldots &	     \ldots &	 4.25 &    5960 &    0.00 &   N &   \ldots &   \ldots &        1 \\
  HD 204942 &          F7V &    8.23 &   84.9  &   -0.14 &		  1 &	 4.25 &    6220 &    0.00 &   N &   \ldots &   \ldots &        1 \\
  HD 205674 &      F3/F5IV &    7.19 &   51.8  &   -0.23 &		  1 &	 4.25 &    6780 &    0.00 &   N &   \ldots &   \ldots &        1 \\
  HD 206554 &           F5 &    7.12 &   65.9  &   -0.23 &		  1 &	 4.00 &    6400 &    0.00 &   N &   \ldots &   \ldots &        1 \\
  HD 206893 &          F5V &    6.69 &   38.3  &   -0.06 &	       1,18 &	 4.25 &    6520 &    0.00 &   N &   \ldots &   \ldots &        1 \\
  HD 207889 &           F5 &    7.20 &   49.6  &   -0.11 &		  1 &	 4.25 &    6520 &    0.00 &   N &   \ldots &   \ldots &        1 \\
  HD 210210 &         F1IV &    6.08 &   88.2  &  \ldots &	     \ldots &	 3.75 &    7100 &    0.00 &   N &   \ldots &   \ldots &        1 \\
  HD 213429 &          F7V &    6.15 &   25.4  &   -0.01 &	       1,19 &	 4.25 &    6040 &    0.00 &   Y &    0.072 &	    6 &        1 \\
  HD 213617 &          F1V &    6.43 &   50.3  &   -0.11 &		  1 &	 4.25 &    7020 &    0.00 &   N &   \ldots &   \ldots &        1 \\
  HD 218980 &            F &    8.58 & (105.0) &  \ldots &	     \ldots &	 4.25 &    7060 &    0.43 &   N &   \ldots &   \ldots &        1 \\
  HD 221853 &           F0 &    7.35 &   68.4  &   -0.05 &	       1,18 &	 4.25 &    6760 &    0.00 &   N &   \ldots &   \ldots &        1 \\       
 \enddata                                                                                                                                             
 \tablecomments{Col.(1): Identification.                                                                                                              
 Col.(2): Spectral type.                                                                                                                              
 Col.(3): V magnitude.                                                                                                                                
 Col.(4): Distance. Parentheses indicate photometric distances, otherwise Hipparcos distances from \citet{vanleeuwen07} are used.               
 Col.(5): Metallicity. Literature data are used; if more than one observation is available the average of the [Fe/H] is quoted.                    
 Col.(6): References for metallicity data.                                                                               
 1) \citet{holmberg07}, 
2) \citet{2006AJ....132..161G}, 
3) \citet{1988ApJ...327..389B}, 
4) \citet{2007PASJ...59..335T}, 
5) \citet{1993A&A...275..101E}, 
6) \citet{1990A&A...227..156B}, 
7) \citet{2007MNRAS.374..664C}, 
8) \citet{1987ApJ...321..967B}, 
9) \citet{1992ApJ...387..170F}, 
10) \citet{1986ApJ...303..724B}, 
11) \citet{1990ApJ...354..310B}, 
12) \citet{1988ApJ...325..749B}, 
13) \citet{2005ApJS..159..141V}, 
14) \citet{2005A&A...442..563M}, 
15) \citet{1990ApJ...351..467B}, 
16) \citet{2003MNRAS.340..304R}, 
17) Metallicity was derived based on $uvby$ \citep{hauck1997} photometric data, using the calibration described by \citet{holmberg07}.
18) \citet{2008A&A...490..297S}, 
19) \citet{2004ApJ...613.1202B}.
 Col.(7): Surface gravity values fixed in the course of fitting stellar atmospheric models.                                                           
 Col.(8): Derived effective temperature.                                                                                                              
 Col.(9): Interstellar extinction.                                                                                                                    
 Col.(10): Multiplicity.                                                                                                                              
 Col.(11): Separation of the components if the object is in multiple system.                                                                          
 Col.(12): References for multiplicity data.                                                                                                   
 1) See Sect.~\ref{agedet}.
2) \citet{2007A&A...464..377F}, 
3) \citet{CCDM}, 
4) \citet{2007AstBu..62..339B}, 
5) \citet{1980PASP...92...98B}, 
6) \citet{2000A&AS..145..215P}.
 Col.(13): Selection criteria: 1) the star was suspected to display IR excess based on earlier IR observations;      
 2) the star was selected because of its kinematic group membership.                                                                                                                                                                                  
 }                                                                                                                                                    
\end{deluxetable}

\clearpage
\LongTables
\begin{deluxetable}{lccccccccccc}                                                                                                                                                   
\tabletypesize{\tiny}                                                                                                                                                               
\tablecaption{MIPS photometry \label{mipstable}}                                                                                                                                    
\tablewidth{0pt}                                                                                                                                                                    
\tablecolumns{12}                                                                                                                                                                   
\tablehead{\colhead{Source ID}  &  \colhead{AOR KEY} &  \multicolumn{3}{c}{24$\mu$m}  & \multicolumn{3}{c}{70$\mu$m} & \multicolumn{3}{c}{160$\mu$m} & \colhead{Notes} \\           
 \cline{3-5} \cline{6-8} \cline{9-11} \\                                                                                                                                        
\colhead{}  &  \colhead{} & \colhead{F$_{24}$ [mJy]}  & \colhead{P$_{24}$ [mJy]}  & \colhead{$\chi_{24}$} &                                                                         
\colhead{F$_{70}$ [mJy]}  & \colhead{P$_{70}$ [mJy]}  & \colhead{$\chi_{70}$} &                                                                                                     
\colhead{F$_{160}$ [mJy]}  & \colhead{P$_{160}$ [mJy]}  & \colhead{$\chi_{160}$}  & \colhead{}}                                                                                     
\startdata      
    HD 3670  &     15010816  &      12.8$\pm$0.5  &  10.4  &   3.9  &    134.9$\pm$10.4  &   1.1  &  12.8  &     77.2$\pm$13.4  &   0.2  &   5.7  &          \\
   HD 14691  &     14996736  &     114.4$\pm$4.5  & 116.9  &  -0.4  &      17.7$\pm$5.0  &  12.7  &   0.9  &            \ldots  &\ldots  &\ldots  &          \\
   HD 15060  &     23050496  &      34.6$\pm$1.3  &  34.6  &  -0.0  &      14.8$\pm$3.5  &   3.7  &   3.1  &            \ldots  &\ldots  &\ldots  &          \\
   HD 15115  &     10885888  &      58.3$\pm$2.3  &  33.5  &   9.7  &    451.9$\pm$32.6  &   3.6  &  13.7  &    217.3$\pm$27.8  &   0.7  &   7.7  &          \\
   HD 15745  &     10886400  &     169.4$\pm$6.7  &  17.3  &  22.3  &    741.0$\pm$52.6  &   1.8  &  14.0  &    230.8$\pm$29.9  &   0.3  &   7.6  &          \\
   HD 16743  &     15002624  &      50.3$\pm$2.0  &  30.1  &   9.1  &    368.8$\pm$26.5  &   3.2  &  13.7  &    174.7$\pm$24.5  &   0.6  &   7.0  &       5  \\
   HD 17390  &     10887168  &      45.3$\pm$1.8  &  41.8  &   1.5  &    255.2$\pm$18.9  &   4.5  &  13.2  &    210.0$\pm$28.3  &   0.9  &   7.3  &          \\
  BD+49{\degr} 896  &     15008000  &       2.4$\pm$0.1  &   2.6  &  -0.5  &       0.5$\pm$4.9  &   0.2  &   0.0  &            \ldots  &\ldots  &\ldots  &          \\
   HD 20759  &     23051008  &      19.0$\pm$0.7  &  18.6  &   0.4  &       9.6$\pm$4.0  &   2.0  &   1.8  &            \ldots  &\ldots  &\ldots  &          \\
   HD 24636  &     23051520  &      42.3$\pm$1.6  &  24.3  &   9.7  &      35.1$\pm$4.7  &   2.6  &   6.8  &            \ldots  &\ldots  &\ldots  &          \\
   HD 25570  &     15002880  &     115.6$\pm$4.6  & 113.7  &   0.3  &    170.5$\pm$13.3  &  12.4  &  11.8  &    115.0$\pm$28.4  &   2.5  &   3.9  &       6  \\
   HD 25953  &     15009792  &      16.7$\pm$0.6  &  16.4  &   0.3  &      -5.0$\pm$4.4  &   1.8  &  -1.5  &     -6.3$\pm$41.3  &   0.3  &  -0.1  &          \\
   HD 27429  &     14996992  &      63.8$\pm$2.5  &  61.7  &   0.6  &       6.8$\pm$6.9  &   6.7  &   0.0  &            \ldots  &\ldots  &\ldots  &          \\
   HD 30447  &     10887680  &      30.1$\pm$1.2  &  12.5  &  13.8  &    289.8$\pm$21.1  &   1.3  &  13.6  &    120.3$\pm$17.6  &   0.2  &   6.8  &          \\
   HD 32195  &     15009536  &      16.4$\pm$0.6  &  12.6  &   4.9  &      17.0$\pm$3.9  &   1.3  &   3.9  &    -21.3$\pm$20.3  &   0.2  &  -1.0  &          \\
   HD 30743  &     15003136  &      66.6$\pm$2.6  &  65.7  &   0.2  &      13.9$\pm$7.5  &   7.2  &   0.8  &      5.4$\pm$17.2  &   1.4  &   0.2  &     5,6  \\
   HD 33081  &     14997248  &      33.8$\pm$1.3  &  32.6  &   0.6  &      35.8$\pm$5.1  &   3.5  &   6.2  &            \ldots  &\ldots  &\ldots  &          \\
   HD 33276  &     15003392  &     198.9$\pm$7.9  & 181.2  &   1.8  &      26.8$\pm$9.2  &  19.6  &   0.7  &     50.9$\pm$42.5  &   4.0  &   1.1  &       6  \\
   HD 34739  &     10888192  &       4.0$\pm$0.1  &   4.1  &  -0.4  &      -1.7$\pm$4.9  &   0.4  &  -0.4  &            \ldots  &\ldots  &\ldots  &   2,3,4  \\
   HD 35114  &     11260160  &      32.2$\pm$1.2  &  25.2  &   4.6  &      20.2$\pm$4.2  &   2.7  &   4.0  &            \ldots  &\ldots  &\ldots  &          \\
   HD 35841  &     15011840  &      18.4$\pm$0.7  &   5.4  &  17.0  &    172.1$\pm$13.6  &   0.5  &  12.5  &     37.5$\pm$14.2  &   0.1  &   2.6  &          \\
   HD 36248  &     14997504  &      16.8$\pm$0.6  &  16.3  &   0.4  &            \ldots  &\ldots  &\ldots  &            \ldots  &\ldots  &\ldots  &       1  \\
   HD 36968  &     15012096  &       8.7$\pm$0.3  &   4.1  &  12.1  &    148.1$\pm$12.5  &   0.4  &  11.7  &    105.0$\pm$16.6  &   0.09 &   6.3  &       5  \\
   HD 37402  &     15011072  &      10.4$\pm$0.4  &  10.3  &   0.1  &      -1.6$\pm$4.2  &   1.1  &  -0.6  &    -19.5$\pm$12.0  &   0.2  &  -1.6  &       5  \\
   HD 38905  &     15012352  &       3.1$\pm$0.1  &   2.9  &   1.3  &      -0.2$\pm$7.4  &   0.3  &  -0.0  &            \ldots  &\ldots  &\ldots  &   2,3,4  \\
   HD 47412  &     14997760  &      38.3$\pm$1.5  &  39.5  &  -0.5  &       5.8$\pm$5.4  &   4.3  &   0.2  &            \ldots  &\ldots  &\ldots  &          \\
   HD 48391  &     15010048  &      15.6$\pm$0.6  &  16.0  &  -0.4  &      17.1$\pm$5.6  &   1.7  &   2.7  &    -26.9$\pm$20.5  &   0.3  &  -1.3  &          \\
   HD 50571  &     15003904  &      70.4$\pm$2.8  &  69.4  &   0.2  &    248.8$\pm$18.7  &   7.5  &  12.8  &    214.6$\pm$36.0  &   1.5  &   5.9  &     6,7  \\
   HD 55003  &     23054848  &      31.6$\pm$1.2  &  32.9  &  -0.7  &       2.5$\pm$3.6  &   3.5  &  -0.2  &            \ldots  &\ldots  &\ldots  &          \\
   HD 56099  &     15004160  &      22.3$\pm$0.8  &  21.8  &   0.4  &      10.6$\pm$7.6  &   2.3  &   1.0  &            \ldots  &\ldots  &\ldots  &     3,4  \\
   HD 58853  &     15004416  &       5.0$\pm$0.2  &   5.1  &  -0.2  &       6.6$\pm$6.5  &   0.5  &   0.9  &            \ldots  &\ldots  &\ldots  &     3,4  \\
   HD 61518  &     15010304  &      15.4$\pm$0.6  &  15.1  &   0.4  &      -9.0$\pm$5.4  &   1.6  &  -1.9  &      3.9$\pm$17.2  &   0.3  &   0.2  &          \\
   HD 67587  &     14998016  &      56.3$\pm$2.2  &  58.3  &  -0.6  &      12.9$\pm$9.6  &   6.3  &   0.6  &            \ldots  &\ldots  &\ldots  &          \\
   HD 69351  &     14998272  &      31.7$\pm$1.2  &  32.8  &  -0.6  &      -1.6$\pm$5.9  &   3.5  &  -0.8  &            \ldots  &\ldots  &\ldots  &          \\
   HD 79873  &     14998528  &      40.7$\pm$1.6  &  41.2  &  -0.2  &      -2.1$\pm$4.8  &   4.4  &  -1.3  &            \ldots  &\ldots  &\ldots  &          \\
   HD 82821  &     15004672  &       7.5$\pm$0.3  &   7.8  &  -0.5  &       4.0$\pm$8.0  &   0.8  &   0.3  &     -4.5$\pm$13.3  &   0.1  &  -0.3  &       5  \\
   HD 86146  &     14998784  &     183.3$\pm$7.3  & 186.2  &  -0.3  &      21.4$\pm$4.6  &  20.2  &   0.2  &            \ldots  &\ldots  &\ldots  &          \\
   PPM 7774  &     15011584  &       4.9$\pm$0.2  &   5.1  &  -0.6  &      -0.3$\pm$7.3  &   0.5  &  -0.1  &    -90.6$\pm$43.5  &   0.1  &  -2.0  &       5  \\
  HD 103257  &     14999040  &      36.7$\pm$1.4  &  35.3  &   0.7  &      12.7$\pm$7.3  &   3.8  &   1.2  &            \ldots  &\ldots  &\ldots  &          \\
  HD 107067  &     15008768  &       8.2$\pm$0.3  &   8.4  &  -0.6  &     -5.4$\pm$10.6  &   0.9  &  -0.6  &    -21.5$\pm$10.4  &   0.1  &  -2.0  &          \\
  HD 108102  &     15009024  &      14.2$\pm$0.5  &  13.9  &   0.3  &      -0.0$\pm$7.0  &   1.5  &  -0.2  &     21.9$\pm$11.6  &   0.3  &   1.8  &          \\
  HD 113337  &     14999296  &      74.7$\pm$2.9  &  71.3  &   0.9  &    178.2$\pm$13.3  &   7.7  &  12.7  &            \ldots  &\ldots  &\ldots  &          \\
  HD 114905  &     14999552  &      40.3$\pm$1.6  &  40.9  &  -0.3  &       3.1$\pm$4.1  &   4.4  &  -0.3  &            \ldots  &\ldots  &\ldots  &          \\
  HD 117360  &     19890432  &      51.0$\pm$2.0  &  55.4  &  -1.6  &       6.6$\pm$6.0  &   6.0  &   0.0  &            \ldots  &\ldots  &\ldots  &          \\
  HD 120160  &     15004928  &      14.6$\pm$0.6  &  14.1  &   0.7  &      30.7$\pm$7.3  &   1.5  &   3.9  &      1.2$\pm$22.7  &   0.3  &   0.0  &          \\
  HD 122106  &     15005184  &      58.8$\pm$2.3  &  59.2  &  -0.1  &       1.8$\pm$7.7  &   6.4  &  -0.5  &            \ldots  &\ldots  &\ldots  &       4  \\
  HD 122510  &     15000064  &      70.6$\pm$2.8  &  70.7  &  -0.0  &       8.8$\pm$5.3  &   7.7  &   0.2  &            \ldots  &\ldots  &\ldots  &          \\
  HD 124988  &     23062016  &      28.8$\pm$1.1  &  28.3  &   0.3  &       9.7$\pm$4.1  &   3.0  &   1.5  &            \ldots  &\ldots  &\ldots  &          \\
  HD 125451  &     15008256  &     126.2$\pm$5.0  & 121.9  &   0.6  &      64.6$\pm$8.1  &  13.2  &   6.3  &     -2.0$\pm$11.2  &   2.7  &  -0.4  &       6  \\
  HD 127821  &     15005440  &      69.8$\pm$2.7  &  69.9  &  -0.0  &    360.3$\pm$25.9  &   7.6  &  13.5  &    295.5$\pm$37.2  &   1.5  &   7.8  &       6  \\
  HD 131495  &     15000320  &      34.1$\pm$1.3  &  35.4  &  -0.7  &      -4.8$\pm$4.8  &   3.8  &  -1.8  &            \ldots  &\ldots  &\ldots  &          \\
  HD 134150  &     10888960  &       2.4$\pm$0.1  &   2.4  &   0.3  &      -6.5$\pm$7.4  &   0.2  &  -0.9  &      3.5$\pm$15.7  &   0.05 &   0.2  &       5  \\
  HD 136580  &     15000576  &      40.1$\pm$1.6  &  40.4  &  -0.1  &      -0.9$\pm$4.2  &   4.4  &  -1.2  &            \ldots  &\ldots  &\ldots  &          \\
  HD 136407  &     15005696  &      66.8$\pm$2.6  &  64.0  &   0.8  &      10.3$\pm$7.9  &   6.9  &   0.4  &     -6.0$\pm$22.8  &   1.4  &  -0.3  &       6  \\
  HD 138100  &     15000832  &      36.5$\pm$1.4  &  37.5  &  -0.5  &       0.8$\pm$4.2  &   4.0  &  -0.7  &            \ldots  &\ldots  &\ldots  &          \\
  HD 139798  &     15008512  &      88.9$\pm$3.5  &  89.9  &  -0.2  &      10.7$\pm$5.9  &   9.8  &   0.1  &    -20.7$\pm$10.6  &   2.0  &  -2.1  &       6  \\
  HD 143840  &     10889472  &      15.4$\pm$0.6  &  14.7  &   0.9  &            \ldots  &\ldots  &\ldots  &            \ldots  &\ldots  &\ldots  &       1  \\
  HD 145371  &     15007744  &       3.3$\pm$0.1  &   3.2  &   0.4  &       0.1$\pm$7.4  &   0.3  &  -0.0  &            \ldots  &\ldots  &\ldots  &   2,3,4  \\
  HD 151044  &     10890240  &      68.3$\pm$2.7  &  63.7  &   1.3  &      95.6$\pm$7.9  &   6.9  &  11.2  &     47.4$\pm$11.7  &   1.4  &   3.9  &       6  \\
  HD 153377  &     15005952  &      18.5$\pm$0.7  &  18.8  &  -0.3  &      -2.0$\pm$5.6  &   2.0  &  -0.7  &     26.9$\pm$24.0  &   0.4  &   1.1  &          \\
  HD 155990  &     15010560  &      14.8$\pm$0.6  &  14.5  &   0.4  &       5.7$\pm$5.1  &   1.5  &   0.8  &            \ldots  &   0.3  &   2.3  &       4  \\
  HD 156635  &     15006208  &      48.2$\pm$1.9  &  50.1  &  -0.7  &       6.9$\pm$6.1  &   5.4  &   0.2  &     56.7$\pm$75.4  &   1.1  &   0.7  &          \\
  HD 170773  &     10890752  &      65.3$\pm$2.6  &  60.5  &   1.5  &    787.9$\pm$56.0  &   6.6  &  13.9  &    692.3$\pm$83.8  &   1.3  &   8.2  &     6,7  \\
  HD 184169  &     10891264  &      10.5$\pm$0.4  &  10.3  &   0.2  &    -12.1$\pm$12.4  &   1.1  &  -1.0  &            \ldots  &\ldots  &\ldots  &   2,3,4  \\
  HD 183577  &     15001600  &      63.0$\pm$2.5  &  61.7  &   0.4  &       4.8$\pm$4.6  &   6.7  &  -0.4  &            \ldots  &\ldots  &\ldots  &          \\
  HD 185053  &     10892032  &      22.8$\pm$0.9  &   9.8  &  13.4  &            \ldots  &\ldots  &\ldots  &            \ldots  &\ldots  &\ldots  &       1  \\
  HD 189207  &     15006464  &      10.1$\pm$0.4  &   9.8  &   0.5  &       5.9$\pm$5.4  &   1.0  &   0.8  &    -46.4$\pm$30.6  &   0.2  &  -1.5  &          \\
  HD 192486  &     15001856  &      43.3$\pm$1.7  &  42.1  &   0.5  &      -5.2$\pm$4.6  &   4.6  &  -2.1  &            \ldots  &\ldots  &\ldots  &          \\
  HD 192758  &     10892800  &      42.1$\pm$1.6  &  23.4  &  10.1  &    452.4$\pm$32.5  &   2.5  &  13.8  &    200.7$\pm$26.3  &   0.5  &   7.5  &          \\
  HD 195952  &     15006720  &      12.9$\pm$0.5  &  13.0  &  -0.2  &       1.4$\pm$8.1  &   1.4  &   0.0  &            \ldots  &\ldots  &\ldots  &     3,4  \\
  HD 199391  &     15006976  &      22.3$\pm$0.9  &  24.4  &  -1.7  &     -7.5$\pm$10.2  &   2.6  &  -0.9  &    130.7$\pm$27.8  &   0.5  &   4.6  &       5  \\
 PPM 171537  &     15011328  &       5.5$\pm$0.2  &   5.3  &   0.6  &      -2.4$\pm$7.7  &   0.5  &  -0.3  &            \ldots  &\ldots  &\ldots  &     3,4  \\
  HD 204942  &     15007232  &      10.9$\pm$0.7  &  11.5  &  -0.6  &    -13.8$\pm$11.0  &   1.2  &  -1.3  &            \ldots  &\ldots  &\ldots  &     3,4  \\
  HD 205674  &     15007488  &      31.2$\pm$1.2  &  24.1  &   4.8  &    232.5$\pm$17.9  &   2.6  &  12.8  &    185.6$\pm$26.4  &   0.5  &   6.9  &          \\
  HD 206554  &     23067648  &      29.0$\pm$1.1  &  29.2  &  -0.1  &       1.4$\pm$6.3  &   3.1  &  -0.2  &            \ldots  &\ldots  &\ldots  &       3  \\
  HD 206893  &     10893312  &      44.2$\pm$1.7  &  40.6  &   1.6  &    265.7$\pm$20.0  &   4.4  &  13.0  &    193.1$\pm$26.2  &   0.9  &   7.3  &          \\
  HD 207889  &     23068160  &      27.0$\pm$1.0  &  25.5  &   1.1  &       4.7$\pm$3.8  &   2.7  &   0.4  &            \ldots  &\ldots  &\ldots  &       3  \\
  HD 210210  &     15002112  &      53.2$\pm$2.1  &  52.9  &   0.1  &       5.0$\pm$4.0  &   5.7  &  -0.1  &            \ldots  &\ldots  &\ldots  &          \\
  HD 213429  &     21840896  &      92.2$\pm$3.6  &  86.3  &   1.3  &      22.2$\pm$4.1  &   9.4  &   3.0  &            \ldots  &\ldots  &\ldots  &          \\
  HD 213617  &     10893824  &      44.3$\pm$1.7  &  41.8  &   1.1  &    119.2$\pm$10.0  &   4.5  &  11.3  &     96.4$\pm$22.9  &   0.9  &   4.1  &       5  \\
  HD 218980  &     13238784  &      11.3$\pm$0.4  &   8.2  &   5.9  &            \ldots  &\ldots  &\ldots  &            \ldots  &\ldots  &\ldots  &       1  \\
  HD 221853  &     10894848  &      78.5$\pm$3.2  &  20.3  &  17.7  &    336.5$\pm$24.7  &   2.2  &  13.5  &    105.3$\pm$20.4  &   0.4  &   5.1  &          \\
\enddata                                                                                                                                                                           
 \tablecomments{Col.(1): Identification.                                                                                                                                            
 Col.(2): AOR Key for MIPS measurement.                                                                                                                                                                                                                                                  
 Using the AOR key one can query additional details for                                                                                                                             
 each observation (e.g. measurement setups) from the Spitzer Data Archive at the Spitzer Science Center.                                                                            
 Col.(3-11). Measured and predicted flux densities with their uncertainties and the significance of                                                                                 
 the excesses at 24/70/160{\micron}. The quoted uncertainties include the calibration uncertainties as well.                                                                           
 Col.(12): Notes. 1: Nebulosity. At 24{\micron} a small aperture with 3.5{\arcsec} radius was used.                                                                                      
 At 70{\micron} and 160{\micron} no photometric values were quoted; 2: A bright nearby source contaminated the aperture photometry at 24{\micron} -- 
 PSF photometry was used;                                                                       
 3: Bright nearby sources contaminated the photometry of our target at 70{\micron}. In order to remove the contribution of these background objects, 
    we fitted PSF to these sources and subtracted their emission before the photometry for our target was performed. 
 4: Nearby bright object at 160{\micron} in the aperture - no photometry is given;                                   
 5: A bright nearby source contaminated the annulus at 160{\micron}. The background source was masked out in the course of photometry; 
 6: The ghost image was subtracted at 160{\micron} before the photometry was performed.                              
 7: The source is marginally extended at 70{\micron} and the photometry was derived using                                                                                                  
 the fitted profile (see Sect.~\ref{spatext}).                                                                                                                                      
}                                                                                                                                                                                   
\end{deluxetable}                                                                                                                                                                   


\clearpage
\begin{deluxetable}{lccc}                                                                                                                                                     
\tabletypesize{\scriptsize}                                                                                                                                                               
\tablecaption{Additional photometric data \label{addphottable}}                                                                                                                                    
\tablewidth{0pt}                                                                                                                                                                    
\tablecolumns{4}                                                                                                                                                                   
\tablehead{\colhead{Source ID}  &  \colhead{Instrument} & \colhead{Wavelength}  & \colhead{F$_{\nu}$} \\
\colhead{}  &  \colhead{} & \colhead{[$\mu$m]}  & \colhead{[mJy]} }
\startdata
HD 151044  &  ISO/ISOPHOT   &  60     & 111.0$\pm$13.0\\
HD 151044  &  ISO/ISOPHOT   &  90     & 101.0$\pm$10.0\\
HD 170773  &  ISO/ISOPHOT   &  60     & 570.0$\pm$35.0 \\
HD 170773  &  ISO/ISOPHOT   &  90     & 771.0$\pm$54.0 \\
HD 213617  &  ISO/ISOPHOT   &  60     & 110.0$\pm$13.0 \\
HD 213617  &  ISO/ISOPHOT   &  90     & 121.0$\pm$14.0 \\
HD 15745   &  IRAM/MAMBO2   &  1200   & 1.3$\pm$0.6\\
HD 25570   &  IRAM/MAMBO2   &  1200   & 0.6$\pm$0.6\\
HD 113337  &  IRAM/MAMBO2   &  1200   & 0.4$\pm$0.3\\
\enddata
\end{deluxetable} 


\begin{deluxetable}{lcccccc}                                                                                                                                                   
\tabletypesize{\scriptsize}                                                                                                                                                               
\tablecaption{Additional spectroscopic information and derived kinematic properties \label{spectable}}                                                                                                                                    
\tablewidth{0pt}                                                                                                                                                                    
\tablecolumns{7}                                                                                                                                                                   
\tablehead{ \colhead{Source ID}  &  \colhead{$\rm v_r$} & \colhead{U} & \colhead{V} & \colhead{W} & \colhead{v$\rm sin{i}$} & \colhead{EW$_{Li}$} \\ 
\colhead{}  &  \colhead{[kms$^{-1}$]} & \colhead{[kms$^{-1}$]} & \colhead{[kms$^{-1}$]} & \colhead{[kms$^{-1}$]} & \colhead{[kms$^{-1}$]} & \colhead{[m$\rm \AA$]}
}                                                                                     
\startdata 
HD 15745   &  2.5$\pm$3.3 &  $-$10.4$\pm$2.5  &  $-$15.3$\pm$1.9   &   $-$7.9$\pm$1.2 & 50 &  $\ldots$ \\ 
HD 16699  & 16.2$\pm$0.2 &  $-$23.5$\pm$1.5  &  $-$15.0$\pm$0.4   &  $-$10.3$\pm$0.3 & 20 &  40$\pm$10 \\    
SAO\,232842  & 15.8$\pm$0.3 &  $\ldots$  &  $\ldots$   &  $\ldots$  & $\ldots$ & 220$\pm$10 \\ 
HD 16743   & 17.3$\pm$8.5 &  $-$23.3$\pm$0.5  &  $-$15.4$\pm$4.6   &  $-$11.3$\pm$7.2 & 100 &  $\ldots$ \\ 
HD 24636   & 14.0$\pm$0.7 &  $-$8.8$\pm$0.3  &  $-$20.5$\pm$0.6   &  $-$1.6$\pm$0.5 & 30 &  50$\pm$10 \\
HD 36968\tablenotemark{a}   & 15.0$\pm$2.0 &  $-$14.8$\pm$2.1  &   $-$6.6$\pm$2.0   &   $-$8.5$\pm$1.3 & $\ldots$ &  $\ldots$ \\ 
HD 170773  &$-$17.5$\pm$1.5 &  $-$22.2$\pm$1.5  &   $-$4.6$\pm$0.2   &  $-$15.0$\pm$0.5 & 50 & $\ldots$  \\     
HD 205674  & $-$1.4$\pm$1.3 &   $-$3.0$\pm$0.8  &  $-$25.2$\pm$1.1   &  $-$14.6$\pm$1.1 & 30 & 20$\pm$10  \\ 
HD 206893  &$-$11.8$\pm$1.6 &  $-$19.2$\pm$0.9  &  $-$7.2$\pm$0.8    &  $-$2.7$\pm$1.1  & 33 & $\ldots$  \\ 
HD 221853  & $-$8.4$\pm$1.8 &  $-$11.9$\pm$0.6  &  $-$22.4$\pm$1.4   &  $-$7.3$\pm$1.5   & 40 & $\ldots$  \\   
\enddata 
\tablenotetext{a}{HD\,36968 was measured in the framework of a previous program that carried out with the 2.3\,m ANU-telescope at the Siding
Spring Observatory (Australia), using the Double Beam Spectrograph \citep[for details see ][]{moor2006}. }
\tablecomments{Col.(1): Identification.
Col.(2): Heliocentric radial velocity. Col.(3-5): Galactic space velocity components of the star. In the calculation of the Galactic space velocity 
we used a right-handed coordinate system (U is positive towards 
the Galactic centre, V is positive in the direction of galactic rotation and W is positive towards the galactic North pole) and 
followed the general recipe described in "The Hipparcos and Tycho Catalogues"  (ESA 1997). 
Col.(6) Projected rotational velocity of the star. Col.(7) Measured lithium equivalent width. }
\end{deluxetable}

\clearpage
\LongTables 
\tabletypesize{\tiny}
\begin{deluxetable}{@{\extracolsep{-3pt}}lcccccccccccc}                                                                          
\tabletypesize{\tiny}
\tablecaption{Synthetic IRS photometry for stars with infrared excess \label{irstable}}                                                              
\tablewidth{0pt} 
\tablecolumns{13}                                                                                              
\tablehead{\colhead{Source ID}  &  \colhead{AOR KEY} &  \multicolumn{11}{c}{Synthetic photometry [mJy]}  \\
\colhead{}  &  \colhead{} & \colhead{8--10$\mu$m}  &\colhead{10--12$\mu$m}  & \colhead{12--14$\mu$m} & \colhead{14--16$\mu$m} & \colhead{16--18$\mu$m}
 & \colhead{18--20$\mu$m} & \colhead{20--23$\mu$m} & \colhead{23--26$\mu$m} & \colhead{26--29$\mu$m} & \colhead{29--32$\mu$m} & \colhead{32-35$\mu$m} }
\startdata                                                                                                                                                                        
    HD 3670  &     15025664  &    71.8$\pm$9.0     &    48.9$\pm$5.0     &    35.8$\pm$4.4     &    26.7$\pm$2.6     &    20.1$\pm$2.0     &    16.6$\pm$2.5     &    14.7$\pm$2.0     &    13.2$\pm$1.0     &    14.1$\pm$1.3     &    15.9$\pm$1.4     &    20.9$\pm$2.6     \\
   HD 15060  &     23050752  &   227.6$\pm$30.4    &   149.9$\pm$15.9    &   109.9$\pm$10.0    &    89.4$\pm$7.5     &    69.5$\pm$6.1     &    53.2$\pm$6.6     &    43.9$\pm$5.8     &    33.3$\pm$2.4     &    26.5$\pm$2.1     &    21.3$\pm$2.5     &    17.5$\pm$3.7     \\
   HD 15115  &     10885632  &   218.0$\pm$36.8    &   148.7$\pm$15.6    &   111.2$\pm$10.0    &    87.0$\pm$7.7     &    70.8$\pm$4.6     &    61.0$\pm$2.3     &    53.7$\pm$2.7     &    54.9$\pm$2.5     &    60.3$\pm$2.8     &    71.8$\pm$5.1     &    93.5$\pm$9.8     \\
   HD 15745  &     10886144  &   123.5$\pm$15.8    &    83.9$\pm$8.3     &    65.2$\pm$4.1     &    67.7$\pm$3.9     &    81.2$\pm$7.5     &   110.2$\pm$9.7     &   143.2$\pm$19.2    &   218.9$\pm$16.7    &   276.8$\pm$19.4    &   335.2$\pm$22.7    &   431.0$\pm$30.5    \\
   HD 16743  &     15018240  &   208.5$\pm$29.0    &   142.5$\pm$15.0    &   105.9$\pm$7.3     &    79.1$\pm$6.0     &    62.9$\pm$3.8     &    54.5$\pm$2.9     &    50.7$\pm$2.0     &    50.4$\pm$1.8     &    52.3$\pm$2.1     &    57.9$\pm$3.0     &    70.4$\pm$5.5     \\
   HD 17390  &     10886912  &   278.2$\pm$30.6    &   196.6$\pm$21.9    &   141.6$\pm$12.9    &   103.7$\pm$8.2     &    80.5$\pm$5.3     &    63.3$\pm$4.5     &    52.2$\pm$4.2     &    42.8$\pm$2.6     &    37.5$\pm$1.7     &    36.2$\pm$1.9     &    38.7$\pm$3.0     \\
   HD 24636  &     23051776  &   163.1$\pm$19.9    &   111.5$\pm$11.1    &    84.2$\pm$7.8     &    68.7$\pm$4.3     &    57.4$\pm$3.0     &    50.2$\pm$3.3     &    46.6$\pm$3.1     &    44.2$\pm$1.4     &    43.6$\pm$1.2     &    44.2$\pm$1.9     &    44.0$\pm$4.3     \\
   HD 25570  &     15018496  &   755.7$\pm$143.7   &   530.2$\pm$56.6    &   383.5$\pm$36.7    &   285.3$\pm$21.6    &   222.8$\pm$16.0    &   178.1$\pm$12.9    &   142.9$\pm$13.5    &   111.2$\pm$7.8     &    92.1$\pm$4.6     &    79.0$\pm$6.3     &    72.2$\pm$6.6     \\
   HD 30447  &     10887424  &    83.0$\pm$11.1    &    55.9$\pm$5.8     &    41.7$\pm$4.3     &    34.0$\pm$2.5     &    29.1$\pm$2.0     &    26.4$\pm$2.5     &    28.5$\pm$3.0     &    35.7$\pm$1.9     &    45.2$\pm$3.9     &    58.5$\pm$5.4     &    80.4$\pm$7.2     \\
   HD 32195  &     15024384  &    86.6$\pm$10.7    &    58.8$\pm$6.7     &    43.4$\pm$4.2     &    31.5$\pm$2.3     &    25.3$\pm$2.3     &    20.6$\pm$2.3     &    17.4$\pm$2.1     &    15.5$\pm$1.1     &    13.5$\pm$1.1     &    11.9$\pm$1.1     &    12.0$\pm$2.0     \\
   HD 33081  &     15013120  &   219.5$\pm$28.6    &   151.5$\pm$16.1    &   109.3$\pm$9.8     &    82.0$\pm$5.9     &    64.3$\pm$4.3     &    51.6$\pm$5.4     &    41.7$\pm$3.8     &    33.5$\pm$2.1     &    27.6$\pm$1.8     &    23.5$\pm$1.8     &    20.2$\pm$1.8     \\
   HD 35114  &     26362112  &   171.5$\pm$25.5    &   110.1$\pm$13.8    &    81.5$\pm$8.4     &    62.3$\pm$5.9     &    50.7$\pm$3.3     &    40.8$\pm$4.1     &    35.9$\pm$3.5     &    29.9$\pm$1.7     &    26.5$\pm$1.9     &    24.1$\pm$2.1     &    22.5$\pm$4.7     \\
   HD 35841  &     15026688  &    36.8$\pm$5.0     &    24.7$\pm$3.0     &    17.7$\pm$1.8     &    13.7$\pm$1.3     &    11.6$\pm$1.3     &    10.7$\pm$1.7     &    13.4$\pm$2.3     &    20.7$\pm$3.0     &    28.4$\pm$3.1     &    37.1$\pm$3.3     &    50.7$\pm$5.4     \\
   HD 36968  &     15026944  &    28.2$\pm$3.4     &    19.4$\pm$2.2     &    14.5$\pm$1.7     &     9.9$\pm$1.4     &     9.4$\pm$1.3     &     7.3$\pm$1.4     &     7.0$\pm$1.7     &     9.3$\pm$1.0     &    12.3$\pm$1.6     &    17.2$\pm$2.4     &    24.9$\pm$3.9     \\
   HD 50571  &     15019520  &   422.2$\pm$28.8    &   311.6$\pm$34.2    &   225.3$\pm$21.0    &   172.7$\pm$12.3    &   134.0$\pm$10.0    &   104.5$\pm$7.0     &    84.5$\pm$6.0     &    68.5$\pm$5.0     &    56.6$\pm$2.5     &    49.1$\pm$3.3     &    50.7$\pm$4.9     \\
  HD 113337  &     15015168  &   475.6$\pm$64.0    &   321.6$\pm$35.7    &   232.5$\pm$21.1    &   176.5$\pm$12.8    &   136.9$\pm$9.6     &   108.5$\pm$6.6     &    87.5$\pm$7.7     &    72.8$\pm$4.5     &    62.4$\pm$1.9     &    57.4$\pm$1.6     &    58.3$\pm$3.8     \\
  HD 120160  &     15020544  &    94.9$\pm$13.4    &    64.1$\pm$6.8     &    46.0$\pm$5.2     &    34.8$\pm$3.0     &    27.4$\pm$2.1     &    22.6$\pm$2.1     &    17.9$\pm$2.2     &    14.9$\pm$1.3     &    12.7$\pm$0.9     &    11.5$\pm$0.9     &    12.1$\pm$1.4     \\
  HD 125451  &     15027968  &       \ldots        &       \ldots        &       \ldots        &   297.1$\pm$18.8    &   235.7$\pm$15.8    &   187.1$\pm$12.5    &   152.4$\pm$12.0    &   121.0$\pm$6.3     &    98.9$\pm$5.8     &    82.0$\pm$5.4     &    73.2$\pm$4.0     \\
  HD 127821  &     15020800  &   475.5$\pm$63.2    &   319.8$\pm$34.7    &   230.8$\pm$21.9    &   176.1$\pm$13.0    &   136.6$\pm$9.7     &   108.9$\pm$7.6     &    88.6$\pm$7.5     &    71.3$\pm$3.5     &    59.3$\pm$2.5     &    53.9$\pm$2.6     &    52.6$\pm$4.3     \\
  HD 151044  &     10889984  &   411.6$\pm$58.7    &   277.6$\pm$26.0    &   209.1$\pm$15.4    &   156.4$\pm$15.3    &   120.2$\pm$9.3     &    95.5$\pm$4.5     &    77.5$\pm$4.7     &    64.7$\pm$4.1     &    54.6$\pm$2.5     &    47.8$\pm$2.4     &    47.0$\pm$2.7     \\
  HD 170773  &     10890496  &   406.5$\pm$53.6    &   276.6$\pm$30.0    &   200.1$\pm$19.1    &   149.5$\pm$12.4    &   117.4$\pm$9.4     &    94.5$\pm$6.3     &    77.5$\pm$6.4     &    63.6$\pm$3.2     &    55.9$\pm$1.8     &    54.6$\pm$1.8     &    62.1$\pm$3.4     \\
  HD 192758  &     10892544  &   157.0$\pm$20.4    &   108.5$\pm$11.2    &    79.6$\pm$7.0     &    61.7$\pm$5.1     &    49.5$\pm$3.3     &    42.7$\pm$2.3     &    39.8$\pm$1.9     &    43.0$\pm$2.4     &    49.1$\pm$3.1     &    61.4$\pm$4.8     &    81.9$\pm$7.2     \\
  HD 205674  &     15022592  &   162.3$\pm$21.3    &   110.1$\pm$11.7    &    80.6$\pm$7.1     &    61.0$\pm$6.2     &    47.5$\pm$4.0     &    37.4$\pm$3.6     &    32.9$\pm$2.8     &    31.0$\pm$1.3     &    30.4$\pm$1.8     &    34.5$\pm$2.6     &    39.9$\pm$2.7     \\
  HD 206893  &     10893056  &   277.5$\pm$36.2    &   188.5$\pm$20.8    &   135.6$\pm$13.0    &   102.9$\pm$8.6     &    80.2$\pm$5.9     &    63.3$\pm$4.9     &    51.7$\pm$5.3     &    42.7$\pm$2.0     &    38.2$\pm$1.3     &    38.0$\pm$0.9     &    42.2$\pm$2.4     \\
  HD 213429  &     15017984  &   575.4$\pm$84.9    &   391.4$\pm$42.8    &   284.5$\pm$23.8    &   218.8$\pm$18.7    &   166.9$\pm$13.2    &   130.1$\pm$11.4    &   105.0$\pm$9.3     &    78.4$\pm$5.7     &    61.6$\pm$5.5     &    48.7$\pm$4.5     &    38.3$\pm$4.9     \\
  HD 213617  &     10893568  &   285.9$\pm$38.2    &   191.0$\pm$21.0    &   137.9$\pm$12.8    &   102.9$\pm$8.3     &    80.9$\pm$5.0     &    66.6$\pm$5.4     &    53.2$\pm$4.2     &    43.8$\pm$2.2     &    38.2$\pm$1.3     &    36.5$\pm$1.3     &    36.3$\pm$3.4     \\
  HD 221853  &     10894592  &   128.5$\pm$18.0    &    87.1$\pm$9.9     &    63.3$\pm$4.9     &    53.5$\pm$3.3     &    48.4$\pm$1.3     &    50.9$\pm$3.0     &    64.3$\pm$8.8     &    92.9$\pm$9.8     &   119.6$\pm$8.7     &   140.1$\pm$6.3     &   170.9$\pm$12.2    \\
  \enddata                                                                                                     
 \tablecomments{Col.(1): Identification.                                                                      
 Col.(2): AOR Key for IRS measurement.                                 
 Col.(3-13). Synthetic photometry. 
 IRS data points are averaged in 11 adjacent bins. We used 2{\micron} wide bins at $\lambda<20${\micron} and 
 3{\micron} wide bins at $\lambda>20$. The quoted uncertainties include 
 both the instrumental noise and the variation of the SED within a bin.   
}                                                                                                            
\end{deluxetable}                                                                                                                                             
\clearpage

\tabletypesize{\scriptsize}
\begin{deluxetable}{lcccc}                                                                                                                                                          
\tablecaption{Disk properties \label{diskprop}}                                                                                                                                     
\tablewidth{0pt}                                                                                                                                                                    
\tablecolumns{5}                                                                                                                                                                    
\tablehead{\colhead{Source ID}  &  \colhead{$\rm T_{dust}$ [K]} &  \colhead{$\rm R_{dust}$ [AU]}                                                                                    
&  \colhead{$\rm f_{dust}$ [$\rm 10^{-4}$]} & \colhead{Reduced $\chi^2$}                                                                                                            
}                                                                                                                                                                                   
\startdata                                                                                                                                                                          
  {\bf HD 3670}  & 53$\pm$1   & 42$\pm$9   &5.4$\pm$0.4  & 0.9 \\
 {\bf HD 15060}  &    $<$ 64  &     $>$51  & $\sim$0.16  & \ldots \\
       HD 15115$^*$  & 61$\pm$1   & 36$\pm$1   &5.1$\pm$0.2  & 3.4 \\
       HD 15745$^*$  & 89$\pm$1   & 18$\pm$1   & 21.9$\pm$0.8  & 2.3 \\
       HD 16743$^*$  & 62$\pm$1   & 47$\pm$2   &3.8$\pm$0.2  & 7.6 \\
       HD 17390  & 48$\pm$1   & 70$\pm$3   &2.1$\pm$0.1  & 2.8 \\
 {\bf HD 24636}  &116$\pm$5   & 10$\pm$1   &1.08$\pm$0.06  & 0.5 \\
       HD 25570  & 51$\pm$3   & 75$\pm$9   &0.53$\pm$0.04  & 0.2 \\
       HD 30447$^*$  & 67$\pm$1   & 33$\pm$2   &9.2$\pm$0.6  & 1.8 \\
 {\bf HD 32195}  & 89$\pm$9   & 12$\pm$3   &0.65$\pm$0.15  & 1.4 \\
 {\bf HD 33081}  & 55$\pm$4   & 45$\pm$7   &0.46$\pm$0.07  & 0.7 \\
 {\bf HD 35114}  & 97$\pm$9   & 12$\pm$2   &0.53$\pm$0.07  & 0.4 \\
       HD 35841  & 69$\pm$1   & 23$\pm$5   & 15.2$\pm$1.0  & 1.4 \\
 {\bf HD 36968}  & 58$\pm$1   & 45$\pm$9   & 13.4$\pm$1.0  & 1.5 \\
       HD 50571  & 45$\pm$2   & 66$\pm$5   &1.5$\pm$0.1  & 0.2 \\
      HD 113337  & 53$\pm$1   & 55$\pm$3   &0.98$\pm$0.07  & 0.5 \\
{\bf HD 120160}  & 57$\pm$4   & 84$\pm$14  &0.81$\pm$0.20  & 0.2 \\
      HD 125451  & 63$\pm$5   & 37$\pm$6   &0.18$\pm$0.02  & 0.2 \\
      HD 127821  & 45$\pm$1   & 66$\pm$3   &2.1$\pm$0.1  & 1.3 \\
      HD 151044  & 57$\pm$2   & 32$\pm$2   &0.77$\pm$0.05  & 0.7 \\
      HD 170773  & 43$\pm$1   & 78$\pm$3   &4.8$\pm$0.2  & 1.3 \\
      HD 192758$^*$  & 61$\pm$1   & 45$\pm$9   &5.7$\pm$0.3  & 3.7 \\
      HD 205674  & 54$\pm$1   & 46$\pm$3   &3.7$\pm$0.3  & 2.4 \\
      HD 206893  & 49$\pm$1   & 49$\pm$2   &2.5$\pm$0.1  & 1.7 \\
{\bf HD 213429}  &    $<$ 62  &     $>$27  & $\sim$0.08  & \ldots \\
      HD 213617  & 55$\pm$1   & 59$\pm$3   &0.96$\pm$0.05  & 0.7 \\
      HD 221853  & 84$\pm$1   & 23$\pm$1   &7.9$\pm$0.4  & 1.6 \\
 \enddata                                                                                                                                                                           
 \tablecomments{Disk parameters, listed in this table, come from a model assuming a single narrow dust 
 ring. Disks marked by asterisks can be better fitted with a two-component model (see Table~\ref{diskprop2}). 
 Col.(1): Identification. Disks discovered in this programme are in boldface.                                                                                        
 Col.(2): Disk temperature.                                                                                                                                                         
 Col.(3): Disk radius.                                                                                                                                                              
 Col.(4): Fractional dust luminosity $f_{\rm dust} = \frac{L_{\rm dust}}{L_{\rm bol}}$.                                                                                                         
 Col.(5): Best reduced $\chi^2$.                                                                                                                                                    
}                                                                                                                                                                                   
\end{deluxetable}                                                                                                                                                                   


\begin{deluxetable}{lccccccc}                                                                                                                                                       
\tablecaption{Disk properties \label{diskprop2}}                                                                                                                                     
\tablewidth{0pt}                                                                                                                                                                    
\tablecolumns{8}                                                                                                                                                                    
\tablehead{\colhead{} & \multicolumn{3}{c}{Warm dust} & \multicolumn{3}{c}{Cold dust} & \colhead{} \\                                                                               
\colhead{Source ID}  &                                                                                                                                                              
\colhead{$\rm T_{dust}$ [K]} &  \colhead{$\rm R_{dust}$ [AU]} &  \colhead{$\rm f_{dust}$ [$\rm 10^{-4}$]} &                                                                         
\colhead{$\rm T_{dust}$ [K]} &  \colhead{$\rm R_{dust}$ [AU]} &  \colhead{$\rm f_{dust}$ [$\rm 10^{-4}$]} &                                                                         
\colhead{Reduced $\chi^2$}                                                                                                                                                          
}                                                                                                                                                                                   
\startdata                                                                                                                                                                          
       HD 15115  &179$\pm$46  &  4$\pm$2   &0.38$\pm$0.08  & 57$\pm$1   & 42$\pm$2   &4.8$\pm$0.2  & 1.6 \\
       HD 15745  &147$\pm$22  &  6$\pm$2   &3.5$\pm$1.6  & 81$\pm$3   & 21$\pm$2   & 18.9$\pm$2.0  & 1.1 \\
       HD 16743  &147$\pm$24  &  8$\pm$3   &0.51$\pm$0.07  & 53$\pm$1   & 63$\pm$4   &3.6$\pm$0.3  & 1.7 \\
       HD 30447  &159$\pm$36  &  6$\pm$3   &0.66$\pm$0.32  & 62$\pm$2   & 38$\pm$3   &8.8$\pm$0.7  & 0.8 \\
      HD 192758  &154$\pm$31  &  7$\pm$3   &0.39$\pm$0.09  & 56$\pm$1   & 53$\pm$11  &5.4$\pm$0.3  & 1.8 \\
 \enddata                                                                                                                                                                           
 \tablecomments{Col.(1): Identification.                                                                                                                                            
 Col.(2): Temperature of warm dust in the inner ring.                                                                                                                               
 Col.(3): Radius of the inner dust ring.                                                                                                                                            
 Col.(4): Fractional dust luminosity of the inner dust ring.                                                                                                                        
 Col.(5): Temperature of cold dust in the outer ring.                                                                                                                               
 Col.(6): Radius of the outer dust ring.                                                                                                                                            
 Col.(7): Fractional dust luminosity of the outer dust ring.                                                                                                                        
 Col.(8): Best reduced $\chi^2$.                                                                                                                                                    
}                                                                                                                                                                                   
\end{deluxetable}                                                                                                                                                                   

\clearpage
\begin{deluxetable}{lcccccccc}                                                                                                                                                     
\tablecaption{Age estimates  \label{agetable}}                                                                                                           
\tablewidth{0pt}                                                                                                                                                                    
\tablecolumns{9}                                                                                                                                                                   
\tablehead{\colhead{Source ID}  & \colhead{$\rm \log{\frac{L_{x}}{L_{bol}}}$} &  \colhead{$\rm \log{R'_{HK}}$}  & \colhead{Ref.} &
           \colhead{Membership} &  \colhead{Ref.} &  \colhead{Age [Myr]} &  \colhead{Dating method} & \colhead{Ref.} 
}	                                                                                                                                                                  
\startdata     
    HD 3670  &  $-$4.39  & \ldots  & \ldots  &  Columba assoc.           & 1                     &  30                        &  1      & \ldots    \\
   HD 15060  & \ldots  & \ldots  & \ldots  &  \ldots                   & \ldots                & 2300$\pm$100               &  6      &  1   \\
   HD 15115  &  $-$4.93  & \ldots  & \ldots  &  $\beta$\,Pic mg.         & 3,4                   &  12                        &  1      & \ldots    \\
   HD 15745  & \ldots  & \ldots  & \ldots  &  $\beta$\,Pic mg.         & 1                     &  12                        &  1      & \ldots    \\
   HD 16743  & \ldots  & \ldots  & \ldots  &  \ldots                   & \ldots                & 10--50                     &  2,3,5  & \ldots    \\
   HD 17390  &  $-$5.15  & \ldots  & \ldots  &  \ldots                   & \ldots                & 1000$^{+300}_{-400}$       &  6      &  1   \\
   HD 24636  &  $-$5.41  & \ldots  & \ldots  &  Tucana-Horologium assoc. & 1                     &  30                        &  1      & \ldots    \\
   HD 25570  &  $-$5.24  & \ldots  & \ldots  &  Hyades?                  & see Sect.~\ref{agedet}& 625$\pm$50                 &  1      & \ldots   \\
   HD 30447  & \ldots  & \ldots  & \ldots  &  Columba assoc.           & 3,4                   &  30                        &  1      & \ldots    \\
   HD 32195  &  $-$3.92  & \ldots  & \ldots  &  Tucana-Horologium assoc. & 4,5                   &  30                        &  1      & \ldots    \\
   HD 33081  & \ldots  & \ldots  & \ldots  &  \ldots                   & \ldots                & 3100$^{+400}_{-500}$       &  6      &  1   \\
   HD 35114  & $-$3.86$^*$  & \ldots  & \ldots  &  Columba assoc.           & 4                     &  30                        &  1      & \ldots    \\
   HD 35841  & \ldots  & \ldots  & \ldots  &  Columba assoc.           & 3,4                   &  30                        &  1      & \ldots    \\ 
   HD 36968  & \ldots  & \ldots  & \ldots  &  Octans assoc.            & 1                     &  20                        &  1      & \ldots    \\
   HD 50571  &  $-$5.27  & $-$4.55   &   1     &  B3 group                 & 1                     &300$\pm$120                 &  1,6    &  3    \\
  HD 113337  &$-$5.07$^*$& \ldots  & \ldots  &  \ldots                   & \ldots                & 40$\pm$20                  &  2      & \ldots    \\
  HD 120160  & \ldots  & \ldots  & \ldots  &  \ldots                   & \ldots                & 1300$\pm$100               &  6      &  1   \\
  HD 125451  &  $-$5.10  &  $-$4.37  &   3     &  Ursa Major mg.           & 2                     &500$\pm$100                 &  1      & \ldots    \\
  HD 127821  &  $-$5.11  & \ldots  & \ldots  &  \ldots                   & \ldots                &220$\pm$50                  &  6      &  2   \\
  HD 151044  & \ldots  &  $-$5.0   &   4     &  \ldots                   & \ldots                & 3000$^{+1300}_{-1000}$     &  2,3    & \ldots    \\
  HD 170773  &$-$4.98$^*$&  $-$4.39  &   1     &  \ldots                   & \ldots                & 200                        &  6      &  3   \\
  HD 192758  & \ldots  & \ldots  & \ldots  &  Argus assoc.             & 3                     &  40                        &  1      & \ldots    \\
  HD 205674  &  $-$5.13  & \ldots  & \ldots  &  ABDor mg.?               & \ldots                & 70--300                    &  1,6    &  3   \\
  HD 206893  &$-$4.87$^*$&  $-$4.47  &   1     &  \ldots                   & \ldots                & 200$^{+1000}_{-200}$       &  6      &  1,3   \\
  HD 213429  & \ldots  &  $-$4.83  &   2     &  \ldots                   & \ldots                & 2200$^{+1300}_{-800}$      &  3,6    &  1   \\ 
  HD 213617  & \ldots  & \ldots  & \ldots  &  \ldots                   & \ldots                & 1200$\pm$300               &  6      &  1   \\ 
  HD 221853  & \ldots  & \ldots  & \ldots  & Local Association         & 3                     & 20--150                    &  1      & \ldots    \\
\enddata                                                                                                                                                                           
 \tablecomments{Col.(1): Identification.                                                                                                                                            
 Col.(2): Fractional X-ray luminosity based on ROSAT data. 
 Asterisks indicate those objects where the correlation between the X-ray source and the star is confirmed by observations with the XMM satellite 
 \citep[XMM-Newton slew survey Source Catalogue, ][]{saxton2008} as well.
 Col.(3): Fractional Ca\,II H\&K luminosity ($\log{R'_{\rm HK}}$). 
 Col.(4): References for $\log{R'_{\rm HK}}$ data: 1) \citet{2006AJ....132..161G}, 2) \citet{gray2003}; 3) \citet{king2005}; 4) \citet{wright2004}.                                                                                                    
 Col.(5). Membership status of the star.
 Col.(6). References for the identification of star as a member of a specific kinematic group in Col.(5): 
 1) this work; 2) \citet{king2003}; 3) \citet{moor2006}; 4) \citet{torres2008}; 5) \citet{zs04b}.
 Col.(7). Estimated age of the star and its formal uncertainty.
 Col.(8). Used age dating methods: 1) stellar kinematic group or cluster membership; 2) isochrone fitting; 3) diagnostic of chromospheric activity indicator; 
 4) diagnostic of coronal activity indicator; 5) lithium abundance; 6) literature data.  
 Col.(9). References for the literature data mentioned in Col.(7-8): 1) \citet{HOLMBERG}; 2) Mo\'or et al., 2010b, in prep.; 3) \citet{rhee2007}.                                                                                                                                                                                                                                                                                                              
}                                                                                                                                                                                   
\end{deluxetable} 

\clearpage


\clearpage

\end{document}